\documentclass[3p]{elsarticle}

\usepackage{amsmath}
\usepackage{amsfonts}
\usepackage{amssymb}
\usepackage{commath}
\usepackage{xcolor}

\renewcommand{\secref}[1]{{section \ref{#1}}}
\renewcommand{\emph}[1]{\textit{\color{red}{#1}}}
\newcommand{\erf}{\textup{erf}}
\newcommand{\sign}{\textup{sign}}
\newcommand{\vcb}[1]{\mathbf{#1}}
\newcommand{\mat}[1]{\mathbb{#1}}
\newcommand{\dint}{\displaystyle\int}

\newcommand{\tens}[1]{\overline{\overline{#1}}}
\newcommand{\spc}[1]{\mathcal{#1}}
\newcommand{\tr}[1]{\textup{T}{#1}}

\journal{Journal of Computational Physics}
\begin{document}

\begin{frontmatter}
\title{High-Order Curvilinear Finite Element Magneto-Hydrodynamics\\ I: A Conservative Lagrangian Scheme}

\author[ELI,IPP,FNSPE]{Jan Nikl\corref{corrauth}}
\cortext[corrauth]{Corresponding author}
\ead{jan.nikl@eli-beams.eu}

\author[FNSPE]{Milan Kuchařík}

\author[ELI,XIAN]{Stefan Weber}

\address[ELI]{ELI Beamlines Centre, Institute of Physics, Czech Academy of Sciences, 25241 Dolní Břežany, Czech Republic}
\address[IPP]{Institute of Plasma Physics, Czech Academy of Sciences, 18200 Prague, Czech Republic}
\address[FNSPE]{Faculty of Nuclear Sciences and Physical Engineering, Czech Technical University in Prague, 11519 Prague, Czech Republic}
\address[XIAN]{School of Science, Xi'an Jiaotong University, 710049 Xi'an, China}

\begin{abstract}
Magneto-hydrodynamics is one of the foremost models in plasma physics with applications in inertial confinement fusion, astrophysics and elsewhere. Advanced numerical methods are needed to get an insight into the complex physical phenomena. The classical Lagrangian methods are typically limited to the low orders of convergence and suffer from violation of the divergence-free condition for magnetic field or conservation of the invariants. This paper is the first part of a new series about high-order non-ideal magneto-hydrodynamics, where a multi-dimensional conservative Lagrangian method based on curvilinear finite elements is presented. The condition on zero divergence of magnetic field and conservation of mass, momentum, magnetic flux and the total energy are satisfied exactly. The curvilinear elements prevent entangling of the computational mesh and its imprinting into the solution. A high-order conservative time integration is applied, where an arbitrary order of convergence is attained for problems of ideal magneto-hydrodynamics. The resistive magnetic field diffusion is solved by an implicit scheme. Description of the method is given and multiple test problems demonstrating properties of the scheme are performed. The construction of the method and possible future directions of development are discussed.
\end{abstract}


\end{frontmatter}


\section{Introduction}
\label{sec_intro}

The magneto-hydrodynamic description is one of the major directions in plasma physics research. It plays a central and indispensable role in such as design of fusion devices or astrophysics and many others. We are particularly interested in the Lagrangian approach, where the coordinate system follows the flow of the fluid. It is mostly favored in areas of modeling where the medium undergoes extreme compression or expansion like inertial confinement fusion \cite{Perkins2017,Clark2015}, Z-pinches \cite{Wu2018}, supernovae collapses \cite{Livne2004}, cosmic jets \cite{Pudritz2012}, prepulse effects of ultra-intense lasers \cite{Nikl2019spie,Holec2018b} or laser ion acceleration beamlines \cite{Psikal2021,Psikal2019spie,Batani2010} as are explored at facilities like ELI Beamlines \cite{Weber2017,ELIBeamlines} and other multi-PW laser systems worldwide \cite{Danson2019}. 

Efficient and accurate numerical tools are then needed for simulations of the physical phenomena occurring within these processes. However, the classical methods for resistive magneto-hydrodynamics (RMHD) are typically limited to low orders of convergence. New high-order methods are slowly appearing for the Eulerian approach \cite{Balsara2015,Susanto2013}, but the literature is scarce for Lagrangian methods. A change of the paradigm comes with application of the finite element method (FEM) \cite{Rieben2007}, which offers generality of the formulation in the mesh topology and also order of the elements. Although the latter option is not pursued by the authors due to the limitations in time integration and conservation properties primarily, another advantage of the design presents itself. The edge-centered elements equipped with an appropriate transformation rule perfectly suit the magnetic field discretization and exactly satisfy the divergence-free constraint, which poses a serious problem in numerical MHD \cite{Toth2000}. Another milestone presents the development of the conservative high-order hydrodynamics based on the curvilinear finite elements \cite{Dobrev2010}, which introduced geometric mass conservation for the curved elements and an arbitrary order of convergence in space given by the polynomial order of the elements. However, conservative time stepping was still limited to the second order only and formulation of MHD was missing. Later, a convenient formulation of conservative Lagrangian MHD was developed for the staggered discretization \cite{Wu2018}. Although limited to the second order and only to the transverse component of the magnetic field in 2D, the methodology is compatible with the finite element hydrodynamics.

This paper continues in these efforts and presents a multi-dimensional conservative high-order resistive magneto-hydrodynamics based on the curvilinear finite elements. Mass, momentum, magnetic flux and the total energy are conserved exactly. In addition, the divergence-free condition for the magnetic field is maintained exactly. Generality of the formulation allows to apply it on unstructured meshes, though the numerical examples are presented only for regular quadrilateral/hexahedral meshes. Together with a conservative high-order time integration, the numerical scheme attains an arbitrary order of convergence for ideal MHD problems theoretically. On the other hand, stability for the problems of resitive magnetodynamics is provided by an implicit scheme of magnetic diffusion. These essential parts form together the foundations of the new series of papers about the high-order curvilinear finite element MHD. The numerical implementation is part of the multi-physics code PETE2 \cite{Nikl2018eps}, which continues our previous efforts \cite{Nikl2018,Holec2018}.

The remainder of the paper is organized as follows. A brief introduction to the physical model and the governing equations of RMHD are given in \secref{sec_phys}. The numerical scheme is then described in \secref{sec_num}, starting from the weak formulation of the problem (\secref{sub_num_weak}) and continuing with the discretization in space (\secref{sub_num_semi}) to the fully discrete model (\secref{sub_num_disc}). Next, the numerical method is validated and benchmarked on a set of test problems in \secref{sec_sim}. A short discussion of benefits and limitations of the model is given in \secref{sec_disc}, followed by concluding remarks in \secref{sec_concl}. The future directions of this series of papers are outlined.


\section{Physical model}
\label{sec_phys}

Within the classical resistive magneto-hydrodynamics, the physical system is described as a magnetized fluid with the mass density $\rho$, mass-averaged velocity $\vec{u}$ and specific internal energy $\varepsilon$. Its dynamics is modeled by the Euler equations, which are complemented by Faraday's law for the self-consistent magnetic field $\vec{B}$, which is assumed to satisfy Gauss's law for magnetism $\nabla\cdot\vec{B}=0$. In the Lagrangian formulation, they take the following integral form \cite{Wu2018}:
\begin{subequations}
\begin{align}
\dod{}{t}\int_V \rho \dif{V} &= 0
,\label{eqn_phys_int_mass}\\
\dod{}{t}\int_V \rho \vec{u} \dif{V} &= \oint_{\partial V} (\tens{\sigma} + \tens{\sigma}_B)\cdot\dif{\vec{S}}
,\label{eqn_phys_int_mom}\\
\dod{}{t}\int_S \vec{B}\cdot\dif{\vec{S}} &= - \oint_{\partial S} \vec{E} \cdot \dif{\vec{l}}
,\label{eqn_phys_int_B}\\
\dod{}{t}\int_V \rho \varepsilon \dif{V} &= \int_V\left( \tens{\sigma}:\nabla\vec{u} + \vec{j}\cdot\vec{E} \right)\dif{V}
,\label{eqn_phys_int_eps}\\
\dod{}{t}\int_V \frac{B^2}{2\mu_0} \dif{V} &= \int_V\left( \tens{\sigma}_B:\nabla \vec{u} - \frac{1}{\mu_0} \vec{B}\cdot\nabla\times\vec{E} \right)\dif{V}
,\label{eqn_phys_int_epsB}
\end{align}
\label{eqn_phys_int}
\end{subequations}
\noindent where the integration is performed in the frame moving with the flow of the fluid ($\vec{E}$ is the fluid-frame electric field). In particular, $V$ is the volume with the elements $\dif{V}$, $\dif{\vec{S}}$ are the (outward) oriented surface elements and $\dif{\vec{l}}$ are the oriented curve elements. The permeability of free space $\mu_0$ appears in the equations, since magnetic fields are considered to originate only from the free currents $\vec{j}$. They are purely solenoidal due to the assumed quasi-neutrality of the fluid, governed by the electrostatic Amp\`{e}re's law $\vec{j}=1/\mu_0\nabla\times\vec{B}$. The model of resistive MHD then closes the system for the electric field $\vec{E}$ by Ohm's law $\vec{E}=\eta\vec{j}$, where $\eta$ is the electric resistivity. Note that also the Joule heating term $\vec{j}\cdot\vec{E}$ is included in the description, exchanging energy between the fluid and magnetic field. Finally, the closure relations determine the value of the stress tensor $\tens{\sigma}$ as a function of the thermodynamic potentials and the symbol $\tens{\sigma}_B$ represents the magnetic stress tensor $\tens{\sigma}_B=1/\mu_0(\vec{B}\otimes\vec{B}-1/2\vec{B}^2 \tens{I})$, where $\tens{I}$ is the unit tensor.

The system of equations \eqref{eqn_phys_int} reveals the fundamental conservation relations, which are respectively corresponding to the following quantities: mass, momentum, magnetic flux and internal energy.
The additional equation for the magnetic energy \eqref{eqn_phys_int_epsB} can be seen as a consequence of Faraday's law and the Lorentz transformation of the laboratory-frame electric field $\vec{E}'$, which becomes $\vec{E} = \vec{E}' + \vec{u}\times\vec{B}$ in the non-relativistic limit ($\abs{\vec{u}} \ll c$, where $c$ is the speed of light). Therefore, this equation can be considered as an auxiliary part of the formulation, but it is essential for the construction of the energy-conserving numerical model as presented in \secref{sec_num}. Together, equations \eqref{eqn_phys_int_mom}, \eqref{eqn_phys_int_eps} and \eqref{eqn_phys_int_epsB} yield conservation of the total energy on a Lipschitz domain $\Omega$:
\begin{multline}
\int_\Omega\left( \rho \varepsilon + \frac{B^2}{2\mu_0} + \frac{1}{2}\rho\vec{u}^2 \right)\dif{V}
= \int_\Omega\left( (\tens{\sigma} + \tens{\sigma}_B) : \nabla\vec{u} + \frac{1}{\mu_0}(\nabla\times\vec{B})\cdot\vec{E} - \frac{1}{\mu_0}\vec{B}\cdot(\nabla\times\vec{E}) + \vec{u} \cdot \nabla\cdot(\tens{\sigma} + \tens{\sigma}_B) \right)\dif{V}
=\\=
\oint_{\partial \Omega} \vec{u} \cdot (\tens{\sigma}+\tens{\sigma}_B) \cdot \vec{n} \dif{S} - \oint_{\partial\Omega} \frac{1}{\mu_0}(\vec{E}\times\vec{B}) \cdot \dif{\vec{S}}
= 0.
\label{eqn_phys_en}
\end{multline}
The last equality assumes the boundary conditions for zero Poynting vector and normal forces ($\vec{n}$ is the outer normal) or a non-moving boundary.

For the numerical solution, the differential formulation is more suitable. The classical Reynolds transport theorem states for a field $\vec{A}=\vec{A}(t,\vec{x})$ on the domain $\Omega(t)$ moving with velocity $\vec{v}$:
\begin{equation}
\od{}{t} \int_{\Omega(t)} \vec{A} \dif{V}
= \int_{\Omega(t)} \left( \dpd{}{t}\vec{A} + (\vec{v}\cdot\nabla)\vec{A} + (\nabla\cdot\vec{v})\vec{A} \right)\dif{V}
= \int_{\Omega(t)} \left( \dod{\vec{A}}{t} + (\nabla\cdot\vec{v})\vec{A} \right)\dif{V}
.\label{eqn_phys_reynolds}
\end{equation}
This formula can be generalized for (partially) solenoidal fields on the moving boundary of a surface $\Gamma(t)$ in the following way \cite{Lax1976}:
\begin{equation}
\od{}{t}\int_{\Gamma(t)} \vec{A} \cdot \dif{\vec{S}}
= \int_{\Gamma(t)} \left(\dpd{}{t}\vec{A} + (\nabla\cdot\vec{A})  \vec{v} + \nabla\times(\vec{A}\times\vec{v}) \right) \cdot\dif{\vec{S}}
= \int_{\Gamma(t)} \od{\vec{A}}{t} \cdot\dif{\vec{S}}
.\label{eqn_phys_reynolds_sol}
\end{equation}
Together with the gauge relation in the form of Gauss' law for the magnetic field $\nabla\cdot\vec{B}=0$, this enables to rewrite the system \eqref{eqn_phys_int} in the differential form for the Lagrangian frame:
\begin{subequations}
\begin{align}
\dod{\rho}{t} &= -\rho\nabla\cdot\vec{u}
,\label{eqn_phys_dif_mass}\\
\rho\dod{\vec{u}}{t} &= \nabla\cdot(\tens{\sigma} + \tens{\sigma}_B)
,\label{eqn_phys_dif_mom}\\
\dod{\vec{B}}{t} &= -\nabla\times\vec{E}
,\label{eqn_phys_dif_B}\\
\rho\dod{\varepsilon}{t} &= \tens{\sigma}:\nabla\vec{u} + \vec{j}\cdot\vec{E}
,\label{eqn_phys_dif_eps}\\
\rho\dod{\varepsilon_B}{t} &= \tens{\sigma}_B:\nabla\vec{u} - \frac{1}{\mu_0}\vec{B}\cdot\nabla\times\vec{E}
,\label{eqn_phys_dif_epsB}
\end{align}
\label{eqn_phys_dif}
\end{subequations}
where $\rho\varepsilon_B=B^2/(2\mu_0)$ is the magnetic energy. This formulation assumes a full spatial dependency of the vector fields $\vec{E}=\vec{E}(t,\vec{x})$ and $\vec{B}=\vec{B}(t,\vec{x})$. However, the model is designed as multi-dimensional and it must be distinguished between the coplanar/collinear and transverse components in 1D and 2D. The reduced system in these lower dimensions is described in \ref{app_phys_dif}.


\section{Numerical model}
\label{sec_num}

The system of equations \eqref{eqn_phys_dif} together with the closure relations is solved numerically by the scheme described in this section. Its construction is based on the finite element method (FEM), which provides flexibility in dimensionality of the problem, topology of the mesh and the choice of the polynomial orders of the discretization. The latter feature gives proportional convergence rates in turn, as shown in \secref{sec_sim}. Moreover, the construction of the finite element spaces is consistent with the solved physical problem of \secref{sec_phys}, satisfying the divergence-free constraint for the magnetic field and point-wise mass conservation by definition, as explained in the following subsections. The latter is related to the fact that curvilinear finite elements are used, which fit the Lagrangian framework and enable to track the motion of free boundaries and discontinuities in detail. However, the most appealing property of the new scheme is simultaneous conservation of all velocity moments, i.e, mass, momentum and energy for arbitrary orders of the elements.


\subsection{Weak formulation}
\label{sub_num_weak}

In order to derive a consistent finite element scheme, the considered system \eqref{eqn_phys_dif} is written in the variational form. The problem is formulated as mixed, where different functional spaces are used for approximation of the physical quantities. The three-dimensional model distinguishes the following spaces:
\begin{itemize}
\item thermodynamic ($\spc{T}$) --  $\spc{T} \subset L_2(\Omega)$,
\item kinematic ($\spc{K}$) -- $\spc{K} \subset (H^1(\Omega))^3$,
\item magnetic ($\spc{M}$) -- $\spc{M}\subset H_{div}(\Omega)$,
\item electric ($\spc{E}$) -- $\spc{E}\subset H_{curl}(\Omega)$.
\end{itemize}
The spaces follow the standard notation, where $L_2$ is the Lebesgue space of square integrable functions, $H^1$ is the Sobolev space of differentiable functions ($\nabla \psi\in (L_2(\Omega))^3, \forall \psi\in H^1(\Omega)$), $H_{div}$ the Sobolev space of divergence-equipped functions ($\nabla\cdot\vec{\Xi}\in L_2(\Omega), \forall \vec{\Xi}\in H_{div}(\Omega)$) and $H_{curl}$ the Sobolev space of curl-equipped functions ($\nabla\times\vec{\xi}\in (L_2(\Omega))^3, \forall \vec{\xi}\in H_{curl}(\Omega)$). The solution of the system is then assumed to lay in the given spaces, specifically, $\vec{u}\in\spc{K}, \vec{B}\in\spc{M}, \vec{E}\in\spc{E}$ and $\varepsilon, \varepsilon_B\in \spc{T}$. Note that the density $\rho$ is not included in the list, since it does not represent a primary variable here due to the geometric conservation law (GCL) for mass on curvilinear elements, as discussed in \secref{sub_num_semi}. After multiplication by the test functions and imposing the smoothness requirements, the weak formulation of the system \eqref{eqn_phys_dif} (without the mass equation \eqref{eqn_phys_dif_mass}) together with Ohm's law can be obtained by integration:
\begin{subequations}
\begin{align} 
\dint_\Omega \rho \dod{\vec{u}}{t}\cdot \vec{\psi} \dif{V} &=
-\dint_\Omega (\tens{\sigma} + \tens{\sigma}_B) : \nabla \vec{\psi} \dif{V}
+\int_{\Gamma_\sigma} \vec{\sigma}_n \cdot \vec{\psi} \dif{S}
, & \forall\vec{\psi}\in\spc{K}
,\label{eqn_num_weak_mom}\\
\dint_\Omega \dod{\vec{B}}{t}\cdot \vec{\Xi} \dif{V} &=
- \dint_\Omega (\nabla\times\vec{E}) \cdot \vec{\Xi} \dif{V}
, & \forall\vec{\Xi}\in\spc{M}
,\label{eqn_num_weak_B}\\
\dint_\Omega \frac{1}{\eta} \vec{E} \cdot \vec{\xi} \dif{V} &=
\dint_\Omega \frac{1}{\mu_0} \vec{B}\cdot(\nabla\times\vec{\xi}) \dif{V}
-\int_{\Gamma_B} \frac{1}{\mu_0}\vec{B}^\tau \cdot \vec{\xi} \dif{S}
, & \forall\vec{\xi}\in\spc{E}
,\label{eqn_num_weak_E}\\
\dint_\Omega \rho \dod{\varepsilon}{t} \varphi \dif{V} &=
\dint_\Omega\left( \tens{\sigma}:\nabla\vec{u} \varphi 
+ \frac{1}{\mu_0}\vec{B}\cdot(\nabla\times\vec{E})\varphi
+ \frac{1}{\mu_0}(\vec{E}\times\vec{B})\cdot\nabla\varphi
\right)\dif{V}
+\notag\\ & \phantom{=}
+\int_{\Gamma_E} \frac{1}{\mu_0} \vec{E}^\tau \cdot \tr{}^\tau{\vec{B}} ~\tr{\varphi} \dif{S}
-\int_{\Gamma_B} \frac{1}{\mu_0} \vec{E} \cdot \vec{B}^\tau ~\tr{\varphi} \dif{S}
, & \forall\varphi\in\spc{T}
,\label{eqn_num_weak_eps}\\
\dint_\Omega \rho\dod{\varepsilon_B}{t} \varphi \dif{V} &= \dint_\Omega\left( \tens{\sigma}_B:\nabla\vec{u}\varphi - \frac{1}{\mu_0}\vec{B}\cdot(\nabla\times\vec{E})\varphi \right)\dif{V}
, & \forall\varphi\in\spc{T}
,\label{eqn_num_weak_epsB}
\end{align}
\label{eqn_num_weak}
\end{subequations}
where $\tr{}$ represents the trace operator for the functions from $\spc{T}$ on the boundary (or its sub-parts $\Gamma_\sigma, \Gamma_E, \Gamma_B\subset\partial\Omega$). Similarly, $\tr{}^\tau$ is the trace of the tangential component of functions from $\spc{M}$. A special treatment must be applied to the energy equations \eqref{eqn_num_weak_eps} and \eqref{eqn_num_weak_epsB} to give them a mathematical sense. The integrals of products of three functions in $L_2(\Omega)$ may not converge in general and an additional constraint on uniform boundedness of $\varphi$ can be imposed to cure the problem, as the functions are piece-wise polynomial in the discrete scheme of \secref{sub_num_semi} and the areas of the elements can be assumed as bounded too. Secondly, the gradient of $\varphi$ is interpreted in the sense of generalized functions, where a finite number of removable discontinuities is present, on which the surface integrals can be evaluated. This is again motivated by the practical construction appearing in \secref{sub_num_semi} and the details can be found there.
The formulation in the lower dimensions can be derived analogously and is presented in \ref{app_num_weak} alongside the choices of the functional spaces. However, the construction is consistent in all dimensions, maintaining identical overall properties.

The boundary conditions considered in \eqref{eqn_num_weak} are the following:
\begin{subequations}
\begin{align}
\vec{u} &= 0 & \text{on}~&\Gamma_u, &
(\tens{\sigma}+\tens{\sigma}_B) \cdot \vec{n} &= \vec{\sigma}_n & \text{on}~&\Gamma_\sigma, &
\overline{\Gamma_u} \cup \overline{\Gamma_\sigma} &=\partial\Omega, &
\Gamma_u \cap \Gamma_\sigma &= \emptyset
,\label{eqn_num_weak_bc_mom}\\
\vec{B}\times\vec{n} &= \vec{B}^\tau & \text{on}~&\Gamma_B, &
\vec{E}\times\vec{n} &= \vec{E}^\tau & \text{on}~&\Gamma_E, &
\overline{\Gamma_B} \cup \overline{\Gamma_E} &= \partial \Omega, & \Gamma_B\cap\Gamma_E &= \emptyset
.\label{eqn_num_weak_bc_E}
\end{align}
\label{eqn_num_weak_bc}
\end{subequations}
The condition on $\Gamma_u$ sets the non-moving (no-slip) boundary, $\Gamma_\sigma$ prescribes the normal forces, $\Gamma_B$ and $\Gamma_E$ determine the tangential components of the magnetic and electric field, respectively. The boundary data are assumed to lay in $L_2$ on the corresponding boundary parts, while it is required that $\vec{E}^\tau\in C^1(\overline{\Gamma_E})$ in order to be consistent with the definition of $\spc{E}$ ($\overline{\Gamma_E}$ is the closure of $\Gamma_E$). 
It should be noted that the boundary conditions on $\Gamma_u$ and $\Gamma_E$ are essential, meaning that they are imposed through the definitions of the corresponding spaces, which are continuous (or their components) to the boundaries. 

The choices of the functional spaces and the resulting weak form \eqref{eqn_num_weak} reveal the essential aspects of the construction. Following the paradigm of the high-order curvilinear finite element hydrodynamics \cite{Dobrev2010}, the thermodynamic potentials are allowed to be discontinuous. This enables to maintain a sharp, non-oscillating solution across physical discontinuities like propagating shocks, for example. Furthermore, the weak formulation of the momentum equation introduces a symmetry between \eqref{eqn_num_weak_mom} and the energy equations \eqref{eqn_num_weak_eps} and \eqref{eqn_num_weak_epsB} in the $\tens{\sigma}:\nabla\vec{\psi}$ and $\tens{\sigma}_B:\nabla\vec{\psi}$ terms. This, in turn, leads to discrete conservation of kinetic energy, as shown in \secref{sub_num_semi}. In addition, the weak form of Joule heating  $\vec{j}\cdot\vec{E}$ provides symmetry between the energy equations \eqref{eqn_num_weak_eps} and \eqref{eqn_num_weak_epsB}, which yields conservation of discrete internal and magnetic energy. However, it is crucial to note that this term is still consistent with the strong Faraday's law \eqref{eqn_num_weak_B}, so the correspondence between the magnetic field and the magnetic energy is preserved. Finally, the choice of the spaces $\spc{M}$ and $\spc{E}$ respects the natural transformation rules of the fields, i.e., constant magnetic flux and constant circulation of electric field over any closed surface/curve in Lagrangian coordinates \cite{Rieben2007}. This point becomes essential for the construction of the scheme on the curvilinear elements in \secref{sub_num_semi}, where constant divergence of $\vec{B}$ and curl of $\vec{E}$ are maintained, despite the deformation. Moreover, the electromagnetic spaces are mutually compatible in the sense of the exact sequences for Sobolev spaces, where differential operators traverse between them. The electrostatic Maxwell's equations \eqref{eqn_num_weak_B} and \eqref{eqn_num_weak_E} follow such de Rham complex, which takes the following form in 3D~\cite{Arnold2006}:
\begin{equation}
L_2 \stackrel{\nabla\cdot}{\longleftarrow}
H_{div} \stackrel{\nabla\times}{\longleftarrow}
H_{curl} \stackrel{\nabla}{\longleftarrow}
H^1
.\label{eqn_num_weak_rham}
\end{equation}
Therefore, it holds $\nabla\times\vec{E}\in\spc{M}$ in Faraday's law \eqref{eqn_num_weak_B} exactly (for appropriately chosen spaces) and the divergence-free structure of $\vec{B}$ is not affected, as $\nabla\cdot\nabla\times\vec{\xi}=0$ for $\xi\in\spc{E}$ is exact too. Generally, the symmetric formulation of Maxwell's equations then provides a good conditioning of the discrete problem and consequently prevents locking of the solution (spatially constant numerical values appearing for over-determined schemes) \cite{Rieben2006,Brezzi1991}.

Similar sequences exist in 2D, where the de Rham diagram becomes:
\begin{equation}
L_2 \stackrel{\nabla\cdot}{\longleftarrow}
H_{div} \stackrel{\nabla_\parallel\times}{\longleftarrow}
H^1
,\qquad
L_2 \stackrel{\nabla_\perp\times}{\longleftarrow}
H_{curl} \stackrel{\nabla}{\longleftarrow}
H^1
.\label{eqn_num_weak_rham_2D}
\end{equation}
The operators for different components (coplanar and transverse) of the fields follow the notation of \ref{app_phys_dif}. From the diagram, it is clear that the divergence-free structure of $\vec{B}_\parallel \in \spc{M}_\parallel \subset H_{div}(\Omega)$ is also maintained and the transformation of $\vec{E}_\parallel\in\spc{E}_\parallel \subset H_{curl}(\Omega)$ is physically consistent too, following the definition of the spaces in \ref{app_num_weak_2D}.

In one dimension, the de Rham complex simplifies even further:
\begin{equation}
(L_2)^2 \stackrel{\nabla_\perp\times}{\longleftarrow}
(H^1)^2
,\label{eqn_num_weak_rham_1D}
\end{equation}
where it should be noted that the curl operator is a mere antisymmetric variation of gradient. Consequently, the 1D Faraday's law \eqref{eqn_app_num_weak_1D_Bp} for $\vec{B}_\perp\in\spc{M}_\perp\subset (L_2)^2$ and $\vec{E}_\perp\in\spc{E}_\perp\subset (H^1)^2$ holds exactly too.


\subsection{Semi-discrete model}
\label{sub_num_semi}

Construction of the numerical scheme for solution of the weak formulation \eqref{eqn_num_weak} continues with definitions of the finite element spaces and inference of a general discrete system in space, whereas the time discretization and specific choices of the functional bases are deferred to \secref{sub_num_disc}.

In order to construct the spatial discretization on the curvilinear finite elements, the notion of the moving mesh must be defined:
\begin{equation}
\Omega(t) = \lbrace \vec{x}(t,\vec{X}) ~|~ \vec{X}\in \Omega_0 \rbrace
,\label{eqn_num_semi_omega}
\end{equation}
where $\vec{x}(t,\vec{X})$ is understood in this context as a time-dependent map relating the material coordinates $\vec{X}\in\Omega_0$ and the fluid element coordinates $\vec{x}\in\Omega(t)$. The material domain $\Omega_0=\Omega(0)$ is taken as the initial one for simplicity, so it can be asserted $\vec{x}(0, \vec{X})=\vec{X}$. The moving volume element coordinates $\vec{x}(t,\vec{X})$ follow the characteristics of the solution, being governed by the equation of motion $\dif{\vec{x}}/\dif{t}=\vec{u}$.

On the (static) Lipschitz domain $\Omega_0$, the tessellation $\Sigma_h$ is considered, where the set of all internal edges is then denoted as $\Upsilon_h$. A conforming discretization with the definitions in \secref{sub_num_weak} is made here, i.e., the finite dimensional subspaces $\spc{T}_h\subset\spc{T}, \spc{K}_h\subset\spc{K}, \spc{M}_h\subset\spc{M}, \spc{E}_h\subset\spc{E}$ are defined on $\Omega_0$. The base functions are then named as $\varphi_i\in\spc{T}_h$ for $i\in\lbrace 1,\ldots N_\spc{T}\rbrace$, $\vec{\psi}_i\in\spc{K}_h$ for $i\in\lbrace 1,\ldots N_\spc{K}\rbrace$, $\vec{\Xi}_i\in\spc{M}_h$ for $i\in\lbrace 1,\ldots N_\spc{M}\rbrace$ and $\vec{\xi}_i\in\spc{E}_h$ for $i\in\lbrace 1,\ldots N_\spc{E}\rbrace$. The primary variables are approximated by the grid functions designated in bold face as follows:
\begin{subequations}
\begin{align}
\vec{x}(t,\vec{X}) &\approx \vec{x}_h(t,\vec{X}) = \textstyle\sum_{i=1}^{N_\spc{K}} \vcb{x}_i(t) \vec{\psi}_i(\vec{X})
, &
\vec{u}(t,\vec{X}) &\approx \vec{u}_h(t,\vec{X}) = \textstyle\sum_{i=1}^{N_\spc{K}} \vcb{u}_i(t) \vec{\psi}_i(\vec{X})
,\label{eqn_num_semi_approx_xue}\\
\vec{B}(t,\vec{X}) &\approx \vec{B}_h(t,\vec{X}) = \textstyle\sum_{i=1}^{N_\spc{M}} \vcb{B}_i(t) \vec{\Xi}_i(\vec{X})
, &
\vec{E}(t,\vec{X}) &\approx \vec{E}_h(t,\vec{X}) = \textstyle\sum_{i=1}^{N_\spc{E}} \vcb{E}_i(t) \vec{\xi}_i(\vec{X})
,\label{eqn_num_semi_approx_BE}\\
\varepsilon(t,\vec{X}) &\approx \varepsilon_h(t,\vec{X}) = \textstyle\sum_{i=1}^{N_\spc{T}} \vcb{e}_i(t) \varphi_i(\vec{X})
, &
\varepsilon_B(t,\vec{X}) &\approx {\varepsilon_B}_h(t,\vec{X}) = \textstyle\sum_{i=1}^{N_\spc{T}} {\vcb{e}_B}_i(t) \varphi_i(\vec{X})
.\label{eqn_num_semi_approx_eeB}
\end{align}
\label{eqn_num_semi_approx}
\end{subequations}

The definitions of the specific finite element spaces is deferred to \secref{subsub_num_disc_spc} to preserve generality of the construction, but it must be noted that they are piece-wise polynomial on the elements in all cases, which satisfy the continuity and smoothness requirements $C^0(\overline{\Omega_e})\cap C^1(\Omega_e)$ for all elements $\Omega_e\in\Sigma_h$. Following the Galerkin approach, the test functions are chosen from the identical spaces as the approximations of the primary variables \eqref{eqn_num_semi_approx}. The resulting vectors, matrices and tensors after integration of the base functions over the domain $\Omega(t)$ are defined in \ref{app_num_semi_mats}. The semi-discrete form of the weak formulation \eqref{eqn_num_weak} is then obtained:
\begin{subequations}
\begin{align}
\dod{\vcb{x}}{t} & = \vcb{u}
,\label{eqn_num_semi_x}\\
\mat{M}_\spc{K} \dod{\vcb{u}}{t} &= -(\mat{F} + \mat{F}_B)\vcb{1} + \vcb{b}_\sigma
,\label{eqn_num_semi_mom}\\
\dod{\vcb{B}}{t} &= -\mat{C}_D \vcb{E}
,\label{eqn_num_semi_B}\\
\mat{M}_\spc{E} \vcb{E} &= \frac{1}{\mu_0} \mat{C}_{\cdot jk} \vcb{B}_j \vcb{1}_k + \mat{X}_B^T\vcb{1}
,\label{eqn_num_semi_E}\\
\mat{M}_\spc{T} \dod{\vcb{e}}{t} &= \mat{F}^T\vcb{u} + \frac{1}{\mu_0} \mat{C}_{ij\cdot} \vcb{E}_i\vcb{B}_j
+ \mat{S}_{ij\cdot}\vcb{E}_i\vcb{B}_j
+ \mat{X}_E \vcb{B} + \mat{X}_B \vcb{E}
+ \vcb{e}_B^c
,\label{eqn_num_semi_eps}\\
\mat{M}_\spc{T} \dod{\vcb{e}_B}{t} &= \mat{F}_B^T\vcb{u} -\frac{1}{\mu_0} \mat{C}_{ij\cdot} \vcb{E}_i\vcb{B}_j
.\label{eqn_num_semi_epsB}
\end{align}
\label{eqn_num_semi}
\end{subequations}
Note that summation is performed over the repeated indexes and the dot index denotes contraction of the tensor over the other two indexes, producing a vector. The vector $\vcb{1}$ is the grid function corresponding to the unity in $\spc{T}_h$.
The newly appearing term $\vcb{e}_B^c$ is the correction term accounting for the definitions of the magnetic energy $\vcb{1}^T \mat{M}_\spc{T} \vcb{e}_B$ and $\vcb{B}^T\mat{M}_\spc{M} \vcb{B}/(2\mu_0)$, as explained in \secref{subsub_num_semi_md}.
The construction of the semi-discrete formulation in the lower dimensions (2D and 1D) is completely analogous, following the weak forms defined in \ref{app_num_weak}, and is not presented here for brevity.


\subsubsection{Consistency properties}
\label{subsub_num_semi_md}

The semi-discrete Faraday's law \eqref{eqn_num_semi_B} utilizes the exact sequence given by the de Rham diagram \eqref{eqn_num_weak_rham}, where $\nabla\times\vec{\xi}\in\spc{M}$ for $\vec{\xi}\in\spc{E}$. Similarly, the operator of the discrete curl $C_D: \spc{E}_h \mapsto \spc{M}_h$ can be defined together with its associated matrix $\mat{C}_D$ \cite{Rieben2007}. Consequently, the following decomposition holds:
\begin{equation}
\mat{C}_{ijk} \vcb{1}_k = (\mat{M}_\spc{M})_{jk} (\mat{C}_D)_{ki}
.\label{eqn_num_semi_Cd}
\end{equation}
Therefore, the mass matrix $\mat{M}_\spc{M}$ can be canceled on both sides of \eqref{eqn_num_weak_B} during the discretization procedure and \eqref{eqn_num_semi_B} is obtained. Thanks to this feature of the spaces, the mass matrix does not need to be inverted numerically, saving computational time and preventing numerical errors of an iterative solution to appear. Moreover, it is clear that functions $\nabla\cdot\vec{\Xi}\neq 0, \vec{\Xi}\in\spc{M}_h$ do not lie in range of $C_D$, so the divergence of $\vec{B}_h$ remains constant. In order to guarantee this property even on the numerical level, an identical quadrature is used for numerical integration of $\mat{C}\cdot\vcb{1}$ and $\mat{M}_\spc{M}$.

It is important to stress the consistency between the Faraday's law \eqref{eqn_num_semi_B} and the magnetic energy equation \eqref{eqn_num_semi_epsB} holds even on the semi-discrete level, even though the decomposition \eqref{eqn_num_semi_Cd} is used. Without motion of the fluid ($\vcb{u}=0$), the change of the magnetic energy $E_B$ is only due to the induced magnetic field:
\begin{multline}
\dod{}{t}E_B 
= \dod{}{t} (\vcb{1}^T \mat{M}_\spc{T} \vcb{e}_B)
= \vcb{1}^T \mat{M}_\spc{T} \dod{\vcb{e}_B}{t}
= -\frac{1}{\mu_0} \mat{C}_{ijk} \vcb{E}_i\vcb{B}_j \vcb{1}_k
=\\
= -\frac{1}{\mu_0} \vcb{B}^T\mat{M}_\spc{M} \mat{C}_D \vcb{E}
= \frac{1}{\mu_0} \vcb{B}^T\mat{M}_\spc{M} \dod{\vcb{B}}{t}
= \dod{}{t} \left(\frac{1}{2\mu_0}\vcb{B}^T\mat{M}_\spc{M} \vcb{B} \right)
,\label{eqn_num_semi_enB}
\end{multline}
where symmetry of the mass matrices is used. When the fluid is in motion, the magnetic mass matrix $\mat{M}_\spc{M}$ is not constant, although $\mat{M}_\spc{T}$ is (see \secref{subsub_num_semi_cons}). This change is due to mutual exchange of momentum between the fluid and magnetic field, but it is not captured perfectly by the transformation of the finite elements for the magnetic field (which are rather designed to conserve magnetic flux, see the next paragraph). Consequently, a marginal discrepancy between the left and right hand side of \eqref{eqn_num_semi_enB} arises, which is given by:
\begin{equation}
\vcb{1}^T \vcb{e}_B^c = \vcb{1}^T \mat{F}_B^T\vcb{u} - \frac{1}{2\mu_0}\vcb{B}^T\dod{\mat{M}_\spc{M}}{t} \vcb{B} \approx 0
.\label{eqn_num_semi_enBc}
\end{equation}
However, the correction term $\vcb{e}_B^c$ is evaluated and added to the right hand side of \eqref{eqn_num_semi_eps} to conserve the total energy exactly, as proved in \secref{subsub_num_semi_cons}. In this sense, consistency between the definitions of magnetic energy $\vcb{1}^T \mat{M}_\spc{T} \vcb{e}_B$ and $\vcb{B}^T\mat{M}_\spc{M} \vcb{B}/(2\mu_0)$ is recovered.

Finally, the transformation properties of the electric and magnetic finite elements naturally mimic the behavior of their physical counterparts. This feature becomes crucial when the computational mesh deforms, while physical consistency of $\vec{E}_h$ and $\vec{B}_h$ is retained without any necessary correction \cite{Rieben2007}. Specifically, magnetic flux and circulation of electric field are invariants of the transformation, which can be formulated differentially as:
\begin{align}
\vec{B}_h\cdot\vec{n}\dif{S}
&= \frac{J}{|J|} \vec{B}'_h\cdot J^{-T}\vec{n}'|J|\dif{S'}
= \vec{B}'_h\cdot \vec{n}' \dif{S'}
,\label{eqn_num_semi_trans_B}\\
\vec{E}_h\cdot\dif{\vec{l}}
&= J^{-T}\vec{E}'_h \cdot J \dif{\vec{l}'}
= \vec{E}'_h \cdot \dif{\vec{l}'}
,\label{eqn_num_semi_trans_E}
\end{align}
where $J_{ij}(t,\vec{X})=\partial \vec{x}_i/\partial \vec{X}_j(t,\vec{X})$ is the Jacobi matrix of the transformation from the material space (with quantities denoted by prime) to the fluid frame. Contravariant transformation is applied on the surface normal vector $\vec{n} = J^{-T}\vec{n}'$ and covariant on the curve element $\dif{\vec{l}} = J \dif{\vec{l}'}$. This proves that the Piola transformations $\vec{B}_h=|J|^{-1} J \vec{B}'_h$ and $\vec{E}_h = J^{-T} \vec{E}'_h$ satisfy the invariance when applied to the fields \cite{Boffi2013}.

A similar situation appears in the lower dimensions and the Piola transformations are applied on the coplanar/collinear components of the fields. However, the transverse components must be considered too, despite the fact that the mesh deformation does not occur in this direction. The length of the normal vector $\vec{n}$ and the curve element $\dif{\vec{l}}$ does not change, but the area of the surface element $\dif{S}$ changes due to the deformation. Consequently, the transformations $\vec{B}_{\perp h}=|J|^{-1}\vec{B}_{\perp h}'$ and $\vec{E}_{\perp h}= \vec{E}_{\perp h}'$ must be applied onto the fields.

Furthermore, a non-uniform motion in the transverse direction also leads to the generation of magnetic fields, which is modeled by the $(\vec{B}_\parallel\cdot\nabla)\vec{u}_\perp$ term in \eqref{eqn_app_phys_dif_1D_Bp} and \eqref{eqn_app_phys_dif_2D_Bp}. In order to write the equations for magnetic field \eqref{eqn_app_num_weak_1D_Bp} and \eqref{eqn_app_num_weak_2D_Bp} in a way similar to \eqref{eqn_num_semi_B}, a discrete operator must be formulated for the term. The construction in 1D is straightforward, following the de Rham complex $L_2\stackrel{\nabla}{\longleftarrow} H^1$, which implies that $\nabla \vec{\psi}_\perp\in \spc{M}_\perp$ for $\vec{\psi}_\perp\in \spc{K}_\perp$ (for appropriately chosen spaces). As the functional spaces of $\spc{M}_\perp$ and $\spc{M}_\parallel$ coincide in 1D up to the vector dimension (see \ref{app_num_weak_1D}), the gradient can be point-wise weighted by $\vec{B}_\parallel$. In 2D, the de Rham diagram \eqref{eqn_num_weak_rham_2D} yields $\nabla \vec{\psi}_\perp\in H_{curl}(\Omega_0)\subset (L_2(\Omega_0))^2$ for $\vec{\psi}_\perp\in \spc{K}_\perp$ and $\spc{M}_\parallel \subset H_{div}(\Omega_0)\subset (L_2(\Omega_0))^2$. The dot product is then performed point-wise in $(L_2(\Omega_0))^2$, obtaining the result in $\spc{M}_\perp\subset L_2(\Omega_0)$.

More formally, the projectors $\Pi_{\spc{M}_\parallel}: \spc{M}_\parallel \to \spc{M}_\perp^2$ and $\Pi_{\nabla \spc{K}_\perp}: \nabla \spc{K}_\perp \to \spc{M}_\perp^2$ must be constructed in 2D (and analogously in 1D). For simplicity, the identity operators are used for appropriately defined spaces, which result in point projection at the degrees-of-freedom of $\spc{M}_{\perp h}$ on the semi-discrete level. Afterwards, the dot contraction from $\spc{M}_\perp^2 \times \spc{M}_\perp^2$ to $\spc{M}_\perp$ is performed point-wise, which is only exact for a (piece-wise) constant argument. Otherwise, a typically acceptable approximation error is introduced by the operation. However, note that only the transverse components are affected, which are not subjects of the divergence-free condition.


\subsubsection{Conservation properties}
\label{subsub_num_semi_cons}

Conservation properties are an essential part of the scheme design. A major role in the construction plays validity of the geometric conservation law (GCL), which can be written in the differential form as \cite{Thomas1979}:
\begin{equation}
\dod{|J|}{t} = |J|\nabla\cdot\vec{u}
.\label{eqn_num_semi_gcl}
\end{equation}
Since the functions from the kinematic space $\spc{K}_h$ are differentiable, the Jacobi matrix $J$ is consistently approximated as $J_h(t,\vec{X})=\nabla_{\vec{X}} \vec{x}_h(t,\vec{X})$, where $\nabla_{\vec{X}}$ is the gradient with respect to the material coordinates. From \eqref{eqn_num_semi_x}, it can be deduced that GCL \eqref{eqn_num_semi_gcl} holds discretely. For further details, see \cite{Dobrev2010}. Consequently, the mass conservation law \eqref{eqn_phys_dif_mass} can be reformulated as $\dif{\rho |J|}/\dif{t}=0$ and it holds point-wise on the discrete level for the density $\rho_h$ defined as:
\begin{equation}
\rho_h(t,\vec{X}) = \rho_0(\vec{X}) / |J_h(t,\vec{X})|
,\label{eqn_num_semi_rho}
\end{equation}
where $\rho_0(\vec{X})=\rho_h(0,\vec{X})$ is the initial profile of density. Note that the symbol $\rho$ is favored in the rest of the work for better readability, but $\rho_h$ is implied on the discrete level.

Another related feature of the scheme is the fact that the thermodynamic mass matrix $\mat{M}_\spc{T}$ and kinematic mass matrix $\mat{M}_\spc{K}$ are constant. Since the mapping follows the characteristics $\vec{x}_h$, material derivatives of the base functions $\lbrace \varphi_i \rbrace_{i=1}^{N_\spc{T}}$ and $\lbrace \vec{\psi}_i \rbrace_{i=1}^{N_\spc{K}}$ are zero (similarly for other base functions). The classical Reynolds theorem \eqref{eqn_phys_reynolds} together with the point-wise mass conservation and definitions of $\mat{M}_\spc{T}$ and $\mat{M}_\spc{K}$ \eqref{eqn_app_num_semi_mats_MtMk} then implies the following:
\begin{equation}
\dod{}{t} (\mat{M}_\spc{T})_{ij}
= \int_{\Omega(t)} \rho \left( \dod{\varphi_i}{t} \varphi_j + \varphi_i \dod{\varphi_j}{t} \right) \dif{V}
=  0
, \qquad
\dod{}{t} (\mat{M}_\spc{K})_{ij}
= \int_{\Omega(t)} \rho \left( \dod{\vec{\psi}_i}{t} \cdot \vec{\psi}_j + \vec{\psi}_i \cdot \dod{\vec{\psi}_j}{t} \right) \dif{V}
= 0
.\label{eqn_num_semi_mats}
\end{equation}
Moreover, the absence of the inter-element constraints in the thermodynamic space $\spc{T}_h\subset L_2(\Omega_0)$ results in a block-diagonal structure of $\mat{M}_\spc{T}$. This means that the matrix can be inverted in each element separately without the need for a global iterative method.

Conservation of the total momentum $P$ is an immediate consequence of \eqref{eqn_num_semi_mom} and the definitions \eqref{eqn_app_num_semi_mats_FFb}, which imply $(\mat{F}^T+\mat{F}_B^T)\vcb{1}_\spc{K}=\vcb{0}$. The vector $\vcb{1}_\spc{K}$ is the grid function corresponding to the unit vector in $\spc{K}_h$ and $\vcb{0}$ is a zero vector of length $N_\spc{T}$. Provided that the boundary is not moving ($\Gamma_\sigma=\emptyset$) or the outer normal force is zero ($\sigma_n=0$), conservation of momentum is proven as follows:
\begin{equation}
\dod{}{t} P =
\dod{}{t}(\vcb{1}_\spc{K}^T\mat{M}_\spc{K} \vcb{u})
=\vcb{1}_\spc{K}^T\mat{M}_\spc{K} \dod{\vcb{u}}{t}
=\vcb{1}_\spc{K}^T(\mat{F}+\mat{F}_B)\vcb{1}
= 0
,\label{eqn_num_semi_cons_mom}
\end{equation}
where the fact that the mass matrix $\mat{M}_\spc{K}$ is constant is used.

The energy conservation property arises from the symmetries between the energy equations \eqref{eqn_num_semi_eps}, \eqref{eqn_num_semi_epsB} and the momentum equation \eqref{eqn_num_semi_mom}. Defining the discrete total energy $E_T$ with help of the mass matrices, the conservation can be proven:
\begin{multline}
\dod{}{t} E_T =
\dod{}{t}\left(
  \vcb{1}^T \mat{M}_\spc{T} \vcb{e}
+ \vcb{1}^T \mat{M}_\spc{T} \vcb{e}_B 
+ \frac{1}{2}\vcb{u}^T \mat{M}_\spc{K} \vcb{u}  \right)
=
\vcb{1}^T \mat{M}_\spc{T} \dod{\vcb{e}}{t}
+ \vcb{1}^T \mat{M}_\spc{T} \dod{\vcb{e}_B}{t} 
+ \vcb{u}^T \mat{M}_\spc{K} \dod{\vcb{u}}{t}
=\\= 
\vcb{1}^T (\mat{F}^T + \mat{F}_B^T) \vcb{u}
- \vcb{u}^T (\mat{F} + \mat{F}_B) \vcb{1}
+ \mat{S}_{ijk}\vcb{E}_i\vcb{B}_j \vcb{1}_k
= 0
,\label{eqn_num_semi_cons_en}
\end{multline}
where symmetry and time-independence of the mass matrices $\mat{M}_\spc{T}$, $\mat{M}_\spc{K}$ is used. In addition, the definition \eqref{eqn_app_num_semi_mats_S} of the tensor $\mat{S}$ implies that $\mat{S}\cdot\vcb{1}=\mat{O}_{\spc{E}\spc{M}}$, where $\mat{O}_{\spc{E}\spc{M}}$ is a zero matrix of size $N_\spc{E}\times N_\spc{M}$. It is also assumed that the Poynting vector across the boundary is equal to zero, so the boundary terms vanish.

In essence, the conservation property can be stated even locally \cite{Dobrev2010}. As the thermodynamic elements are discontinuous, the test function can be chosen as the characteristic function of an arbitrary sub-part of the mesh (e.g. a single element) instead of the unitary field and the associated vector $\vcb{1}$. The local conservation can be proved analogously to  \eqref{eqn_num_semi_cons_en} with the only difference that the fluxes over the sub-domain boundary would not vanish due to the (global) boundary conditions. Still, the change of the energy over an arbitrary sub-domain has a divergence form, which can be understood as the local conservation property \cite{Abgrall2017, Abgrall2020}. However, this is only true as far as the consistency between $\vcb{1}^T \mat{M}_\spc{T} \vcb{e}_B$ and $\vcb{B}^T\mat{M}_\spc{M} \vcb{B}/(2\mu_0)$ holds according to \secref{subsub_num_semi_md}. Strictly speaking, the contribution of the correction term $\vcb{e}_B^c$ to the energy equation \eqref{eqn_num_semi_cons_en} does not satisfy this property, despite being negligible.


\subsection{Discrete model}
\label{sub_num_disc}

The construction of the numerical scheme after formulation of the general semi-discrete system \eqref{eqn_num_semi} proceeds with temporal and specific spatial discretization. The time integration is divided into two parts, magnetodynamic and ideally magneto-hydrodynamic. The magnetodynamic part described in \secref{subsub_num_disc_md} is solved implicitly (due to its parabolic nature) on a static mesh, whereas the part representing ideal magneto-hydrodynamics is solved by the explicit scheme of \secref{subsub_num_disc_mhd} (due to its hyperbolic nature) and the mesh position is advanced. The schemes are coupled together in a multi-step method described in \secref{subsub_num_disc_time}, which preserves the conservation properties of the scheme described in \secref{subsub_num_semi_cons}.


\subsubsection{Magnetodynamics}
\label{subsub_num_disc_md}

The magnetodynamic part of the scheme advances the discrete magnetic field $\vcb{B}$ and electric field $\vcb{E}$ without motion of the computational mesh. The time  discretization is performed between the discrete time levels $n$ and $n+1$, which are designated by the upper index of the grid functions. 

Within resistive MHD, eddy currents cause diffusion of the magnetic field and electric field consequently. Therefore, a (semi-)implicit time integration is desired for solution of the resulting parabolic system when the time step $\Delta t$ is comparable with the diffusion time $\tau_B$, i.e., $\Delta t \gtrsim \tau_B=\mu_0 L^2/\eta$, where $L$ is a characteristic length. However, the physical system might be nearly ideal ($\Delta t \ll \tau_B$) and an expensive implicit solution can be replaced by an explicit approach. For this reason, the parameter $\alpha$ is introduced to enable switching between the schemes:
\begin{equation}
\alpha =
\begin{cases}
0 & \text{forward Euler (explicit, first order)},\\
1/2 & \text{Crank-Nicolson (semi-implicit, second order, A-stable)},\\ 
1 & \text{backward Euler (fully implicit, first order, L-stable)}.
\end{cases}
\label{eqn_num_disc_alpha}
\end{equation}

The magnetodynamic system, composed of the Faraday's law \eqref{eqn_num_semi_B} and Ohm's law \eqref{eqn_num_semi_E}, is not solved in its primary mixed form, but transformed to an equation for the electric field by substitution of \eqref{eqn_num_semi_B} to \eqref{eqn_num_semi_E}:
\begin{subequations}
\begin{align}
\left( \mat{M}_\spc{E} + \frac{\alpha\Delta t}{\mu_0} \mat{D} \right) \vcb{E}^{n+\alpha}
&= \frac{1}{\mu_0} \mat{C}_{\cdot jk} \vcb{B}_j^{n} \vcb{1}_k
+ \mat{X}_B^T\vcb{1}
,\label{eqn_num_disc_md_E}\\
\frac{1}{\Delta t} \vcb{B}^{n+1} &= 
\frac{1}{\Delta t} \vcb{B}^{n}
-\mat{C}_{D} \vcb{E}^{n+\alpha}
,\label{eqn_num_disc_md_B}
\end{align}
\label{eqn_num_disc_md_EB}
\end{subequations}
where the electric field on the intermediate time level is interpolated as $\vcb{E}^{n+\alpha} = \alpha \vcb{E}^{n+1} + (1-\alpha) \vcb{E}^{n}$. The newly appearing diffusion matrix $\mat{D}$ is defined together with the rest of the matrices in \ref{app_num_semi_mats}. However, it is consistent with the semi-discrete form and can be equivalently understood as the following composition of the matrices:
\begin{equation}
\mat{D}_{ij} = \mat{C}_{ikl} (\mat{C}_D)_{kj} \vcb{1}_l
.\label{eqn_num_disc_md_D}
\end{equation}
This assertion can be viewed as a consequence of the compatibility of the spaces mentioned in \secref{subsub_num_semi_md}. After \eqref{eqn_num_disc_md_E} is solved numerically, the resulting electric field is inserted into \eqref{eqn_num_disc_md_B} to obtain the magnetic field. This procedure guarantees that $\vcb{B}^{n+1}$ remains divergence-free even on the numerical level due to the properties of $\mat{C}_D$ discussed in \secref{subsub_num_semi_md}.

For numerical solution of the diffusion equation \eqref{eqn_num_disc_md_E}, the Auxiliary-space Maxwell Solver (AMS) is used as the preconditioner \cite{Kolev2009b}, which relies on a parallel implementation of the algebraic multigrid method (AMG) \cite{Henson2002,Ruge1987}. Thanks to this choice, optimal convergence can be attained even for fine resolutions. In this case, the main solver is a parallel implementation of the conjugate gradient method.

When the new values of the fields are known, the explicit contribution to the energy equations \eqref{eqn_num_semi_eps}, \eqref{eqn_num_semi_epsB} can be evaluated:
\begin{subequations}
\begin{align}
\eval[2]{\dod{\vcb{e}}{t}}_{\text{Joule}} &= 
\mat{M}_\spc{T}^{-1} \left( 
+\frac{1}{\mu_0} \mat{C}_{ij\cdot} \vcb{E}_i^{n+\alpha} \vcb{B}_j^{n+1/2}
+ \mat{S}_{ij\cdot}\vcb{E}_i^{n+\alpha} \vcb{B}_j^{n+1/2}
+ \mat{X}_E\vcb{B}^{n+1/2}
+ \mat{X}_B\vcb{E}^{n+\alpha}
\right)
,\label{eqn_num_disc_md_eps}\\
\eval[2]{\dod{\vcb{e}_B}{t}}_{\text{Joule}} &= \mat{M}_\spc{T}^{-1} \left( - \frac{1}{\mu_0} \mat{C}_{ij\cdot} \vcb{E}_i^{n+\alpha} \vcb{B}_j^{n+1/2} \right)
.\label{eqn_num_disc_md_epsB}
\end{align}
\label{eqn_num_disc_md_epss}
\end{subequations}
It should be stressed that $\mat{M}_\spc{T}$ can be pre-inverted in an element-wise fashion, as mentioned in \secref{subsub_num_semi_cons}. Therefore, \eqref{eqn_num_disc_md_epss} can be evaluated on the element level and the local update of the magnetic energy \eqref{eqn_num_disc_md_epsB} can be reused in \eqref{eqn_num_disc_md_eps} due to the symmetry between the equations, so solution of the auxiliary equation for $\vcb{e}_B$ has nearly zero computational costs and preserves the symmetry numerically.

The choice of the time-centered magnetic field in \eqref{eqn_num_disc_md_epss} is motivated by the consistency with \eqref{eqn_num_disc_md_B}, following the discussion of the semi-discrete form in  \secref{subsub_num_semi_md}. When discretized this way, consistency of the magnetic field $\vcb{B}$ with the magnetic energy $\vcb{e}_B$ holds discretely:
\begin{multline}
\eval[1]{(E_B^{n+1} - E_B^{n})}_{\text{Joule}}
= \eval[1]{(\vcb{1}^T \mat{M}_\spc{T} \vcb{e}_B^{n+1} - \vcb{1}^T \mat{M}_\spc{T} \vcb{e}_B^n )}_{\text{Joule}}
= - \frac{\Delta t}{\mu_0} \mat{C}_{ijk} \vcb{E}_i^{n+\alpha} \vcb{B}_j^{n+1/2} \vcb{1}_k
=\\= \frac{1}{2\mu_0} ((\vcb{B}^{n+1})^T-(\vcb{B}^n)^T) \mat{M}_\spc{M} (\vcb{B}^{n+1} + \vcb{B}^n)
= \frac{1}{2\mu_0} ((\vcb{B}^{n+1})^T \mat{M}_\spc{M} \vcb{B}^{n+1}- (\vcb{B}^{n})^T \mat{M}_\spc{M} \vcb{B}^n)
,\label{eqn_num_disc_md_en}
\end{multline}
where symmetry of $\mat{M}_\spc{M}$ is used and zero boundary terms are assumed.


\subsubsection{Ideal magneto-hydrodynamics}
\label{subsub_num_disc_mhd}

Unlike \secref{subsub_num_disc_md}, the part of the scheme dedicated to ideal magneto-hydrodynamics is solved explicitly and operates with an already known magnetic field, which enters the scheme only indirectly through the Lorentz force matrix $\mat{F}_B$. This part of the scheme, complementary to \eqref{eqn_num_disc_md_EB} and \eqref{eqn_num_disc_md_epss}, is given by:

\begin{subequations}
\begin{align}
\eval[2]{\dod{\vcb{x}}{t}}_{\text{hydro}} &= \vcb{u}^\ast
,\label{eqn_num_disc_mhd_x}\\
\eval[2]{\dod{\vcb{u}}{t}}_{\text{hydro}} &= -\mat{M}_\spc{K}^{-1}((\mat{F} + \mat{F}_B)\vcb{1}
+ \vcb{b}_\sigma)
,\label{eqn_num_disc_mhd_mom}\\
\eval[2]{\dod{\vcb{e}}{t}}_{\text{hydro}} &= \mat{M}_\spc{T}^{-1}
(\mat{F}^T\vcb{u}^\ast  + \vcb{e}_B^c)
,\label{eqn_num_disc_mhd_eps}\\
\eval[2]{\dod{\vcb{e}_B}{t}}_{\text{hydro}} &= \mat{M}_\spc{T}^{-1}\mat{F}_B^T\vcb{u}^\ast
.\label{eqn_num_disc_mhd_epsB}
\end{align}
\label{eqn_num_disc_mhd}
\end{subequations}
The system is nearly identical to the original method of the high-order curvilinear finite element hydrodynamics \cite{Dobrev2010} and is only augmented by addition of the Lorentz force $\mat{F}_B\vcb{1}$ to the momentum equation \eqref{eqn_num_disc_mhd_mom} and the equation for the magnetic energy \eqref{eqn_num_disc_mhd_epsB}. The definition of the velocity $\vcb{u}^*$ varies depending on the time integration algorithm. For classical multi-step methods it is set equal to the starting (intermediate) time level, but time symmetry and energy conservation are lost this way. By contrast, the special choice of $\vcb{u}^*$ made in \secref{subsub_num_disc_time} leads to fulfilling of the energy conservation. In all cases, momentum conservation is not affected by the choice of time stepping due to linearity of the expressions and the proof made for the semi-discrete form in \secref{subsub_num_semi_cons} holds discretely too.


\subsubsection{Time integration}
\label{subsub_num_disc_time}

The time discretization of the coupled system \eqref{eqn_num_disc_md_EB}, \eqref{eqn_num_disc_md_epss} and \eqref{eqn_num_disc_mhd} can follow different strategies depending on the coupling of the system, which is approximately characterized by the magnetic Reynolds number $\textup{Re}_m=u L \mu_0 / \eta$. In cases dominated by diffusion ($\textup{Re}_m \ll 1$) or by convection ($\textup{Re}_m \gg 1$), operator splitting can be applied to the system and an arbitrarily high order of the time integration scheme can be used for the explicit hydrodynamic part \eqref{eqn_num_disc_mhd}. 
For this purpose, energy-conserving implicit-explicit (IMEX) methods are used \cite{Sandu2021}. A special case is the second order RK2-Average scheme, which allows to combine the magnetodynamics and ideal magneto-hydrodynamics. The improved inter-coupling between the two parts favours this method in the case of a strong coupling ($\textup{Re}_m \sim 1$).

Within the high-order conservative IMEX methods, the vector of the state variables is split into  implicit and explicit parts. The implicit part $V$ is dedicated to velocity and conserves kinetic energy due to its symplectic nature. The explicit part $Y$ accommodates the rest of the quantities:
\begin{equation}
V = [~ \vcb{u} ~], \qquad
Y = [~ \vcb{x}, ~\vcb{e}, ~\vcb{e}_B ~]^T
.\label{eqn_num_disc_time_vy}
\end{equation}
The calculation uses the derivatives $V'=V'(t,Y)$ and $Y'=Y'(t,V,Y)$ given by \eqref{eqn_num_disc_mhd} with the Butcher tables, which can be found for the third (RK3hc) and fourth order schemes (RK4hc) in \cite{Sandu2021}. However, it must be stressed that the lower-triangular form of the Butcher tables for the selected class of implicit methods together with the only dependency of $V'$ on the explicit state $Y$ allows to solve the velocity equation \eqref{eqn_num_disc_mhd_mom} explicitly.

A special case of the IMEX methods is the RK2-Average scheme, originally used for the high-order curvilinear finite element hydrodynamics \cite{Dobrev2010}. Here, it is extended for the Lorentz force contribution to the momentum equation \eqref{eqn_num_disc_mhd_mom} and the Joule heating terms \eqref{eqn_num_disc_md_epss}. It can be derived from the classical RK2 method, sharing with it the second order of convergence. The first step of the method calculates the quantities on the intermediate time level $n+1/2$ as follows:
\begin{subequations}
\begin{align}
\vcb{u}^{n+1/2} &= \vcb{u}^{n} - \frac{\Delta t}{2} \mat{M}_\spc{K}^{-1}((\mat{F}^n + \mat{F}_B^n)\vcb{1} + \vcb{b}_\sigma^n)
,\label{eqn_num_disc_time_rkavg_1_u}\\
\vcb{x}^{n+1/2} &= \vcb{x}^{n} + \frac{\Delta t}{2} \vcb{u}^{n+1/2}
,\label{eqn_num_disc_time_rkavg_1_x}\\
\vcb{e}^{n+1/2} &= \vcb{e}^{n} + \frac{\Delta t}{2}\mat{M}_\spc{T}^{-1}
(\mat{F}^n)^T\vcb{u}^{n+1/2}
+ \frac{\Delta t}{2} \eval[2]{\dod{\vcb{e}}{t}}_{\text{Joule}}^{n,\Delta t/2,\alpha=1}
,\label{eqn_num_disc_time_rkavg_1_e}\\
\vcb{e}_B^{n+1/2} &= \vcb{e}_B^{n} + \frac{\Delta t}{2}\mat{M}_\spc{T}^{-1}
(\mat{F}_B^n)^T\vcb{u}^{n+1/2}
+ \frac{\Delta t}{2} \eval[2]{\dod{\vcb{e}_B}{t}}_{\text{Joule}}^{n,\Delta t/2,\alpha=1}
,\label{eqn_num_disc_time_rkavg_1_eB}
\end{align}
\label{eqn_num_disc_time_rkavg_1}
\end{subequations}
where the superscript of the Joule heating terms defined by \eqref{eqn_num_disc_md_epss} denotes the time level for construction of the matrices, length of the time step and value of the $\alpha$ parameter, respectively. The second step of the method takes the following form:
\begin{subequations}
\begin{align}
\vcb{u}^{n+1} &= \vcb{u}^{n} - \Delta t \mat{M}_\spc{K}^{-1}((\mat{F}^{n+1/2} + \mat{F}_B^{n+1/2})\vcb{1} + \vcb{b}_\sigma^{n+1/2})
,\label{eqn_num_disc_time_rkavg_2_u}\\
\vcb{x}^{n+1} &= \vcb{x}^{n} + \Delta t \bar{\vcb{u}}^{n+1/2}
,\label{eqn_num_disc_time_rkavg_2_x}\\
\vcb{e}^{n+1} &= \vcb{e}^{n} + \Delta t\mat{M}_\spc{T}^{-1}
(\mat{F}^{n+1/2})^T\bar{\vcb{u}}^{n+1/2}
+ \Delta t \eval[2]{\dod{\vcb{e}}{t}}_{\text{Joule}}^{n+1/2,\Delta t,\alpha=1/2}
+ \vcb{e}^c_B
,\label{eqn_num_disc_time_rkavg_2_e}\\
\vcb{e}_B^{n+1} &= \vcb{e}_B^{n} + \Delta t\mat{M}_\spc{T}^{-1}
(\mat{F}_B^{n+1/2})^T\bar{\vcb{u}}^{n+1/2}
+ \Delta t \eval[2]{\dod{\vcb{e}_B}{t}}_{\text{Joule}}^{n+1/2,\Delta t,\alpha=1/2}
,\label{eqn_num_disc_time_rkavg_2_eB}
\end{align}
\label{eqn_num_disc_time_rkavg_2}
\end{subequations}
where the intermediate velocity is defined as $\bar{\vcb{u}}^{n+1/2}=1/2(\vcb{u}^{n+1}+\vcb{u}^{n})$. The correction term $\vcb{e}^c_B$ is calculated between the time levels $n$ and $n+1$. The construction in 1D/2D is analogous and deferred to \ref{app_num_disc_time}.

As mentioned already, the RK2-Average scheme conserves the total energy $E_T$. Following the lines of the proof in \secref{subsub_num_semi_cons} for the semi-discrete form, the conservation can be proven:
\begin{multline}
E_T^{n+1} - E_T^n
=
\vcb{1}^T \mat{M}_\spc{T} (\vcb{e}^{n+1} + \vcb{e}_B^{n+1} - \vcb{e}^{n} - \vcb{e}_B^{n})
+ (\vcb{u}^{n+1})^T \mat{M}_\spc{K} \vcb{u}^{n+1}
- (\vcb{u}^{n})^T \mat{M}_\spc{K} \vcb{u}^{n}
=\\
= \vcb{1}^T ((\mat{F}^{n+1/2})^T + (\mat{F}_B^{n+1/2})^T) \bar{\vcb{u}}^{n+1/2}
- (\bar{\vcb{u}}^{n+1/2})^T (\mat{F}^{n+1/2} + \mat{F}_B^{n+1/2}) \vcb{1}
+ \mat{S}_{ijk}\vcb{E}_i^{n+1/2}\vcb{B}_j^{n+1/2} \vcb{1}_k
= 0
.\label{eqn_num_disc_time_en}
\end{multline}

The time step control adopts the algorithm of the high-order curvilinear finite element hydrodynamics, where a CFL (Courant--Friedrichs--Lewy) criterion is based on the minimal singular value of $J_h$ at the integration points \cite{Dobrev2010}. In order to account for the magnetic stress contribution, the condition is modified in that the speed of sound $c_s$ is replaced by the magneto-sonic velocity $v_f^2=c_s^2 + v_A^2$, where $v_A=|\vec{B}|/\sqrt{\mu_0 \rho}$ is the Alfvén velocity. The characteristics with slope equal to $v_f$ represent the fastest propagating modes in classical MHD \cite{Torrilhon2003}, guaranteeing stability for the scheme in turn.


\subsubsection{Artificial viscosity}
\label{subsub_num_disc_visc}

For stabilisation of the numerical scheme, artificial viscosity is used similarly to the high-order curvilinear hydrodynamics \cite{Dobrev2010}. Its role is to transform kinetic energy to internal at places of strong compression to prevent oscillatory behaviour and deterioration of the mesh \cite{Caramana1998a}. The artificial viscosity enters the numerical scheme through the definition of the stress tensor $\tens{\sigma} = -p \tens{I} + \tens{\sigma}_a$, where $p$ is thermodynamic pressure and $\tens{\sigma}_a$ is the artificial viscosity tensor. The physical part of the tensor is simplified to the isotropic pressure only for all cases considered here, but the method is general and a non-artificial viscosity can be present.
Multiple methods for construction of $\tens{\sigma}_a$ were proposed in \cite{Dobrev2010,Kolev2009} and are not repeated here for brevity, but all of them rely on the tensor $\nabla\vec{u}_\parallel$. The multi-dimensional formulation of the viscosity is used here to take into account the transverse components of the velocity $\vec{u}_\perp$ in the lower dimensions. The gradient $\nabla \vec{u}$ is calculated and complemented to a square tensor by zeros, assuming that the transverse derivatives vanish for the slab geometry. The rest of the procedure is identical with the original method and is performed in the plane/line. The only exception is the modification of the characteristic velocity in the CFL condition, where the magneto-sonic waves velocity $v_f$ replaces the sound speed velocity, analogously to \secref{subsub_num_disc_time}.

It should be noted that non-linearity does not arise only from the coupling between the momentum and internal energy equations like in the classical hydrodynamics, but also from the combination of the momentum and magnetic field equation in 1D and 2D. The convective term $(\vec{B}_\parallel\cdot\nabla)\vec{u}_\perp$ is present in \eqref{eqn_app_phys_dif_1D_Bp} and \eqref{eqn_app_phys_dif_2D_Bp}, which is responsible for the coupling. However, an analogous technique to the RK2-Average time discretization of the energy equation \eqref{eqn_num_disc_time_rkavg_1_e} and \eqref{eqn_num_disc_time_rkavg_2_e} is applied in this case, so the averaged velocities $\vcb{u}^{n+1/2}_\perp$ and $\bar{\vcb{u}}^{n+1/2}_\perp$ are used in \eqref{eqn_app_num_disc_time_rkavg_1_Bh} and \eqref{eqn_app_num_disc_time_rkavg_2_Bh}. Similarly, the IMEX methods update the magnetic field within the state $Y$ with known implicit velocity from the state $V$. The pre-accelerated velocities then give a stabilizing estimate of the mechanical work $\tens{\sigma}:\nabla\vec{u}$ and the magnetic dynamo effect $(\vec{B}_\parallel\cdot\nabla)\vec{u}_\perp$.


\subsubsection{Specific spatial discretization}
\label{subsub_num_disc_spc}

The section about semi-discrete form (\secref{sub_num_semi}) already defined the finite element spaces for the spatial discretization, but only in a general manner, with respect to their conformity with the functional spaces. The generality of the description allows to use various finite element families for different geometries on unstructured meshes. However, the numerical tests of \secref{sec_sim} narrow the scope of the practical verification of numerical properties to only  regular hexahedral meshes with the following finite element spaces equipped with an isoparametric mapping:
\begin{itemize}
\item thermodynamic -- $\spc{T}_h = \lbrace \varphi \in \spc{T} ~|~ 
\eval[0]{\varphi}_{\Omega_e} \in O^{p}(\Omega_e) \quad \forall \Omega_e \in \Sigma_h \rbrace$,
\item kinematic -- $\spc{K}_h = \lbrace \vec{\psi} \in \spc{K} ~|~ 
\eval[0]{\vec{\psi}}_{\Omega_e} \in (O^{p+1}(\Omega_e))^3, 
\eval[0]{\vec{\psi}}_{\Gamma_e} \in (Q^{p+1}(\Gamma_e))^3
\quad \forall \Omega_e \in \Sigma_h, \Gamma_e \subset \partial \Omega_e \rbrace$,
\item magnetic -- $\spc{M}_h = \lbrace \vec{\Xi} \in \spc{M} ~|~ 
\eval[0]{\vec{\Xi}}_{\Omega_e} \in {RT}^{q+1}_{3D}(\Omega_e) \quad \forall \Omega_e \in \Sigma_h \rbrace$,
\item electric -- $\spc{E}_h = \lbrace \vec{\xi} \in \spc{E} ~|~ 
\eval[0]{\vec{\xi}}_{\Omega_e} \in {N\!D}^{q+1}_{3D}(\Omega_e) \quad \forall \Omega_e \in \Sigma_h \rbrace$,
\end{itemize}
where $O^p$ and $Q^p$ are products of polynomials up to the order $p$ in 3D and 2D, respectively. The sets $\Gamma_e$ represent the faces of the element $\Omega_e$. The polynomial spaces $RT^p_{3D}$ and $N\!D^p_{3D}$ represent the Nédélec $H_{div}$ and $H_{curl}$ conforming spaces in 3D, respectively \cite{Nedelec1980}, having the maximal polynomial order $p$. The construction in the lower dimensions is analogous and left for \ref{app_num_disc_spc}.

\begin{figure}[hbtp]
\centering
\includegraphics[width=.3\textwidth]{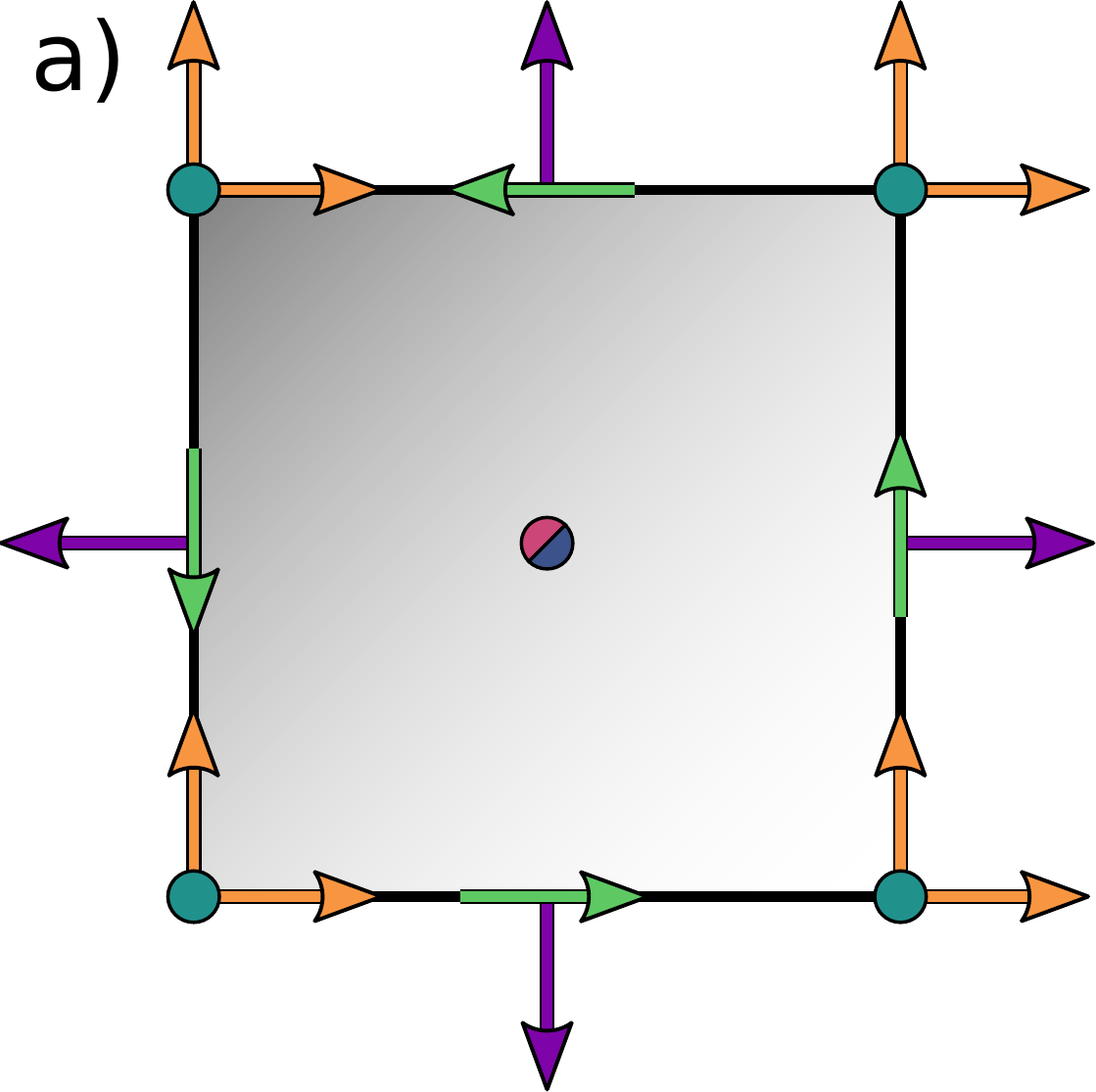}
\includegraphics[width=.3\textwidth]{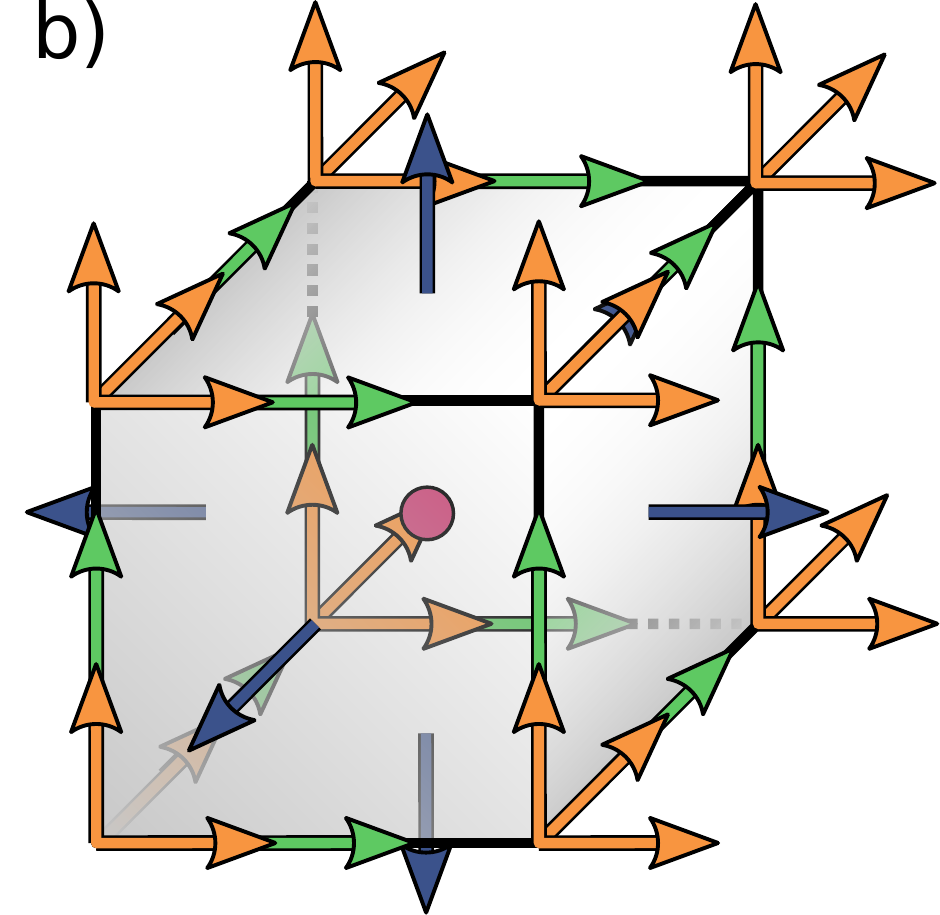}
\caption{The magneto-hydrodynamic finite elements $T0M0$ in 2D (a) and 3D (b). The colors represent different degrees-of-freedom: {\color[HTML]{cc4778}red} -- thermodynamic, {\color[HTML]{f89540}orange} -- kinematic, {\color[HTML]{3b528b}blue} -- (transverse) magnetic field, {\color[HTML]{5ec962}green} -- (coplanar) electric field, {\color[HTML]{7e03a8}purple} -- coplanar magnetic field (only (a)), {\color[HTML]{21918c}cyan} -- transverse electric field (only (a)).}
\label{fig_num_disc_T0M0}
\end{figure}

Despite the fact that the construction of the semi-discrete model in \secref{sub_num_semi} does not require it, the polynomial orders are derived from only the parameters $p$ and $q$ for optimal conditioning. The orders in the hydrodynamic part are related through the term $\mat{F}^T\vcb{u}$, where the kinematic elements with the polynomial order higher by one are optimal for regularization of the energy equation \eqref{eqn_num_semi_eps}, as the velocity is differentiated. The orders in the electromagnetic part are identical, because of the compatibility between the spaces, as explained in \secref{subsub_num_semi_md}. Henceforth, the elements of this kind are denoted as $TpMq$. The lowest order $T0M0$ elements are depicted in \figref{fig_num_disc_T0M0}.

Finally, it must be noted that the classical Lagrange elements are used for the kinematic quantities from $\spc{K}_h$, but Bernstein polynomials are preferred for the thermodynamic space $\spc{T}_h$. This choice is motivated by positivity of the approximated physical quantities, that is specific internal energy and specific magnetic energy. Since these polynomials form a positive basis, the interpolated functions are always positive when the discrete values of the corresponding grid functions are. Consequently, the mass matrices $\mat{M}_\spc{T}$ are positive definite. From the practical point of view, this feature prevents overshoots of the thermodynamic potentials near discontinuities like propagating shocks, for example \cite{Glaubitz2019,Abgrall2010}.


\section{Numerical tests}
\label{sec_sim}

A series of tests was performed to validate construction of the  proposed scheme and assess its performance. Propagation of magneto-hydrodynamic waves is examined in \secref{sub_sim_torrilhon}. Convergence is measured on the smooth problems of ideal magneto-hydrodynamics in \secref{sub_sim_taylorgreen} and resistive magnetodynamics in \secref{sub_sim_magdiff}. The temporal convergence is tested in \secref{sub_sim_advdiff}. Furthermore, \secref{sub_sim_rotor} and \secref{sub_sim_blast} show the capabilities of the scheme in 2D and 3D on physically relevant problems.

The numerical implementation utilizes the MFEM library for assembly of the finite elements, providing flexibility and scalability \cite{mfem-library,Anderson2019}. The visualizations were made with the GLVis tool for fast and accurate rendering of high order finite elements \cite{glvis-tool}.


\subsection{MHD waves}
\label{sub_sim_torrilhon}

The first test problem mainly concentrates on validation of the numerical scheme by the problem of MHD waves propagation. In contrast to the rest of the test cases, the problem is defined in one dimension, but both, colinear and transverse, components of the fields are involved. 

The problem is adopted from \cite{Torrilhon2003} and is defined as a Riemann problem with the following left and right states:
\begin{equation}
\left( \begin{array}{c}
\rho_l \\
\vec{u}_l \\
p_l \\
(\vec{B}_l)_x \\
(\vec{B}_l)_y \\
(\vec{B}_l)_z \\
\end{array} \right)
= 
\left( \begin{array}{c}
3 \\
\vec{0} \\
3 \\
1.5 \\
1 \\
0 \\
\end{array} \right)
, \qquad
\left( \begin{array}{c}
\rho_r \\
\vec{u}_r \\
p_r \\
(\vec{B}_r)_x \\
(\vec{B}_r)_y \\
(\vec{B}_r)_z \\
\end{array} \right)
=
\left( \begin{array}{c}
1 \\
\vec{0} \\
1 \\
1.5 \\
\cos(\chi) \\
\sin(\chi) \\
\end{array} \right)
.
\label{eqn_sim_torrilhon_ic}
\end{equation}
Note that the values are given in relative units, where the magnetic field is normalized by $\sqrt{\mu_0}$. The twist angle of magnetic field is set to $\chi=1.5$, which is close enough to $\pi/2$ to seed a large variety of magneto-hydrodynamic waves  \cite{Torrilhon2003}. A solution is sought on the domain  $(-1,1)$, where the interface is located at the origin of the coordinate system. The boundary conditions for zero velocity are applied in agreement with the initial conditions.

The numerical solution is computed up to the final time $0.4$ with different orders of the finite elements. The time integration scheme RK2-Average is used for $T0M0$ and $T1M1$ elements, RK3hc(A.$\alpha$) for $T2M2$ and RK4hc(A.$\alpha$) for $T3M3$ (consult \secref{subsub_num_disc_time}). In all cases, the CFL constant is set to $0.5$. As the problem involves propagating discontinuities, it is sensitive to the choice of the artificial viscosity model. The best results were obtained with the model based on a full eigenvector decomposition of the velocity gradient \cite{Dobrev2010,Kolev2009}, which is also used in the following sections. Its linear coefficient is set to the value $0.25$ and the quadratic to $1$. The resolution keeps a constant number of degrees-of-freedom (DOFs) to give comparable results. Specifically, the mesh consists of 480 $T0M0$, 240 $T1M1$, 160 $T2M2$ or 120 $T3M3$ elements.

\begin{figure}[hbtp]
\centering
\includegraphics[width=.49\textwidth]{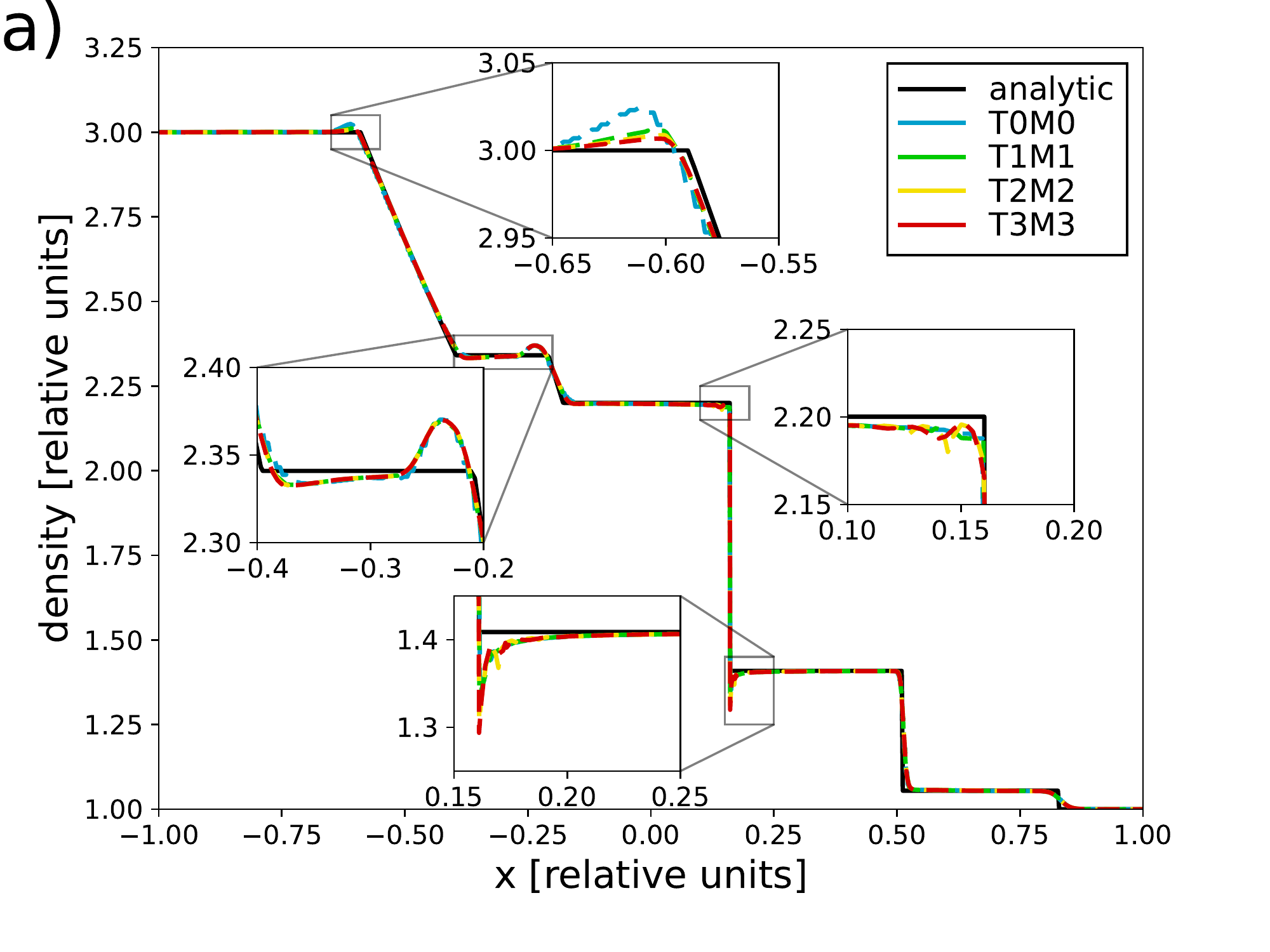}
\includegraphics[width=.49\textwidth]{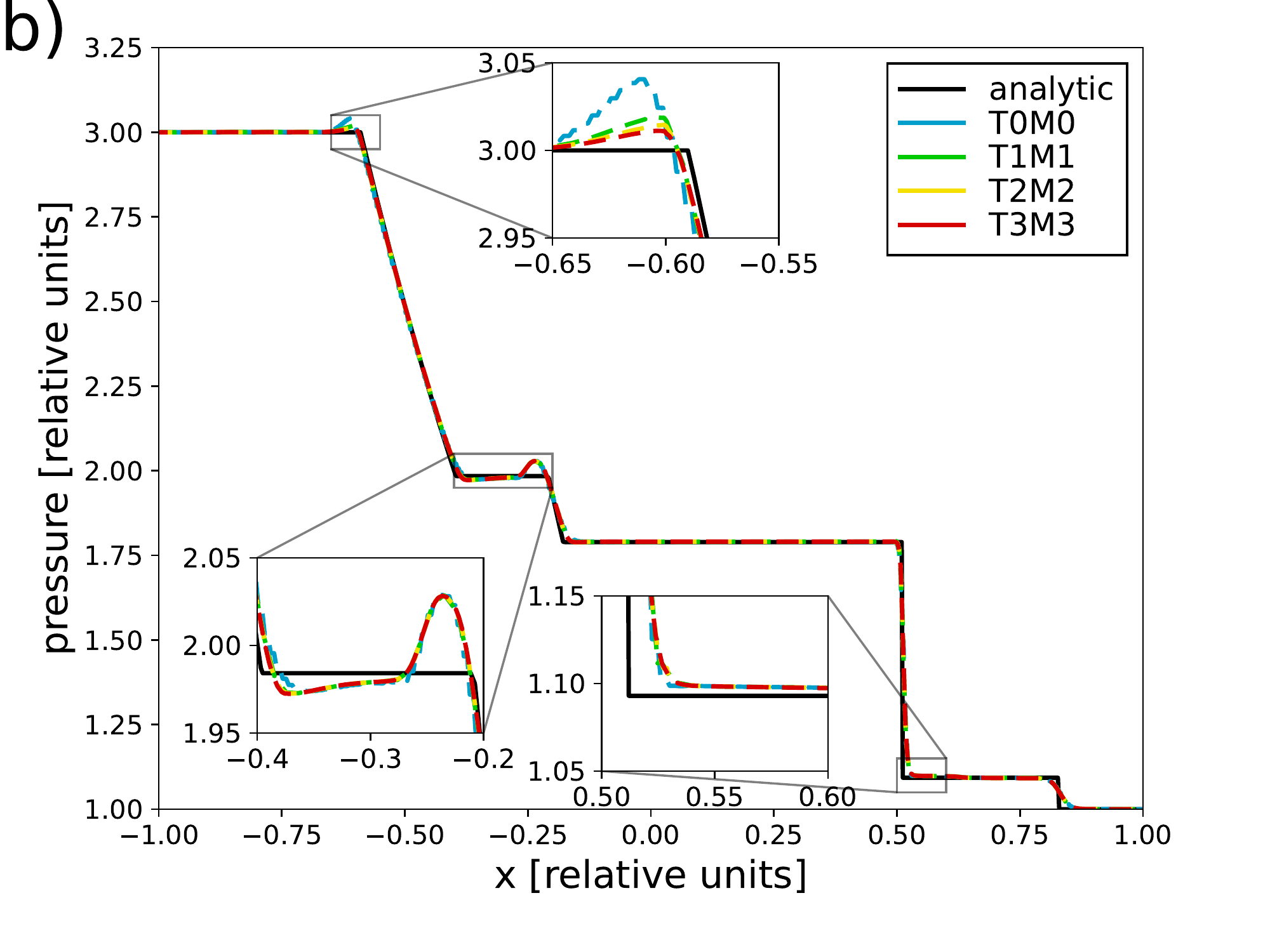}
\\
\includegraphics[width=.49\textwidth]{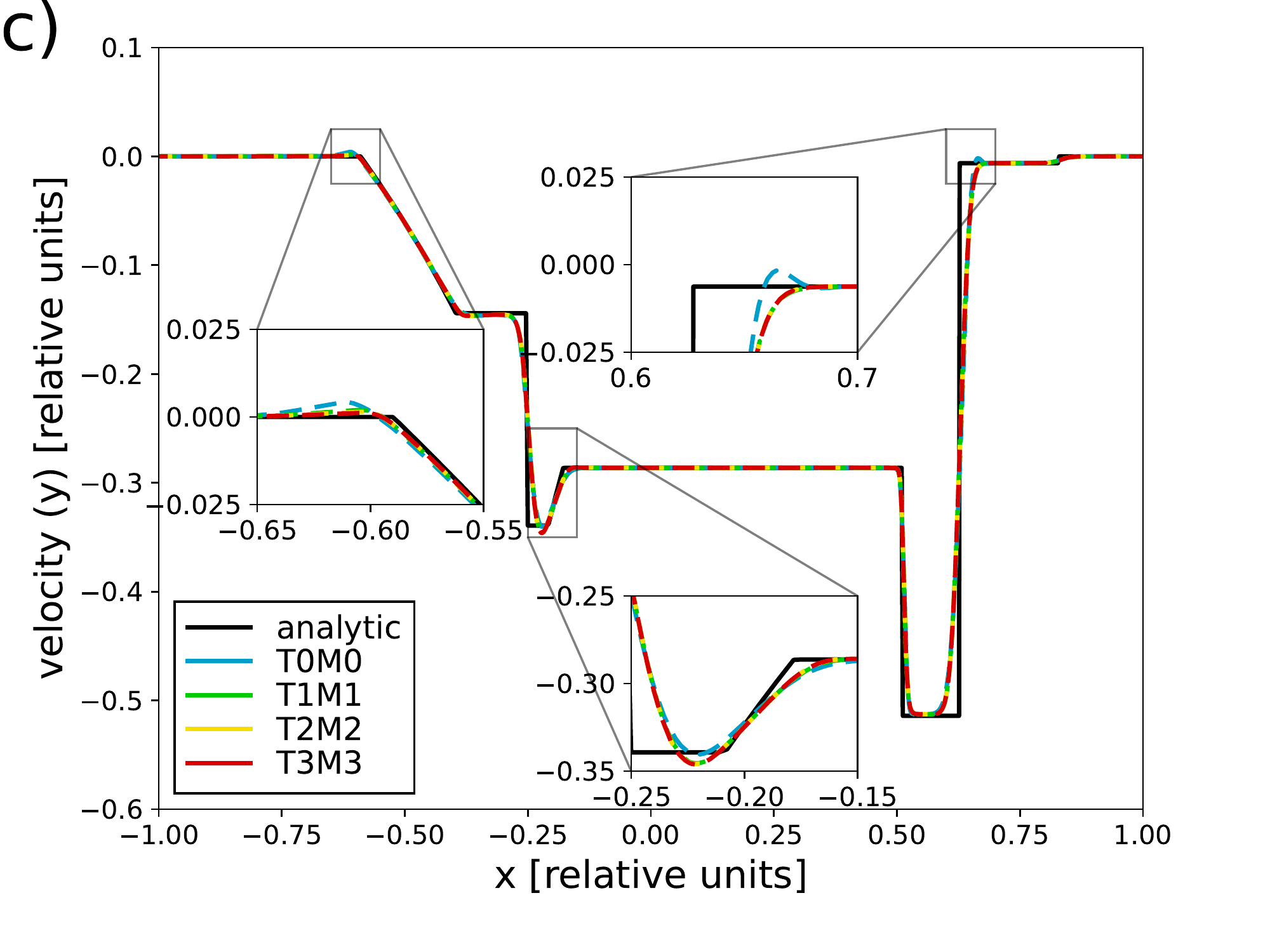}
\includegraphics[width=.49\textwidth]{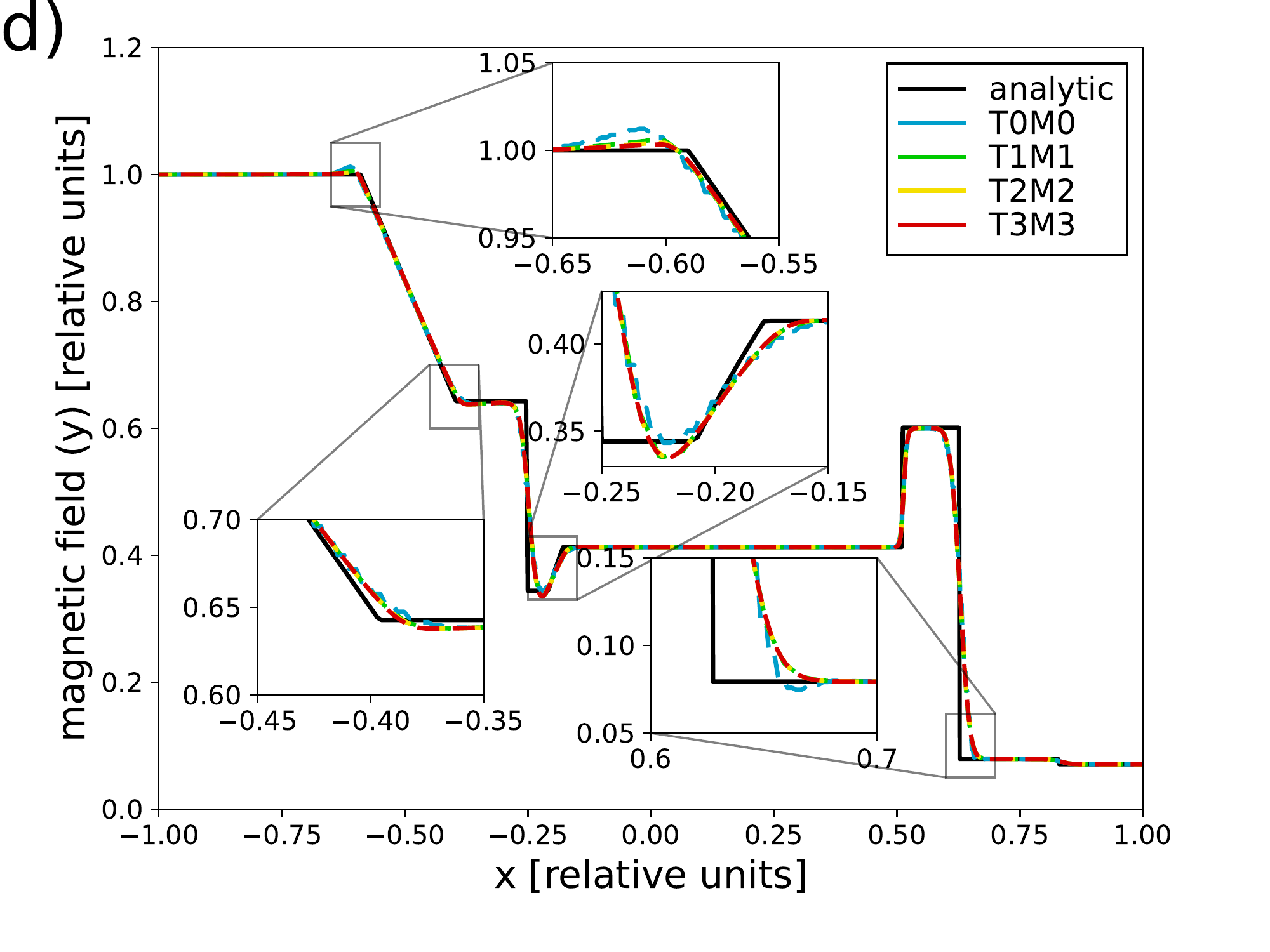}
\\
\includegraphics[width=.49\textwidth]{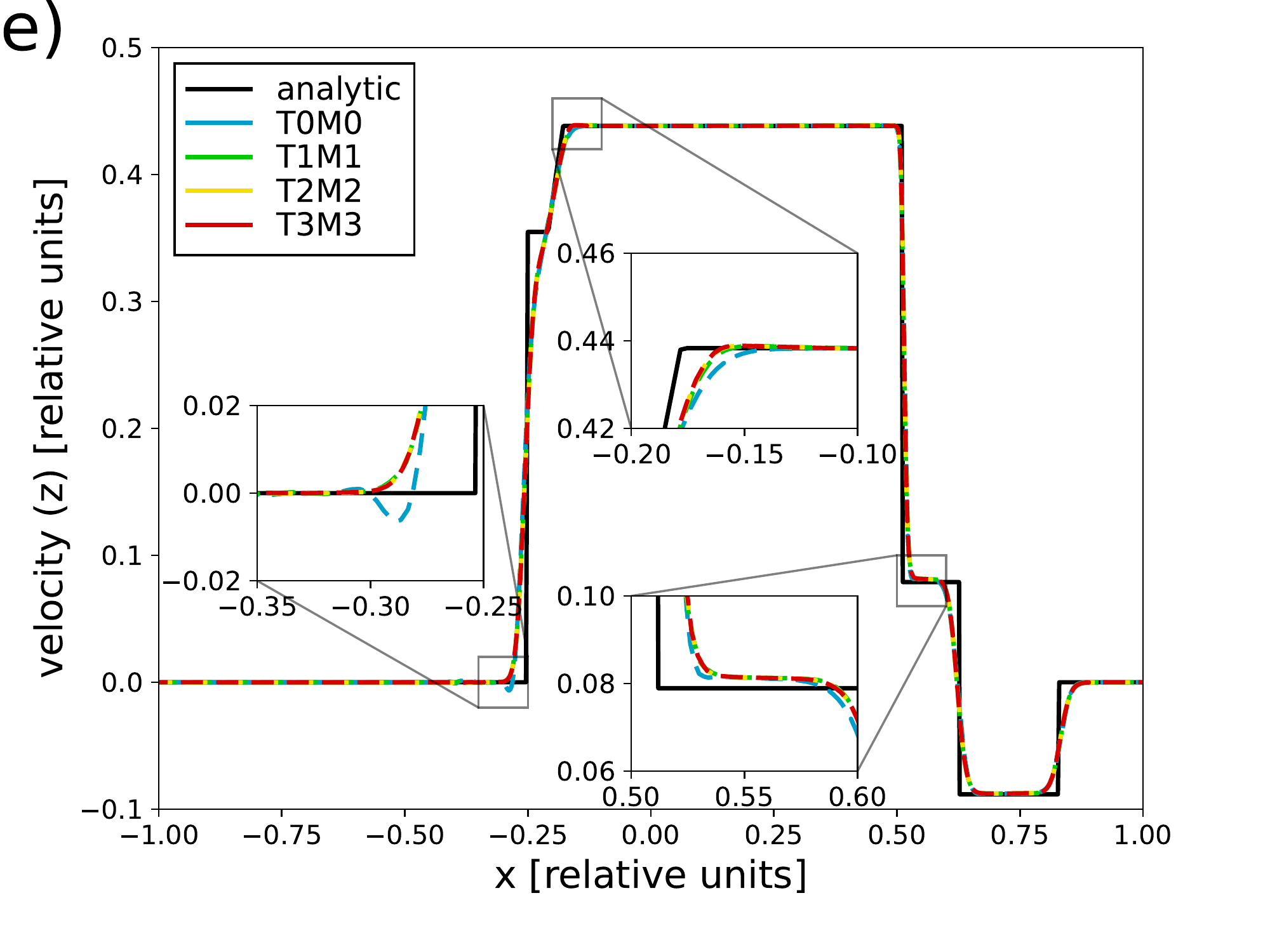}
\includegraphics[width=.49\textwidth]{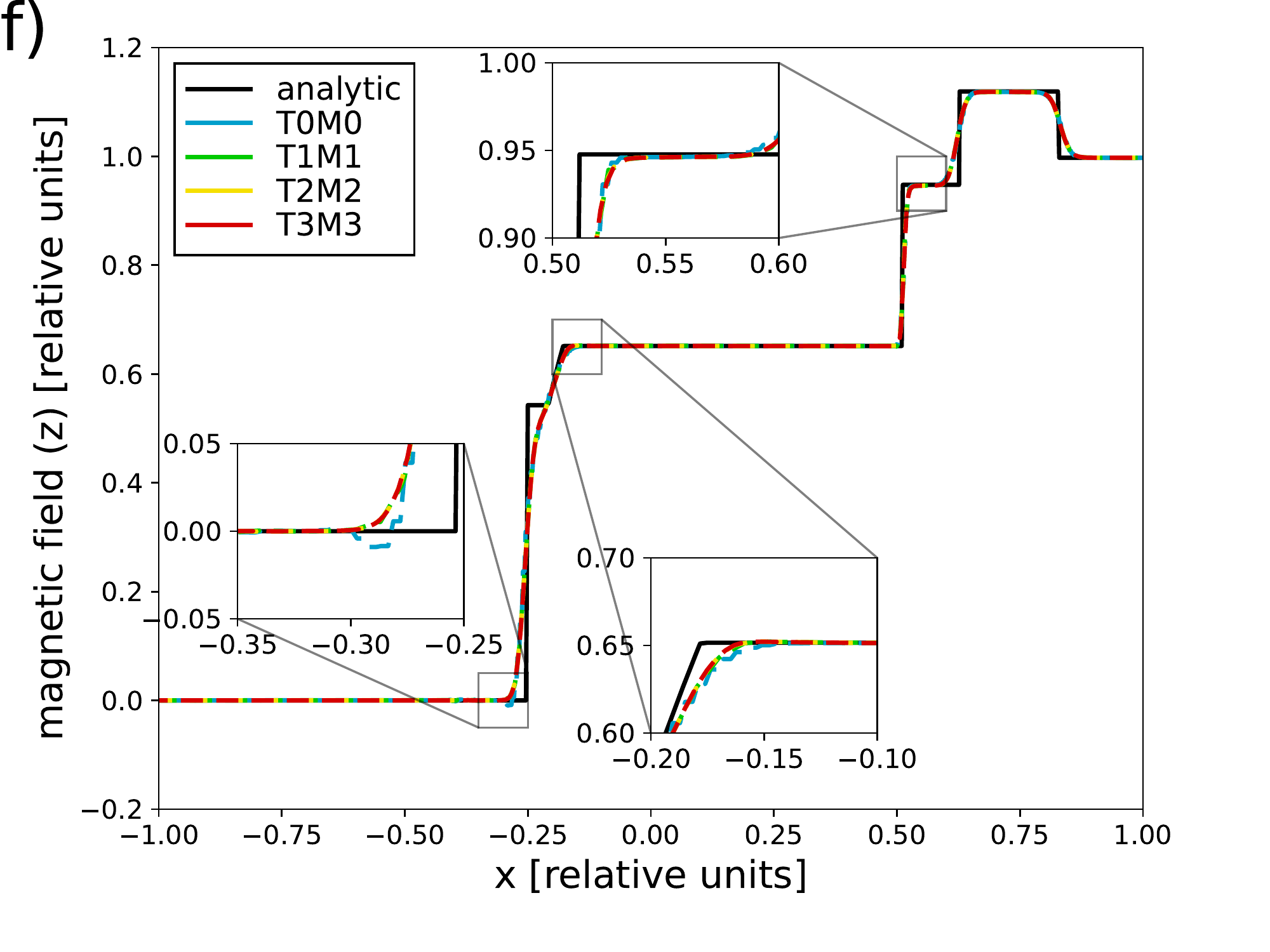}
\caption{Numerical solution of the MHD wave problem: a) density b) pressure c) velocity ($y$ component) d) magnetic field ($y$ component) e) velocity ($z$ component) f) magnetic field ($z$ component). The legend denotes the finite elements used. See the accompanying text for details.}
\label{fig_sim_torrilhon_sol}
\end{figure}

The results of the simulations are presented in \figref{fig_sim_torrilhon_sol} together with the analytic solution \cite{Torrilhon2003,Torrilhon2002}. As already mentioned, the solution involves multiple linear waves. Going from the left, it is a fast rarefaction wave, rotational Alfvénic wave, slow rarefaction wave, contact discontinuity, slow shock wave, rotational Alfvénic wave and fast shock wave. Note that not all waves appear in all figures. For example, the rotational waves are not associated with any normal compression and thus they are only visible in the plots of the transverse velocity and magnetic field. To summarize the results, the curves for all finite elements except the piece-wise constant $T0M0$ are nearly identical. This can be seen as an implication of the Lax theorem limiting the convergence to the first order only at discontinuities. However, it is important that precision is not lost for higher order elements, even though their amount is lower. The zeroth order elements exhibit small over/undershoots near the rotational waves mainly. This phenomenon can be probably attributed to the staggered-like discretization of the transverse magnetic field and velocity according to \ref{app_num_weak_1D}. A slight difference between the elements of different orders can be recognized at the head of the fast rarefaction wave (around $x\approx -0.6$), where the solution is continuous. As can be expected, the overshoot decreases with an increasing order of the elements. On the other hand, the so-called wall heating effect near the contact discontinuity (around $x\approx 0.15$) is  slightly stronger for the higher order elements. This effect originates from the beginning of the simulation, when artificial viscosity was very active and the higher order elements naturally tended to oscillate more strongly due to Runge's phenomenon.


\subsection{MHD Taylor-Green vortex}
\label{sub_sim_taylorgreen}

In order to assess convergence of the scheme, the classical steady-state solution of incompressible, inviscid Navier--Stokes equations known as the Taylor-Green vortex was extended to the coplanar ideal MHD. The magnetic field is specifically chosen as $\vec{B}=\beta \sqrt{\mu_0}\vec{u}$ with $\beta$ being an arbitrary constant, which automatically guarantees divergence-free structure of the field. The stationary solution is then given by the equations:
\begin{subequations}
\begin{align}
\vec{u}_x(x,y) &= \sin(\pi x) \cos(\pi y)
,\label{eqn_sim_taylorgreen_sol_ux}\\
\vec{u}_y(x,y) &= -\cos(\pi x) \sin(\pi y)
,\label{eqn_sim_taylorgreen_sol_uy}\\
p(x,y) &= 1 + \frac{1 - \beta^2}{4} (\cos(2\pi x) + \cos(2\pi y)) - \frac{\beta^2}{2} (\sin^2(\pi x)\cos^2(\pi y) + \cos^2(\pi x)\sin^2(\pi y))
.\label{eqn_sim_taylorgreen_sol_p}
\end{align}
\label{eqn_sim_taylorgreen_sol}
\end{subequations}
Note that unlike the classical formulation for plain hydrodynamics~\cite{Dobrev2010}, the second term appears in \eqref{eqn_sim_taylorgreen_sol_p} to compensate for the magnetic pressure.

Following the methodology of \cite{Dobrev2010}, the solution is extended to the compressible (magneto-)hydrodynamics by means of the method of manufactured solution. In essence, the following energy source term is added to the energy equation to compensate for the mechanical work:
\begin{equation}
S_e = \frac{3}{8}\pi ( \cos(3 \pi x) \cos(\pi y) - \cos(\pi x) \cos(3 \pi y) )
.\label{eqn_sim_taylorgreen_se}
\end{equation}
The equation of ideal gas is considered here with the Poisson constant $\gamma=5/3$, atomic weight $A=1$ and proton number $Z=1$. The density profile is homogeneous with $\rho\equiv 1$. The problem is solved on a unit square with the boundary conditions for zero velocities. Unlike other MHD tests, the model of artificial viscosity is not applied, since the problem is smooth and the subtle work of the artificial viscosity would only deviate the solution from the stationary regime.

\begin{figure}[hbtp]
\centering
\includegraphics[width=.49\textwidth]{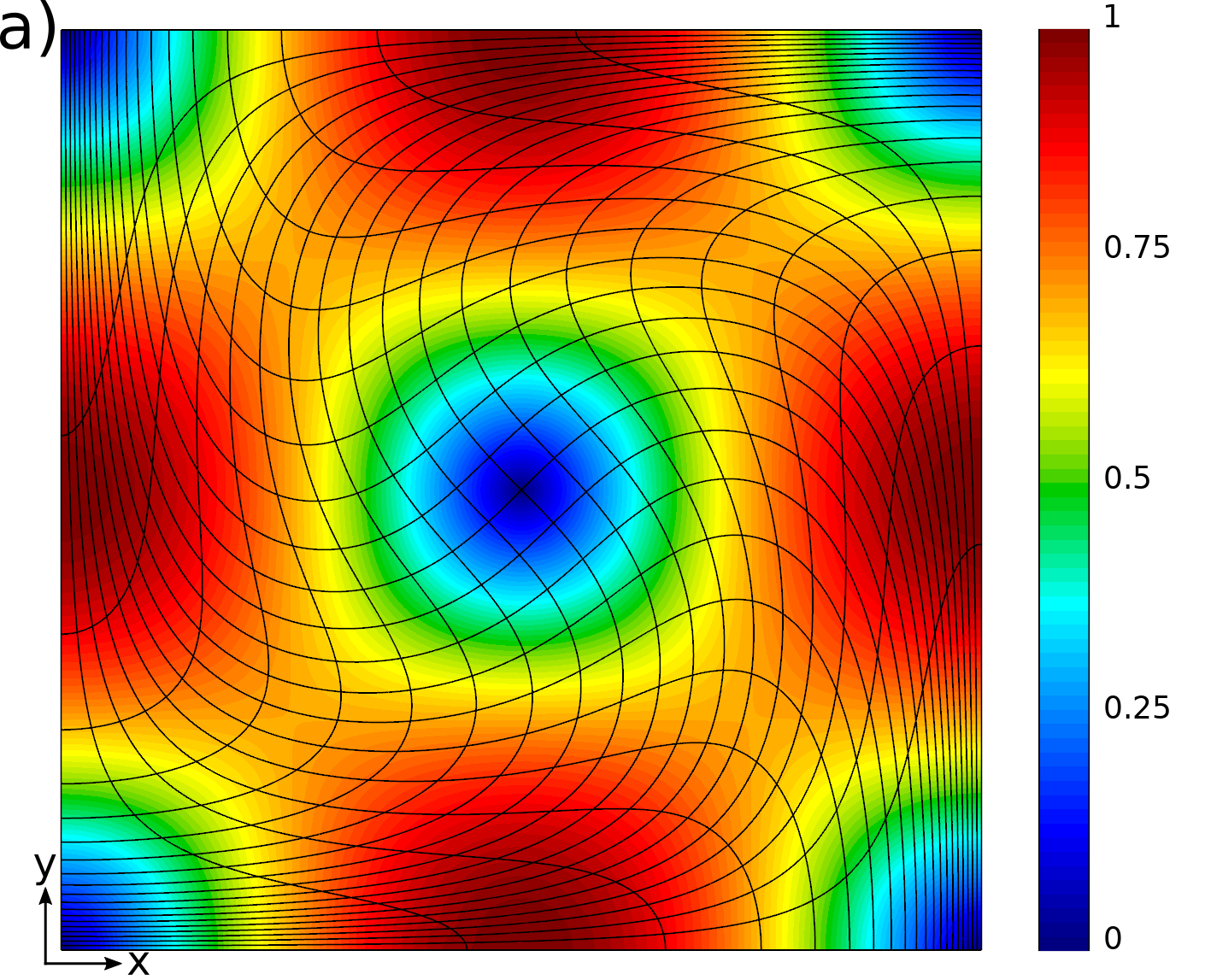}
\includegraphics[width=.49\textwidth]{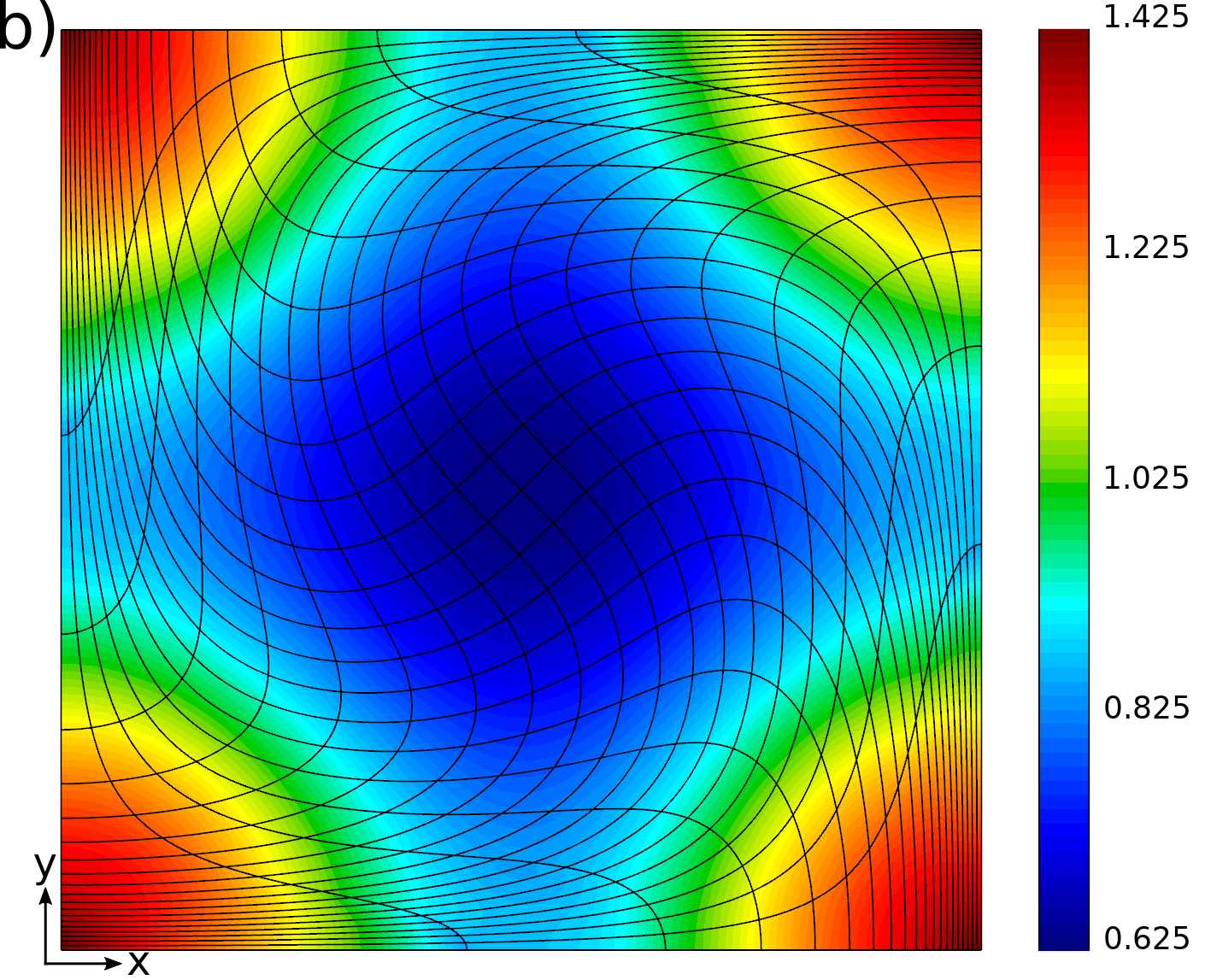}
\caption{Numerical solution of the magneto-hydrodynamic Taylor-Green vortex problem with $\beta=0.5$ at the final time $t=0.75$: a) magnitude of velocity [relative units] b) pressure [relative units]. The computational mesh has $20\times 20$ finite elements of the $T2M2$ family.}
\label{fig_sim_taylorgreen_sol}
\end{figure}

The numerical solution at the final time $t=0.75$ (in relative units) is plotted in \figref{fig_sim_taylorgreen_sol}, where the computational mesh has $20$ quadratic/cubic T2M2 finite elements in each dimension. The factor of the magnetic field is set to $\beta=0.5$, corresponding to the ratio 4:1 between the dynamic and magnetic pressure. The time integration used RK3hc(A.$\alpha$) scheme with the CFL factor 0.5 (see \secref{subsub_num_disc_time}). From the results, the deformation of the curvilinear finite elements due to the mass flow is apparent, but the sufficiently high order of interpolation preserves a good accuracy of the solution without any significant imprints of the mesh. 

\begin{figure}[hbtp]
\centering
\includegraphics[width=.49\textwidth]{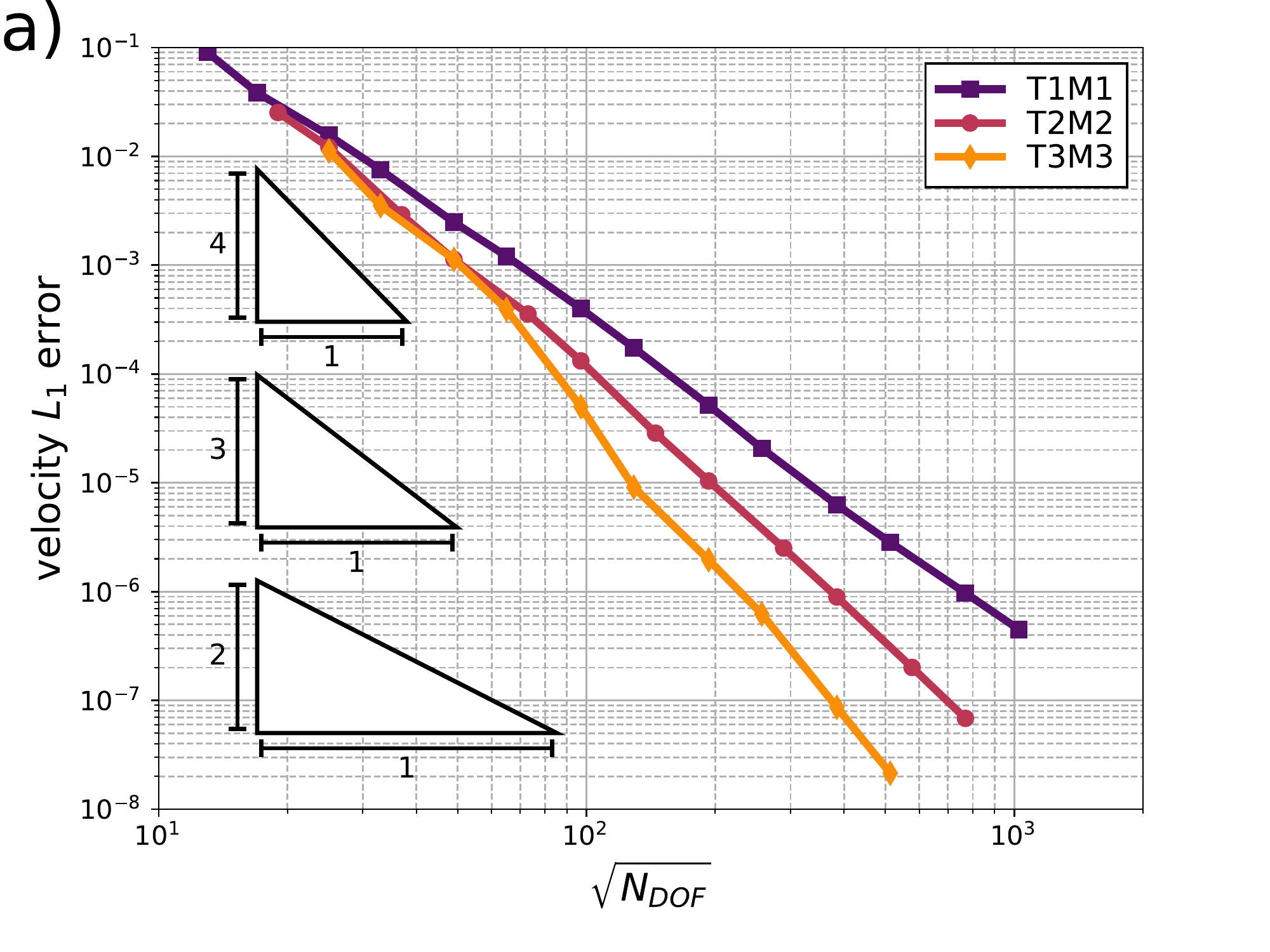}
\includegraphics[width=.49\textwidth]{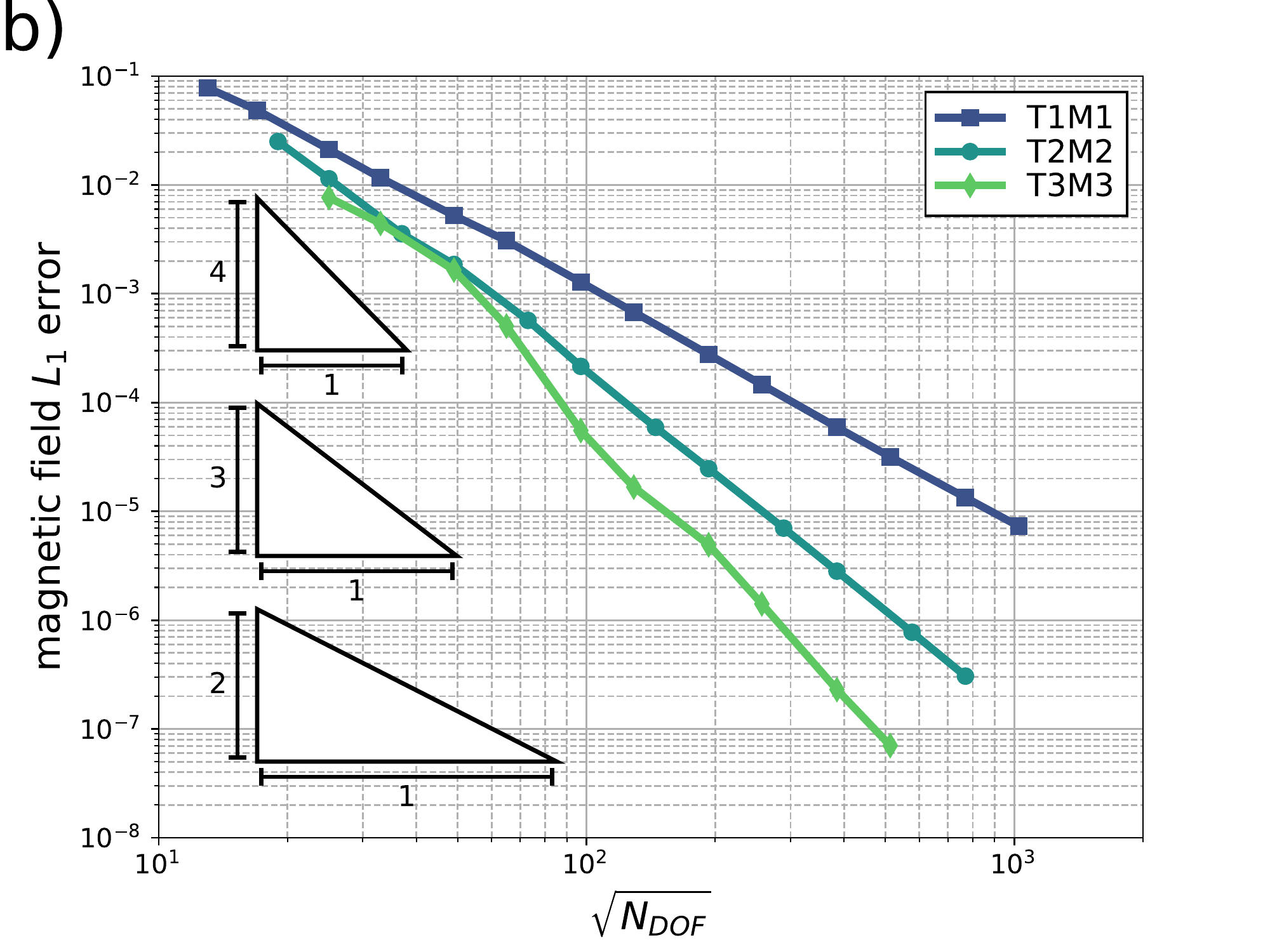}
\caption{Dependencies of the velocity (a) and magnetic field (b) $L_1$ integral error on the number of degrees-of-freedom for the magneto-hydrodynamic Taylor-Green vortex problem with $\beta=0.5$ at the final time $t=0.75$. The finite elements used are denoted in the legend.}
\label{fig_sim_taylorgreen_conv}
\end{figure}

Resilience of the curvilinear finite elements to mesh deformation is also confirmed by the convergence plots in \figref{fig_sim_taylorgreen_conv} for varying number degrees-of-freedom (and the time step due to the applied CFL condition). In particular, the symbol $N_{DOF}$ designates the number of degrees-of-freedom for the $H^1$-conforming basis, which can be related to the number of elements along each axis $N_x, N_y$ through $N_{DOF}=((p+1) N_x+1)((p+1) N_y+1)$ for $TpMp$ elements. The time integration schemes are chosen to match the spatial order of elements, where RK2-Average, RK3hc(A.$\alpha$) and RK4hc(A.$\alpha$) are used used for $T1M1, T2M2$ and $T3M3$ elements respectively (see \secref{subsub_num_disc_time}). The plots clearly indicate convergence rates with the order at least proportional to the order of the thermodynamic/magnetic basis. Moreover, the velocities exceed this rate with the average slope approximately $2.76$ for $T1M1$, $3.48$ for $T2M2$ and $4.29$ for $T3M3$ elements. This points to the fact that the magnetic pressure contributes to the momentum equation \eqref{eqn_phys_dif_mom} unlike the classical problem ($\beta=0$). The order of the magnetic pressure term then surpasses that of the thermal pressure. When the magnetic contribution is not negligible compared the thermal ($\beta \gtrsim 1$), the velocity can benefit from this increase due to its higher by one interpolation order compared to the thermodynamic basis (see the definitions in \secref{subsub_num_disc_spc}). Therefore, the resulting order of convergence ranges between the thermodynamic and kinematic orders of interpolation, where the increase diminishes for higher orders of elements and low values of the $\beta$ parameter.


\subsection{Magnetic diffusion}
\label{sub_sim_magdiff}

Following the previous section, convergence of the proposed numerical scheme is evaluated, but rather the magnetodynamic part is investigated in this case. An initial profile of magnetic field $\vec{B}(\vec{x})=\delta \vec{B} \sqrt{\mu_0}\exp(-|\vec{x}|^2/\sigma_0^2)$ is considered, which is being diffused by means of resistive eddy currents, while the rest of the quantities is constant initially. Despite the fact that the problem is not fully physically realistic, as the magnetic field forms a magnetic monopole, the divergence is conserved (see \secref{subsub_num_semi_md}) and the problem has a simple analytic solution asymptotically:
\begin{align}
\vec{E} &= \frac{\eta}{\mu_0} \nabla\times\vec{B}
, \qquad
\dpd{\vec{B}}{t} = -\nabla\times\vec{E} 
= \frac{\eta}{\mu_0} (\Delta \vec{B} - \nabla( \nabla\cdot\vec{B}))
,\label{eqn_sim_magdiff_EB}\\
\vec{B}(t,\vec{x}) &= \delta\vec{B}_0 \sqrt{\mu_0} \left(\frac{\sigma_B}{\sigma(t)}\right)^{3/2} \exp\left( - \frac{|\vec{x}|^2 }{\sigma(t)^2}\right)
,\qquad
\sigma(t) = \sqrt{\sigma_0^2 + 4 \eta t/\mu_0}
.\label{eqn_sim_magdiff_B}
\end{align}
It can be noticed that the divergence ($\nabla\cdot\vec{B}$) contributes to the solution according to \eqref{eqn_sim_magdiff_EB}, but only through higher orders of the expansion in $\eta_m t / \sigma_0^2$, where $\eta_m=\eta/\mu_0$ is the magnetic diffusivity. Therefore, these terms can be truncated for a sufficiently small ratio. Furthermore, the magnetic Reynolds number $R_m=u \sigma_0 / \eta_m$ must be kept small to prevent convection of the field. The velocity can be approximated from the momentum equation, giving a similar expression to the previous one $R_m \sim t |\delta \vec{B}|^2 / \eta_m$. Finally, the effect of Joule heating must be suppressed, which is proportional to $t/\eta E^2\sim t \eta_m |\delta \vec{B}|^2 / \sigma_0^2$. However, when compared to the magnetic pressure $|\delta \vec{B}|^2/2$, the factor $\eta_m t / \sigma_0^2$ is obtained again. When all these conditions are satisfied, the problem is dominated only by the (physical) magnetic diffusion process.

\subsubsection{Convergence analysis}
\label{subsub_sim_magdiff_conv}

The asymptotic problem is solved on the domain $(-1,1)^d$, where $d$ is the dimension. The width of the profile is $\sigma_0=0.1$ to avoid boundary effects, since the boundary condition on zero tangential magnetic field is applied. The diffusivity is set to $\eta_m=1$ and the magnetic field $\delta\vec{B}=(1, 1, 1)$. To satisfy the stated criteria, the final time is only as short as $t=10^{-10}$. The computation is performed with the RK2-Average scheme and the implicit magnetic diffusion scheme. The time step is set to $\Delta t=10^{-12}$ in order to use a representative number of time steps.

\begin{figure}[hbtp]
\centering
\includegraphics[width=.49\textwidth]{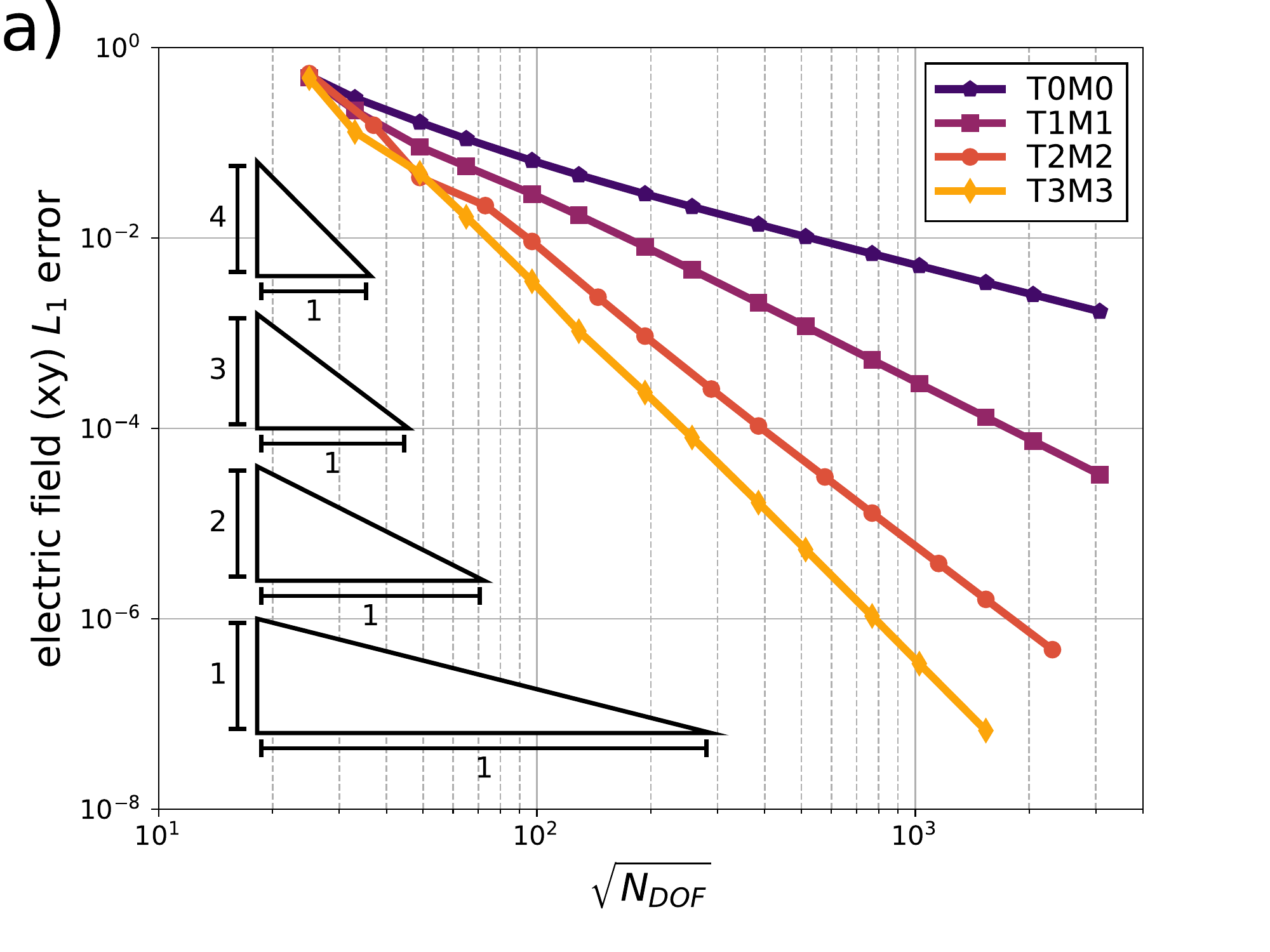}
\includegraphics[width=.49\textwidth]{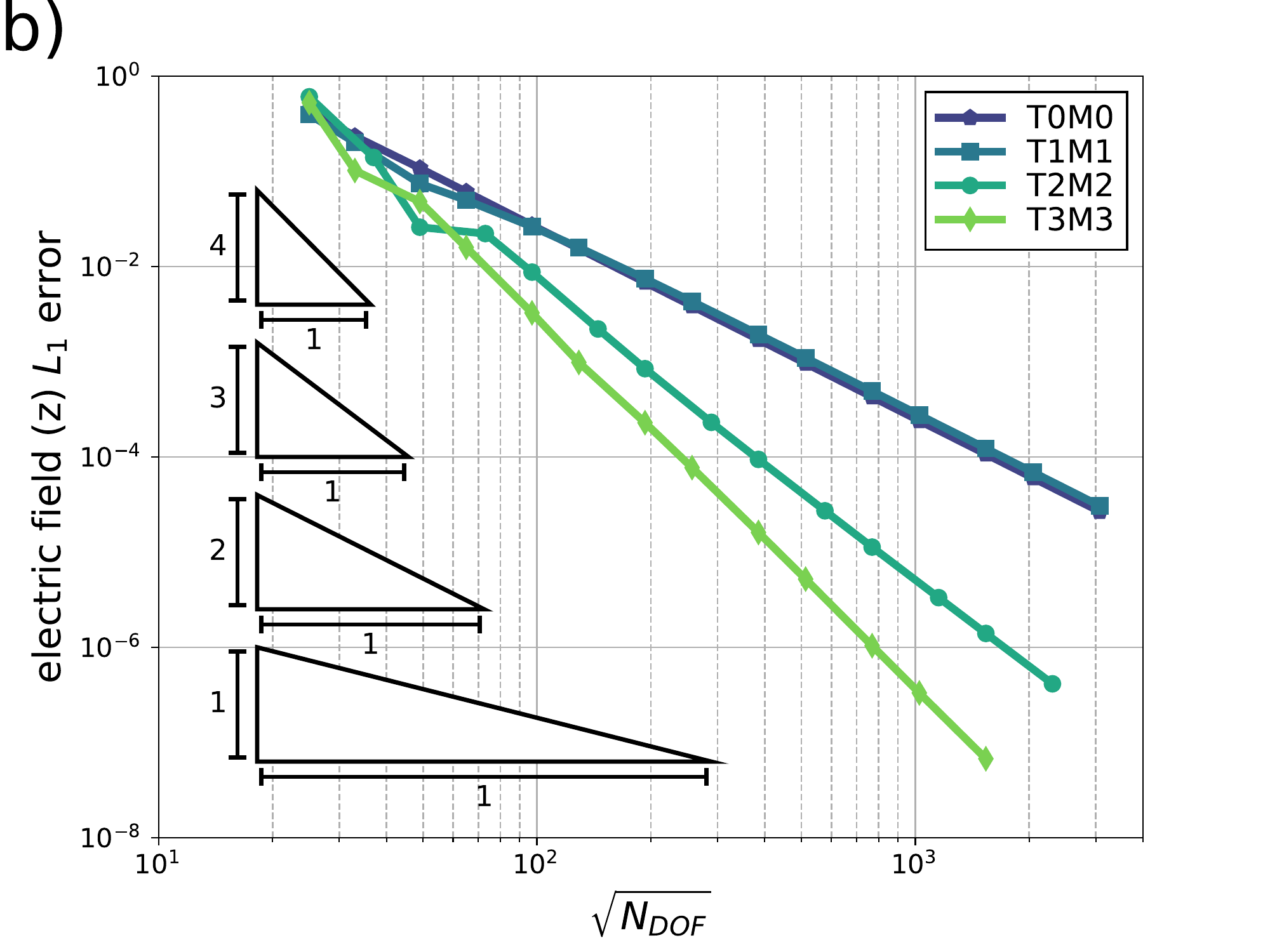}
\caption{Electric field in $xy$-plane (a) and $z$-axis (b) $L_1$ error dependencies on the number of degrees-of-freedom for the magnetic diffusion problem in 2D. The asymptotic setting of the problem is considered with the final time $t=10^{-10}$. The finite elements used are denoted in the legend.}
\label{fig_sim_magdiff_conv}
\end{figure}

Convergence of the electric field in 2D is plotted in \figref{fig_sim_magdiff_conv} for different finite elements. The results show a clear proportionality of the order of convergence to the polynomial order of the electric field elements $p+1$ for $TpMp$ elements. An exception from the rule is posed by the $T0M0$ element in the out-of-plane component. In this configuration, the solution for the electric field benefits from the staggered-like discretization (see \secref{subsub_num_disc_spc}) and converges with the second order. However, the effect is not propagated to convergence of the coplanar magnetic field, which has an insufficient order of the elements and converges only with the first order (not shown).

\begin{figure}[hbtp]
\centering
\includegraphics[width=.49\textwidth]{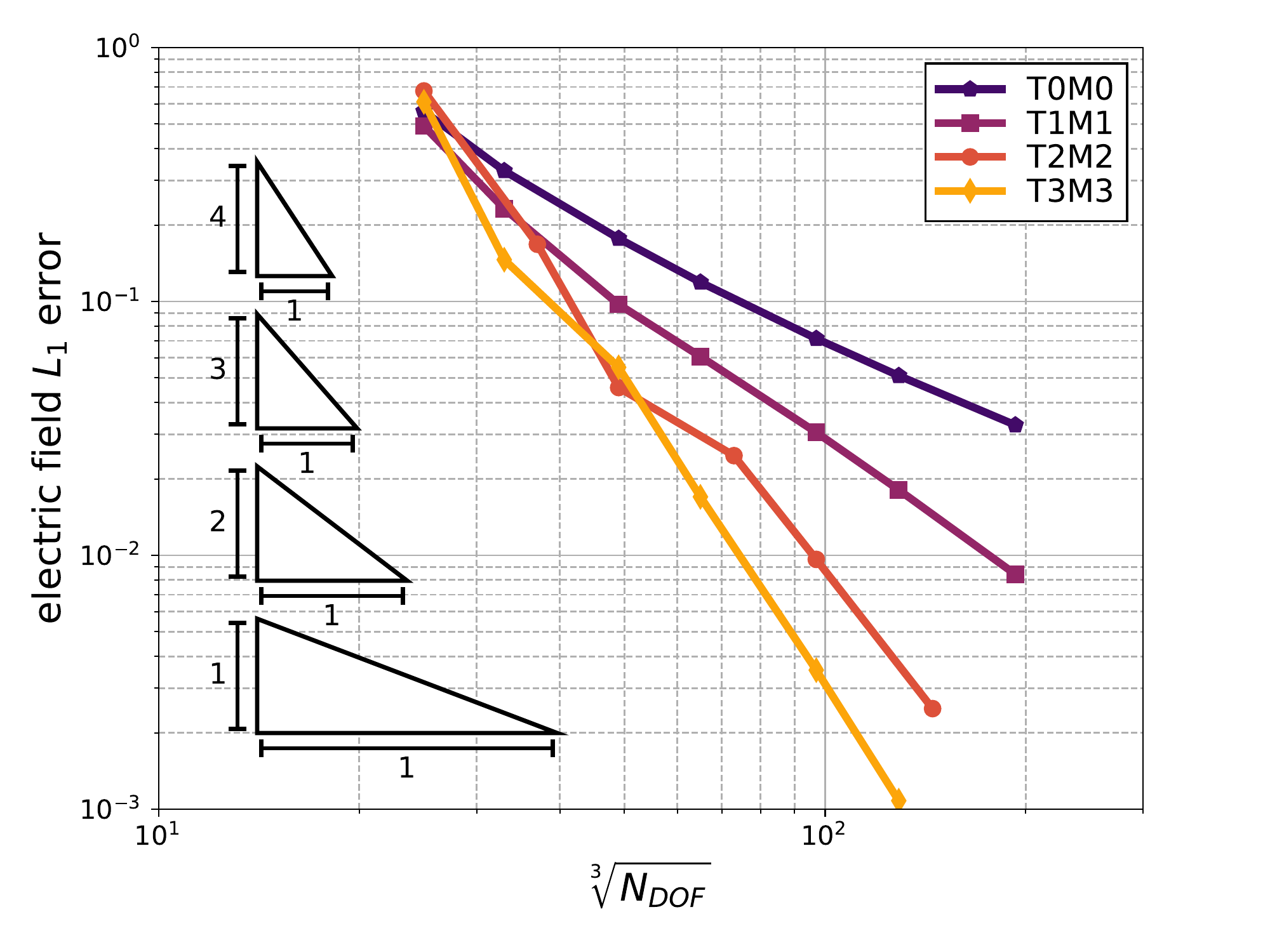}
\caption{Electric field $L_1$ error dependencies on the number of degrees-of-freedom for the magnetic diffusion problem in 3D. The asymptotic setting of the problem is considered with the final time $t=10^{-10}$. The finite elements used are denoted in the legend.}
\label{fig_sim_magdiff_conv3D}
\end{figure}

The results for the 3D case are presented in \figref{fig_sim_magdiff_conv3D}. The trends closely resemble the coplanar electric field in 2D, where the order of convergence is given by the polynomial order of the electric field elements. The increased convergence rate observed for the out-of-plane component in 2D is not replicated here.

\subsubsection{Energy conservation}
\label{subsub_sim_magdiff_en}

In addition to the asymptotic convergence analysis, an elucidating insight is obtained from the non-asymptotic case, where conservation of energy can be studied. The parameters are kept identical, but the final time is increased to $t=0.02$. Moreover, the width of the initial Gaussian profile is increased to $\sigma_0=0.3$, as the boundary effects do not contribute to the energy balance and the boundary conditions can be tested this way. Unlike the previous case, where internal energy of the homogeneous medium did not play a role, it becomes essential for evaluation of the energy exchange process and it is set equal to the analytic integral of the magnetic energy initially. Finally, the conservative time-stepping using RK2-Average scheme is applied with the step $\Delta t = 4\cdot10^{-4}$, despite the fact that the hydrodynamic motion is still minuscule ($R_m\approx 0.02$).

\begin{figure}[hbtp]
\centering
\includegraphics[width=.49\textwidth]{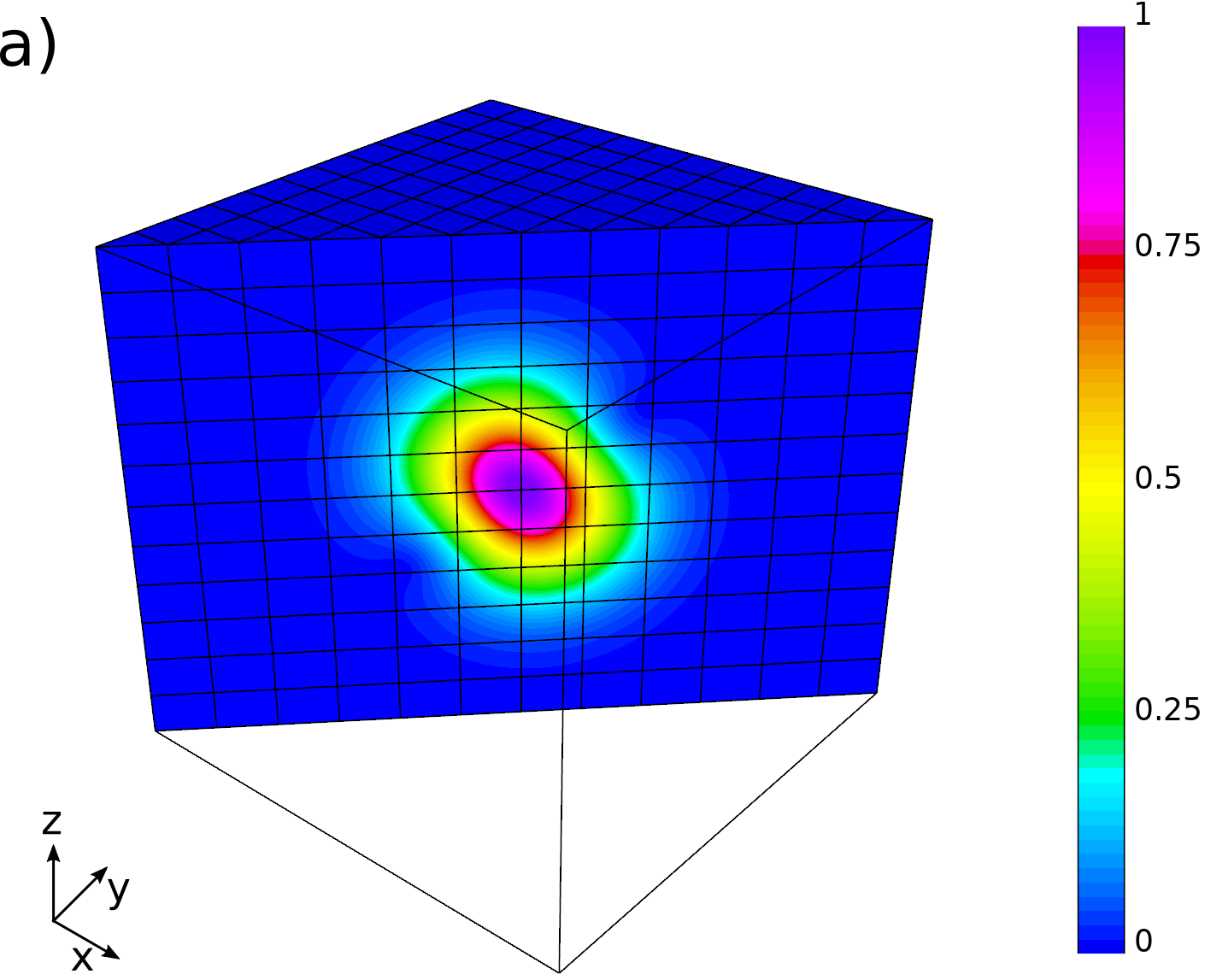}
\includegraphics[width=.49\textwidth]{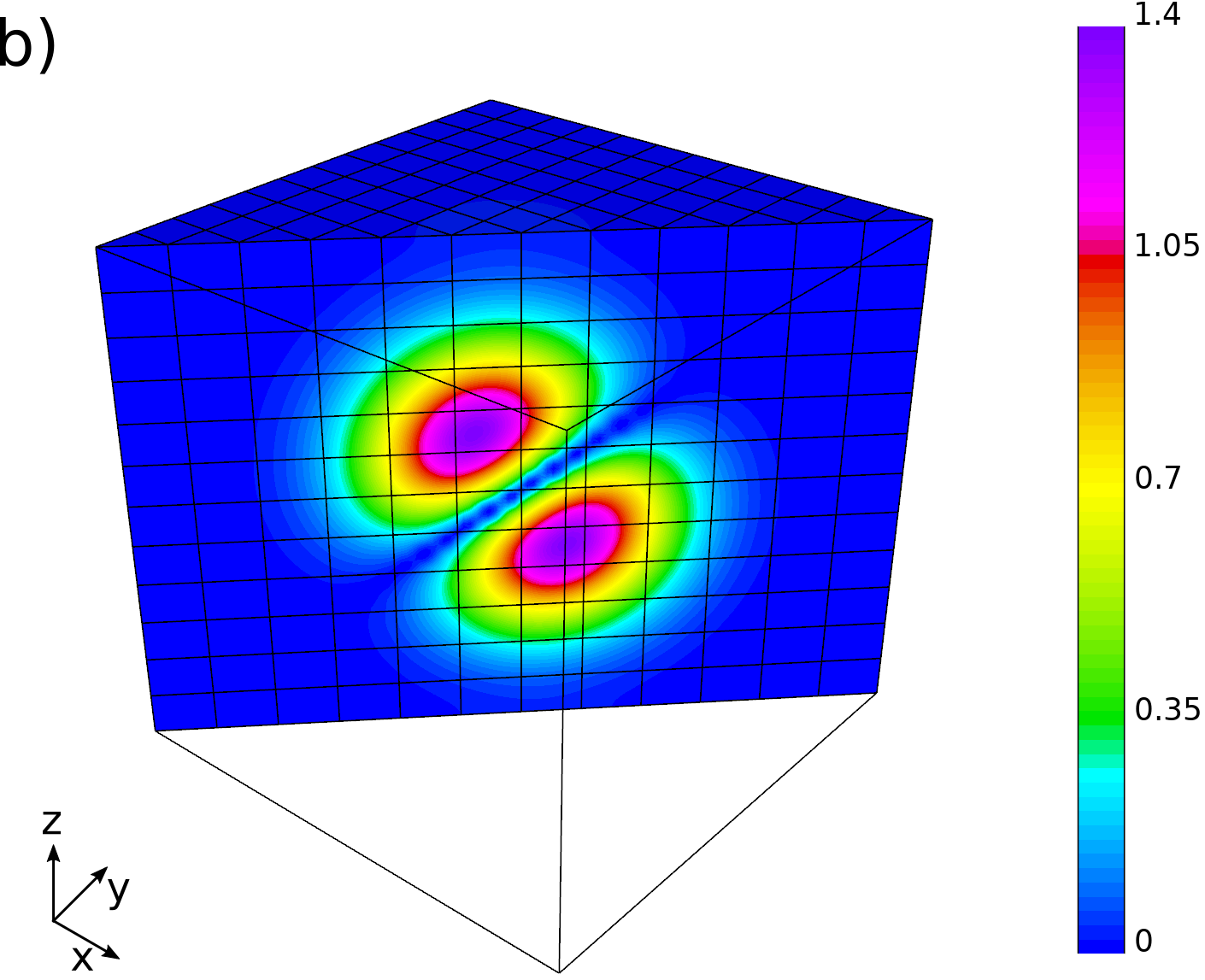}
\caption{Profiles of electric field (a) and magnetic field (b) in the non-asymptotic magnetic diffusion problem at the final time $t=0.02$ (in relative units). The computational mesh consists of $12\times12\times12$ finite elements of the $T3M3$ family.}
\label{fig_sim_magdiff_en_3D}
\end{figure}

The given parameters correspond to the dimensionless factor $\eta_mt/\sigma_0^2\doteq 0.22$, which indicates an advanced state of diffusion. The solution in 3D at the final time is shown in \figref{fig_sim_magdiff_en_3D}. Despite the low number of elements, details of the solution are captured relatively well due to the high order elements used. An asymmetry induced by the divergence term in \eqref{eqn_sim_magdiff_EB} is apparent too, pointing to the fact that the asymptotic solution \eqref{eqn_sim_magdiff_B} is no longer applicable.

\begin{figure}[hbtp]
\centering
\includegraphics[width=.49\textwidth]{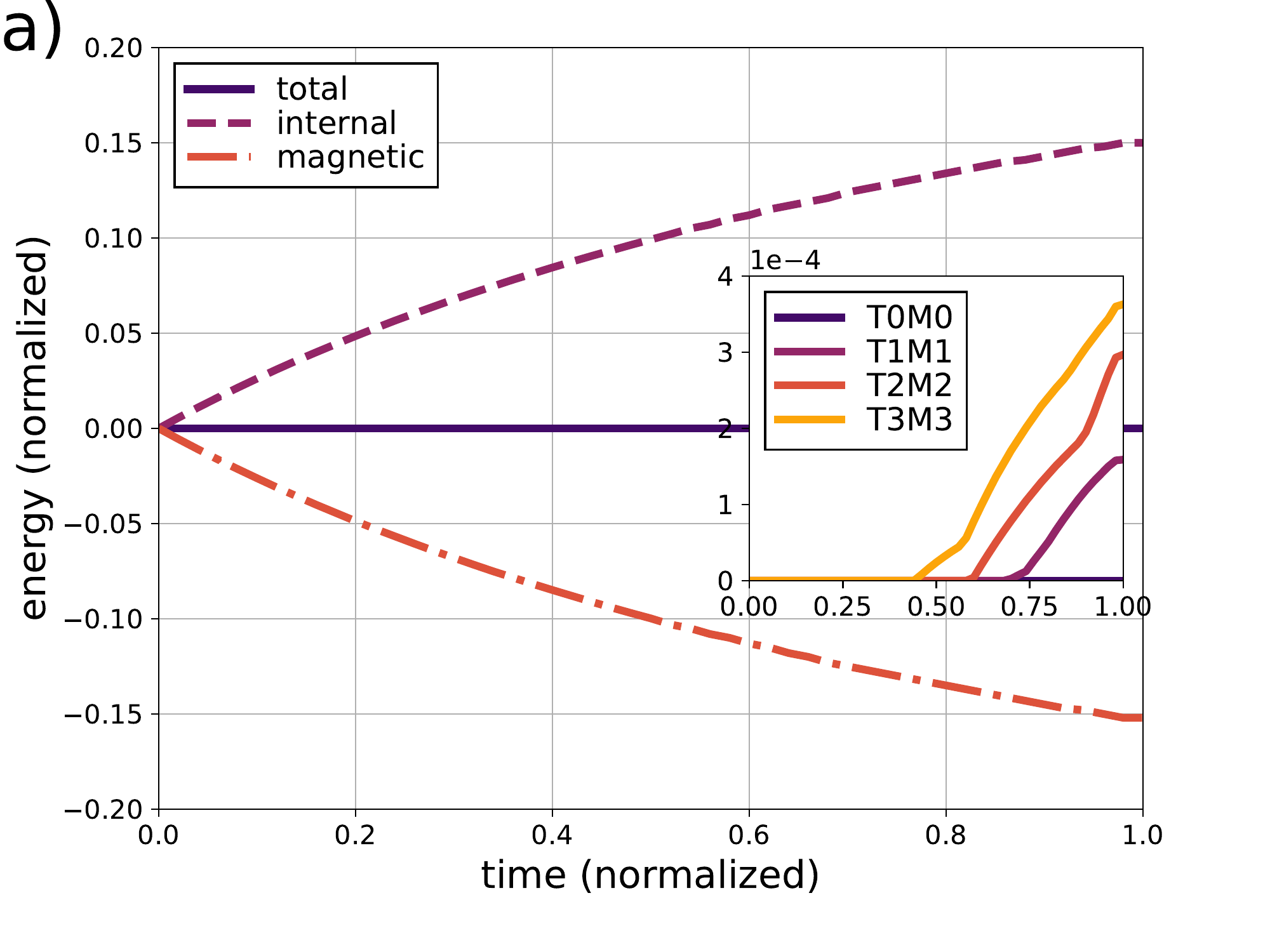}
\includegraphics[width=.49\textwidth]{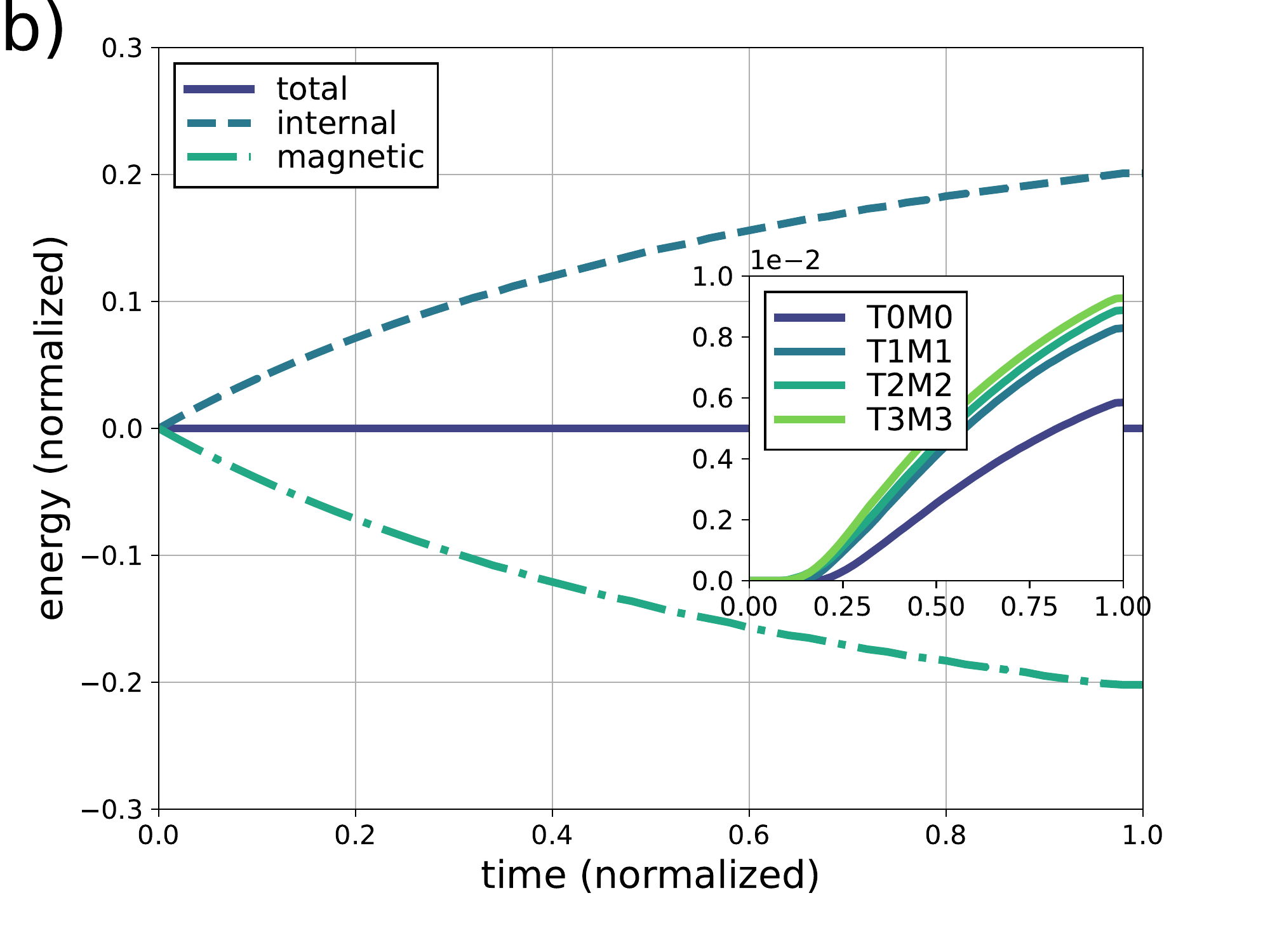}
\caption{Evolution of total, internal and magnetic energy in the magnetic diffusion problem in 2D (a) and 3D (b). The energies are relative to their initial values and normalized to the value of total energy. Time is normalized to the final time $t=0.02$. The inset plot shows the total energy deviation for different finite elements for simulations \textit{without} the Poynting vector term. The resolution is 48 $T0M0$, 24 $T1M1$, 16 $T2M2$ and 12 $T3M3$ finite elements in each dimension (the internal and magnetic energy curves are nearly identical and are not distinguished here).}
\label{fig_sim_magdiff_en_en}
\end{figure}

\begin{figure}[hbtp]
\centering
\includegraphics[width=.49\textwidth]{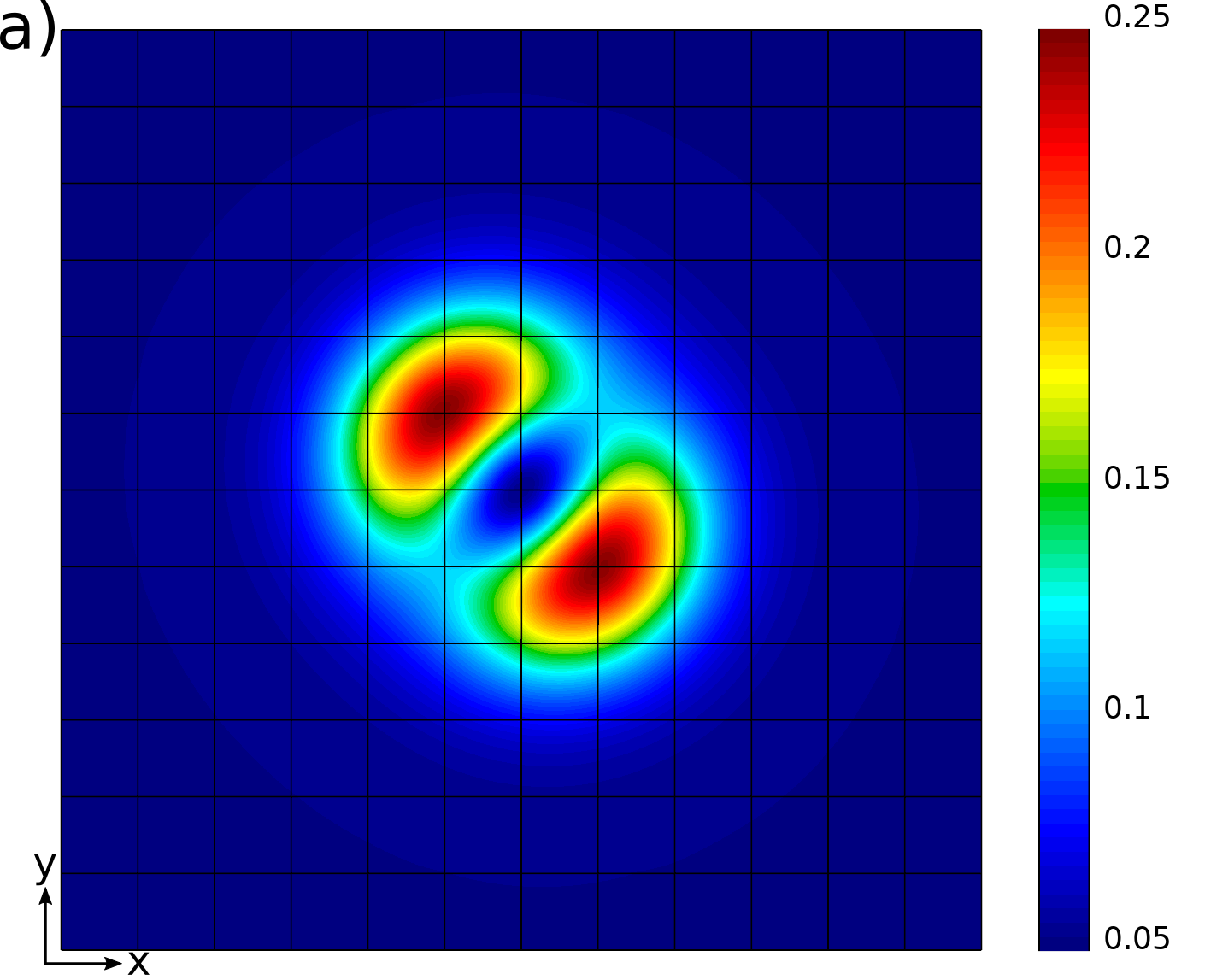}
\includegraphics[width=.49\textwidth]{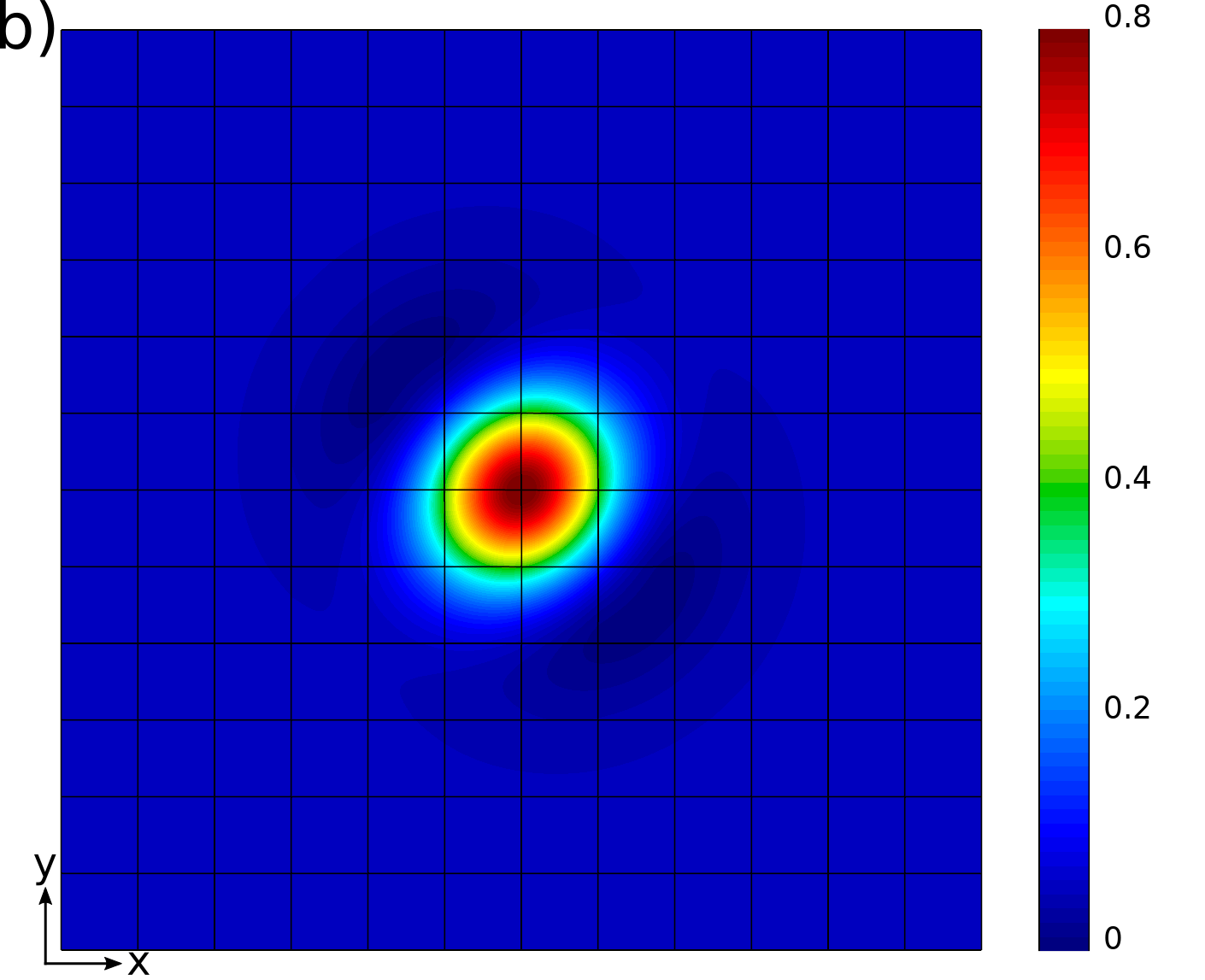}
\caption{Specific internal energy profiles (in relative units) for the non-asymptotic magnetic diffusion problem in 2D: a) \textit{with} and b) \textit{without} the Poynting vector term. The resolution is $12\times 12$ elements of the $T3M3$ family. The snapshots are taken at the final time $t=0.02$.}
\label{fig_sim_magdiff_en_eps}
\end{figure}

Departure from the linear regime is also confirmed by the energy plots in \figref{fig_sim_magdiff_en_en}. The Joule heating process, converting the magnetic to internal energy, slowly decelerates and ends with the total amount of energy exchanged being about $15~\%$ in 2D and $20~\%$ in 3D. The total energy is conserved throughout the process due to the construction of the scheme (see \secref{subsub_num_semi_cons}). However, this is only true when the Poynting vector term is included in \eqref{eqn_num_semi_eps}. Although it is neglected in the electrostatic approximation, it arises from the dual formulation of  Joule heating in this case. When the term $\mat{S}_{ij\cdot}\vcb{E}_i\vcb{B}_j$ is omitted, energy conservation is violated, as can be observed on the inset plots of \figref{fig_sim_magdiff_en_en}. This effect is more pronounced for the higher order elements, which better resolve the gradients. Although the overall effects are quantitatively insignificant (of the order $10^{-4}$ in 2D and $10^{-3}$ in 3D), the spatial profiles of specific internal energy differ completely, as can be seen in \figref{fig_sim_magdiff_en_eps}. Essentially, the heating term is then proportional to $\sim \vec{B}^2$ rather than to $\sim \vec{E}^2$ as it is supposed to be. Therefore, the maximum is located at the center of the domain instead of the peripheral area. Moreover, regions of cooling (decreasing energy) can be noticed. This stresses the importance of the Poynting vector term in \eqref{eqn_num_semi_eps}, which cannot be omitted.

\subsection{MHD advection-diffusion problem}
\label{sub_sim_advdiff}

The previous tests concentrated on verification of the properties of the ideal magnetohydrodynamics and resistive magnetodynamics separately. The purpose of this test is to evaluate performance of the scheme on a coupled advection-diffusion problem. Moreover, the temporal convergence (in the CFL factor $C_{CFL}$) is measured in addition to the spatial convergences from \secref{sub_sim_taylorgreen} and \secref{sub_sim_magdiff}. However, design of such problem is challenging, since even a simple Riemann problem has a complex solution, as shown in \secref{sub_sim_torrilhon}. Therefore, the physical settings of the problem are simplified to only non-ideal rotational Alfvénic wave, while the longitudinal motion is suppressed, i.e. the magnetic pressure is compensated by the thermal one. Furthermore, the conditions are maintained isothermal over the course of the simulation.

In particular, the initial conditions for the magnetic field along $y$ axis and the electric field along $z$ axis are following:
\begin{align}
B_y(x) &= \frac{B_o + B_c}{2} + \frac{B_o - B_c}{2} \erf\left(\frac{\abs{x} - x_0}{\sqrt{4\eta_m t_0}}\right)
,\label{eqn_sim_advdiff_By0}\\
E_z(x) &= \sign(x) \frac{B_o - B_c}{2} \sqrt{\frac{\eta_m}{t_0 \pi}} \exp\left(-\frac{(\abs{x}-x_0)^2}{4\eta_m t_0}\right)
,\label{eqn_sim_advdiff_Ez0}
\end{align}
where the magnetic diffusivity is set to $\eta_m=\eta/\mu_0=0.004$, the central magnetic field is $B_c=0.01$ and the outer magnetic field $B_o=0$ (with corresponding boundary conditions). Due to absence of longitudinal motion, the magnetic field along $x$ axis $B_x=0.2$ and the density $\rho_0=1$ are constant. The simulations are performed on the domain $(-1.5, 1.5)$ till the final time $t=1$. As the profiles are smooth, artificial viscosity is not applied.

\begin{figure}[hbtp]
\centering
\includegraphics[width=.5\textwidth]{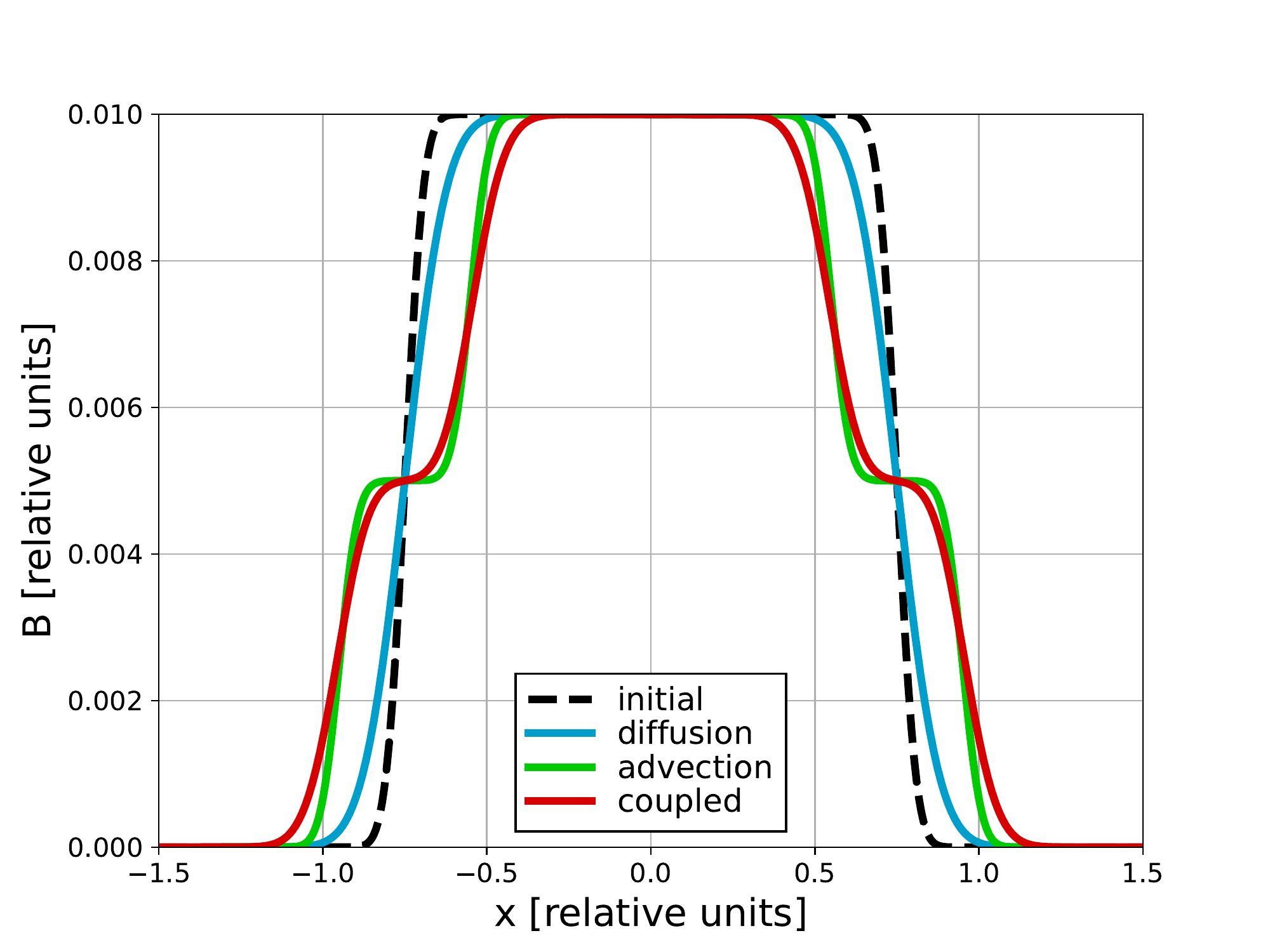}
\caption{Spatial profiles of the magnetic field in the advection-diffusion problem: initial profile (dashed black), analytic diffusion solution (blue), analytic advection solution (green), numerical solution ($C_{CFL}=0.25$) of the coupled advection-diffusion problem (yellow). The resolution is $300$ finite elements of $T3M3$ type.}
\label{fig_sim_advdiff_sol}
\end{figure}

\begin{figure}[hbtp]
\centering
\includegraphics[width=.48\textwidth]{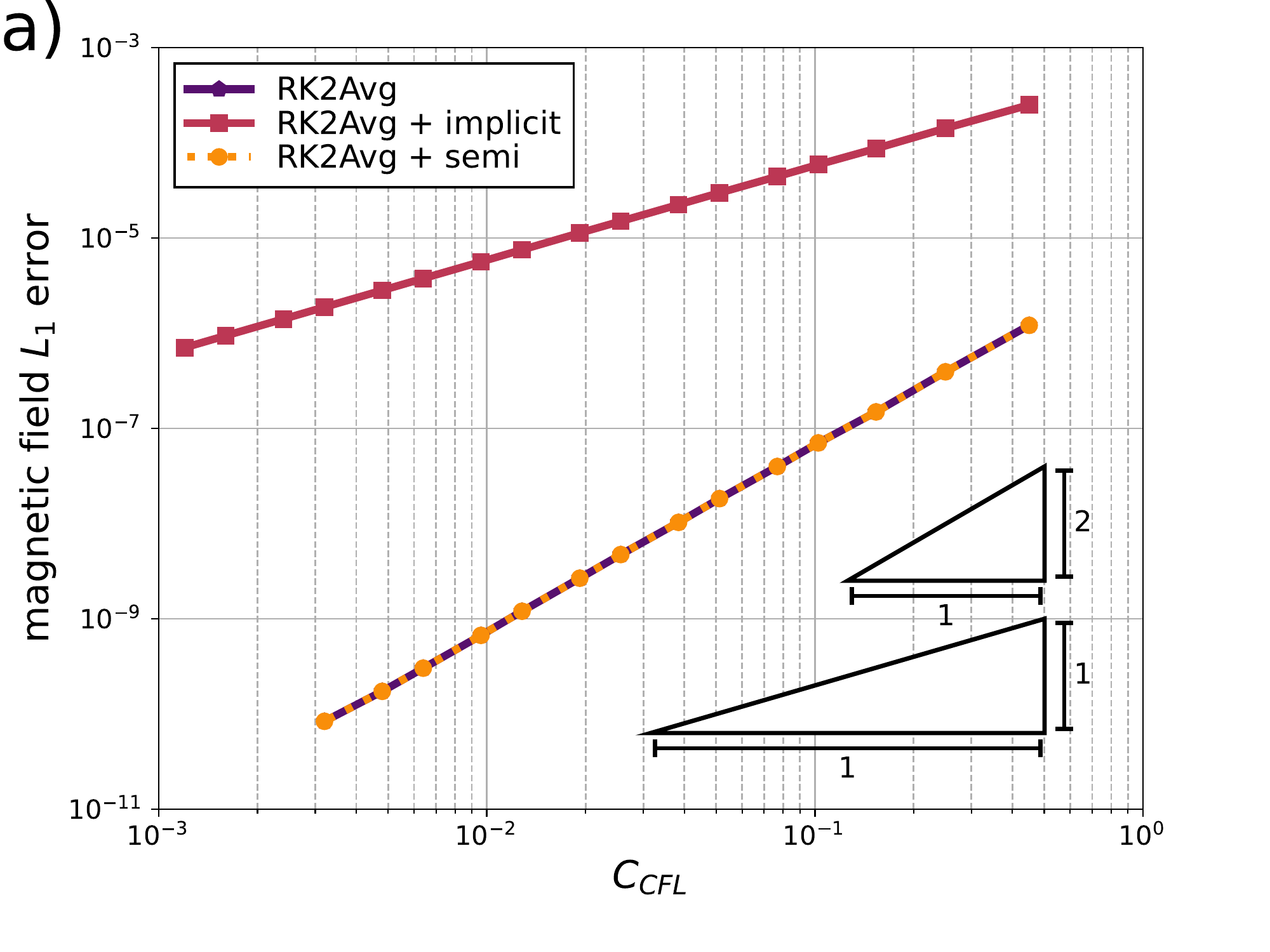}
\includegraphics[width=.48\textwidth]{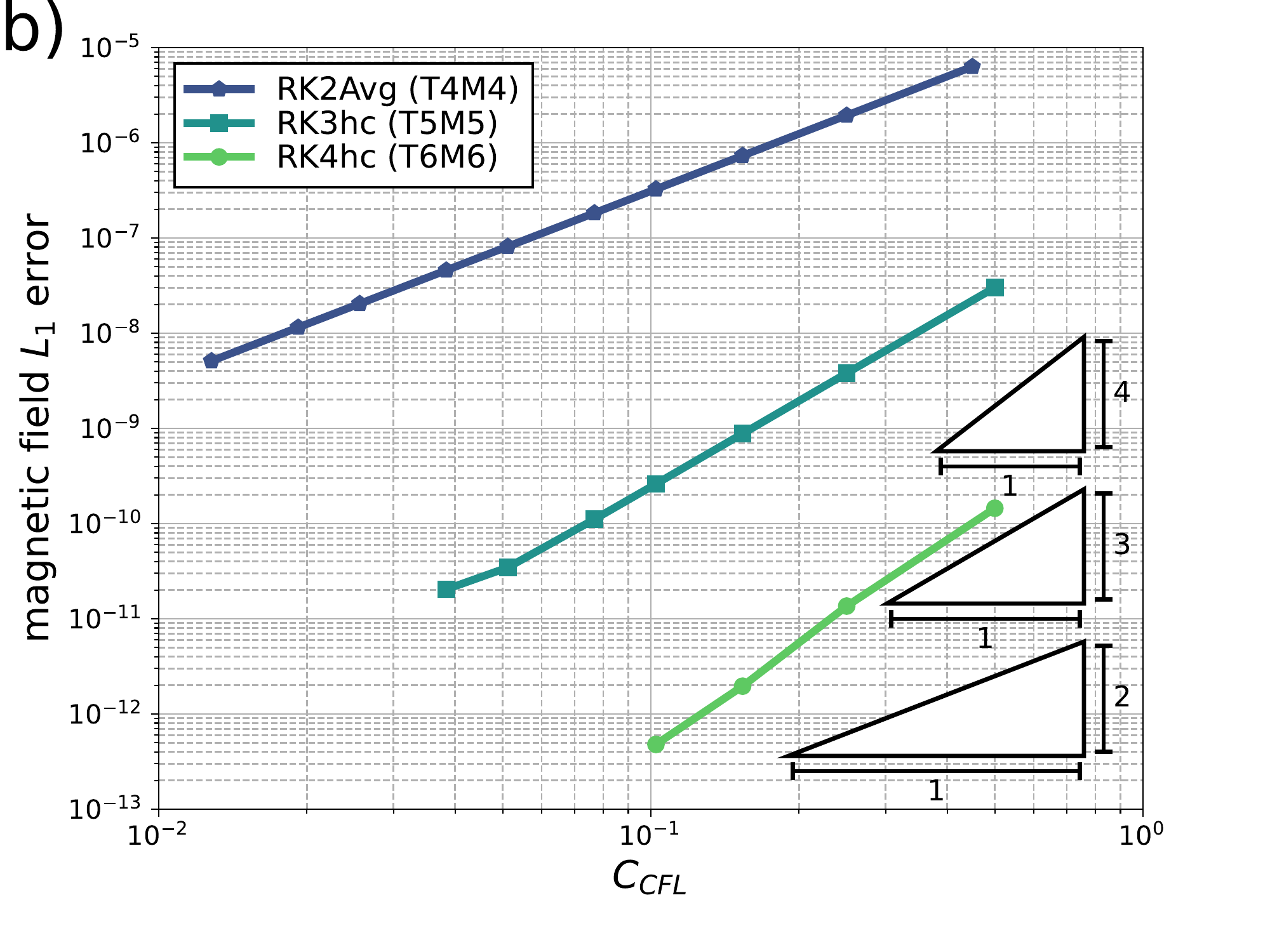}
\caption{Temporal convergence of the magnetic field to the analytic solution in the advection-diffusion problem: a) in the diffusion limit b) in the advection limit. The legend denotes the applied time integration scheme (see the accompanying text for details). The resolution is $300$ finite elements of $T4M4$, $T5M5$ and $T6M6$ type for the RK2-Average, RK3hc, RK4hc schemes, respectively.}
\label{fig_sim_advdiff_advdiff}
\end{figure}

It can be recognized that \eqref{eqn_sim_advdiff_By0} and \eqref{eqn_sim_advdiff_Ez0} are analytic solutions of the magnetic diffusion of a rectangular profile with the step located at $x_0=0.75$. Consequently, the problem has an analytic solution in the diffusion limit ($B_x=0$), where only the time parameter can be prolonged from $t_0=0.25$. The initial conditions and solution are depicted in \figref{fig_sim_advdiff_sol}. The numerical results shown in \figref{fig_sim_advdiff_advdiff} confirm that the RK2-Average scheme converges with the second order. This behavior is identical with the splitted variant of RK2-Averaged scheme with the semi-implicit magnetodynamic scheme ($\alpha=1/2$), which is applied on each of the time levels (before the hydrodynamic update). Contrary, the splitted implicit scheme ($\alpha=1$) attains only the first order of convergence. The relatively high resolution ($300$ finite elements of the $T4M4$ type) is chosen to reach the regime, where the time integration error dominates.

The solution in the advection limit ($\eta_m=0$) has the form of two counter-propagating waves on each side with the Alfvén velocity $v_A=B_x/\sqrt{\rho_0\mu_0}$. The $y$ axis components of the magnetic field and velocity read:
\begin{align}
B_y(x,t) &= \frac{B_o + B_c}{2} + \frac{B_o - B_c}{4}\left(
	\erf\left(\frac{\abs{x} - x_0 - v_A t}{\sqrt{4\eta_m t_0}}\right)
	+\erf\left(\frac{\abs{x} - x_0 + v_A t}{\sqrt{4\eta_m t_0}}\right)
\right)
,\label{eqn_sim_advdiff_By}\\
u_y(x,t) &= \sign(x)\frac{B_c - B_o}{4\sqrt{\rho_0\mu_0}}\left(
	\erf\left(\frac{\abs{x} - x_0 - v_A t}{\sqrt{4\eta_m t_0}}\right)
	-\erf\left(\frac{\abs{x} - x_0 + v_A t}{\sqrt{4\eta_m t_0}}\right)
\right)
.\label{eqn_sim_advdiff_uy}
\end{align}
The results of the simulations in \figref{fig_sim_advdiff_advdiff} clearly indicate that the convergence rate is given by the order of the IMEX method, where the RK2-Average scheme attains the second order, RK3hc the third order and RK4hc nearly the fourth order.

\begin{figure}[hbtp]
\centering
\includegraphics[width=.5\textwidth]{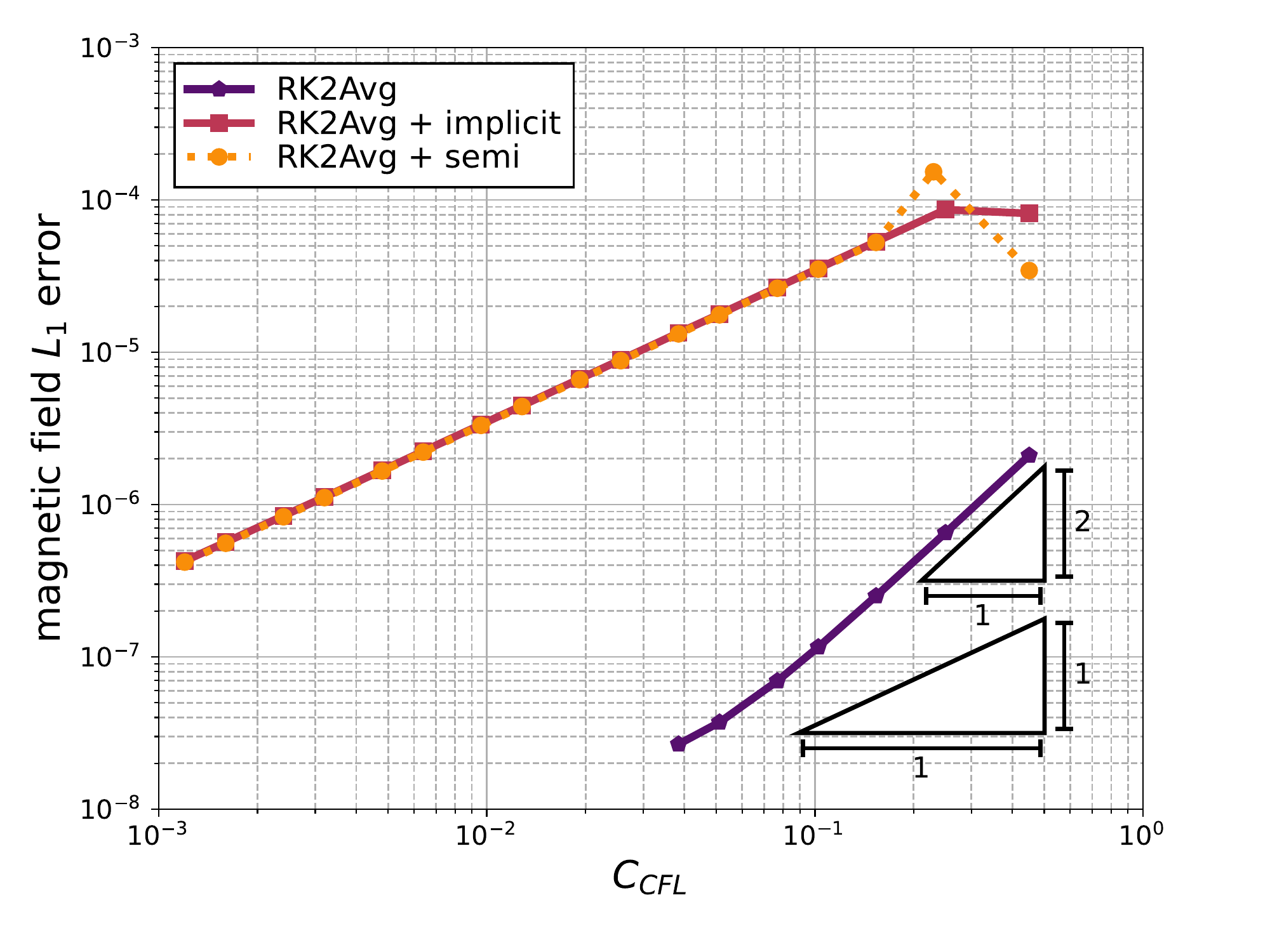}
\caption{Temporal convergence of the magnetic field to the reference solution in the advection-diffusion problem in the intermediate regime. The legend denotes the applied time integration scheme (see the accompanying text for details). The resolution is $300$ finite elements of $T4M4$. The reference numerical solution has the CFL factor $C_{CFL}=4\cdot 10^{-4}$ and is computed with the RK2-Average scheme.}
\label{fig_sim_advdiff_mix}
\end{figure}

The considered advection-diffusion problem does not have a simple analytic solution for the intermediate regime and the reference numerical solution with a small CFL factor ($C_{CFL}=4\cdot 10^{-4}$) is used instead. For the given parameters, the magnetic Reynolds number can be estimated as $R_m\sim v_A^2 t/\eta_m= 10$ for the final time $t=1$, corresponding to a tight coupling. The convergence plots in \figref{fig_sim_advdiff_mix} show that the performance of the semi-implicit splitted scheme degrades to the first order for coupled problems. However, the full RK2-Average scheme keeps its second order of convergence. This justifies the construction of the scheme in \secref{subsub_num_disc_time}.

It can be also noticed that the splitted semi-implicit scheme exhibits a deviation from the convergence curve for high values of $C_{CFL}$. A closer analysis reveals that the semi-implicit scheme is prone to an oscillatory behavior under specific conditions, which is known as "ringing" \cite{Rieben2006}. In this case, it is induced by the separation of the counter-propagating waves, which leads to a rapid collapse of the electric field at the center of the slopes. The RK2-Average scheme is not vulnerable to such MHD driving mechanism, but it may still exhibit non-physical effects for highly heterogeneous media \cite{Rieben2006,Bochev2003} or rapidly time-varying fields \cite{White2009}, for example.


\subsection{MHD rotor}
\label{sub_sim_rotor}

The next problem considered is the magneto-hydrodynamic rotor, originating from a simplified model of star formation~\cite{Mouschovias1980}. A rotating magnetized cylinder emits torsional Alfvénic waves, dissipating its angular momentum through the process. This presents a classical test problem for Eulerian coplanar MHD~\cite{Balsara1999}, but rarely for Lagrangian codes, where the computational mesh entangles during the rotation rapidly. However, the proposed numerical scheme can benefit from the curvilinear nature of the finite elements avoiding this catastrophic process.

The problem is considered on the 2D computational domain $(-1,1)\times(-1,1)$, where the cylinder (with a small slope at the edge) is embedded in an initially static ambient medium. The initial profiles of density and velocity are defined as follows:
\begin{subequations}
\begin{align}
\rho(x,y) &=
\begin{cases}
10 & r(x,y) \leq r_0 ,\\
1 & r(x,y) \geq r_1 ,\\
1 + 9 f(r) & r_0 < r(x,y) < r_1 ,
\end{cases}
\label{eqn_sim_rotor_ic_rho}\\
\vec{u}(x,y) &=
\begin{cases}
\vec{0} & r(x,y) \geq r_1 ,\\
f(r) u_0 / \max(r, r_0) 
\left( \begin{array}{c}
-y \\
x
\end{array} \right) & r(x,y) < r_1,
\end{cases}
\label{eqn_sim_rotor_ic_u}\\
f(r) &= (r_1 - r) / (r_1 - r_0)
,\label{eqn_sim_rotor_ic_f}\\
r(x,y) &= \sqrt{x^2 + y^2}
.\label{eqn_sim_rotor_ic_r}
\end{align}
\label{eqn_sim_rotor_ic}
\end{subequations}
The parameters of the radii and velocity are set to $r_0=0.1, r_1=0.115, u_0=2$. The initial pressure is homogeneous everywhere to start from a static equilibrium with $p\equiv 1$, where the ideal gas equation of state is considered with $\gamma=1.4$ and $A=1, Z=0$. In addition, a homogeneous magnetic field is initially imposed along the horizontal axis with magnitude $B_x=5/2 \sqrt{\mu_0/\pi}$.

\begin{figure}[hbtp]
\centering
\includegraphics[width=.49\textwidth]{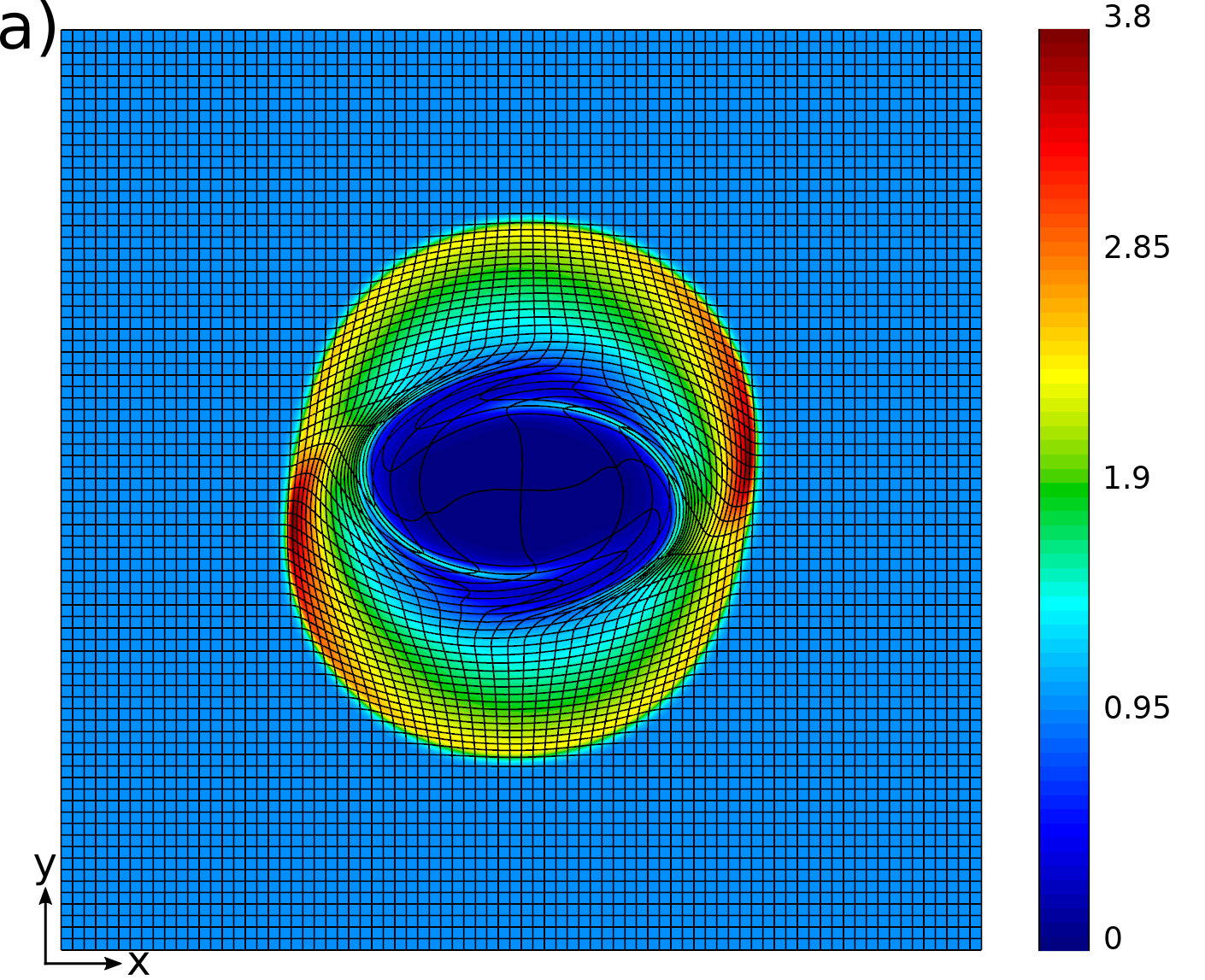}
\includegraphics[width=.49\textwidth]{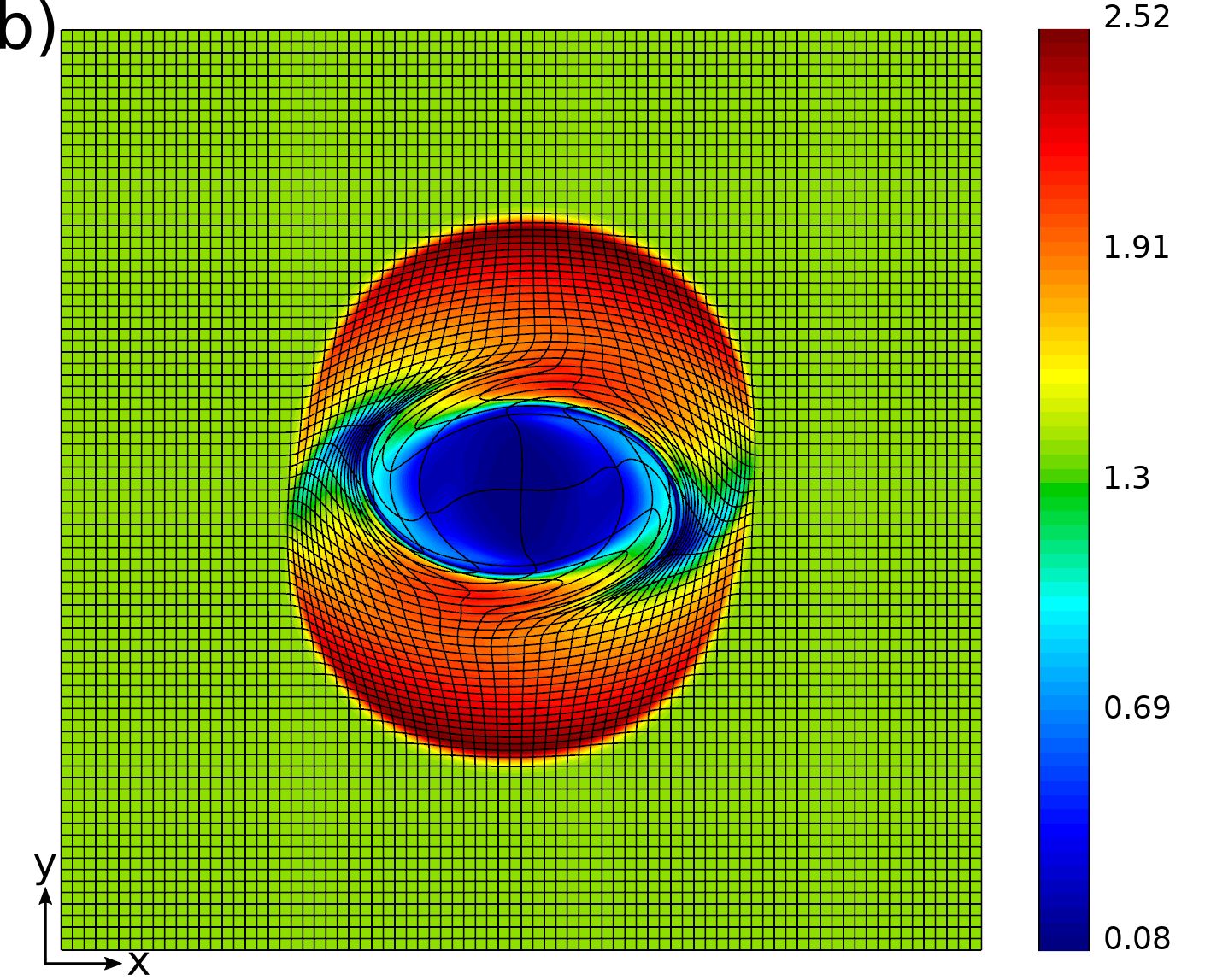}
\caption{Numerical solution of the MHD rotor problem at the final time $t=0.15$: a) thermal pressure [relative units] b) magnitude of magnetic field [relative units]. The resolution of the computational mesh is $80\times80$ finite elements of the $T2M2$ family.}
\label{fig_sim_rotor_sol}
\end{figure}

The results for the final time $t=0.15$ are presented in \figref{fig_sim_rotor_sol} for a computational mesh composed of $80\times80$ finite elements from the $T2M2$ family. The time integrating scheme RK3hc(A.$\alpha$) with CFL equal $0.5$ is used (see \secref{subsub_num_disc_time}). As singularities and strong torsion are present in the solution, the artificial viscosity based on the full eigenvector decomposition is applied \cite{Dobrev2010,Kolev2009}, where the linear and quadratic coefficients are both equal to $1$. The numerical solution shows that even a relatively low number of $T2M2$ finite elements is capable to capture the details without almost any mesh imprint, despite the strong distortion of the mesh, which winds up on the revolving rotor. The tensor artificial viscosity also manifests its strength here, especially in the inner circle around the rotor, where the matter becomes strongly pressed by the magnetic field rather through torsion than normal compression. The shear motion and subsequent transverse compression is not discriminated in the spectral decomposition and the profiles remain smooth and without any artifacts.


\subsection{MHD blast}
\label{sub_sim_blast}

The final problem presented is the magneto-hydrodynamic blast~\cite{Zachary1994}, which was first studied in the astrophysical context as a simplified model of magnetized cosmic jets \cite{Kossl1990}. Similarly to the classical hydrodynamic problem of Sedov blast~\cite{Sedov1993}, a high amount of energy is concentrated in a single point initially, launching a strong blast wave to the ambient medium. The difference is given by the fact that an imposed magnetic field then collimates the blast and competes through the magnetic pressure with the thermal one. 
The symmetry of the problem around the axis of the magnetic field pre-determines the problem for a study of the geometrical effects in multiple dimensions.

The 3D case is considered, where the simulation box spans over $(-1,1)^3$ and is filled with an ambient gas of density $\rho\equiv 1$, where the ideal gas EOS is considered with $\gamma=5/3$ and $A=1, Z=0$. The energy equal to the unity is deposited at the center of the domain (as a Dirac function integrally projected onto the mesh), while the background temperature is negligible. As a part of the initial condition, the magnetic field is imposed along the $x$ axis with magnitude $B_x=\sqrt{4/3\mu_0}$.

\begin{figure}[hbtp]
\centering
\includegraphics[width=.49\textwidth]{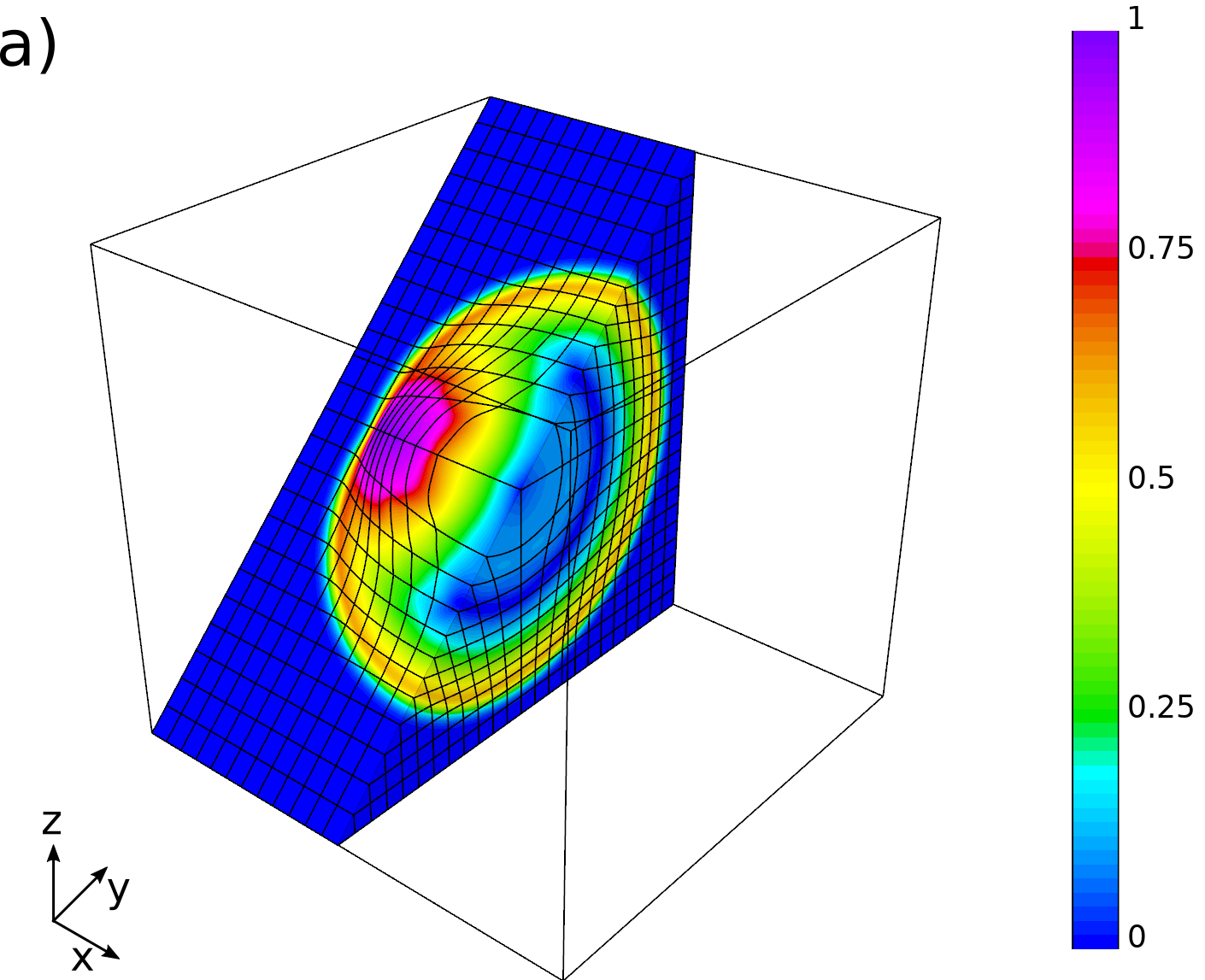}
\includegraphics[width=.49\textwidth]{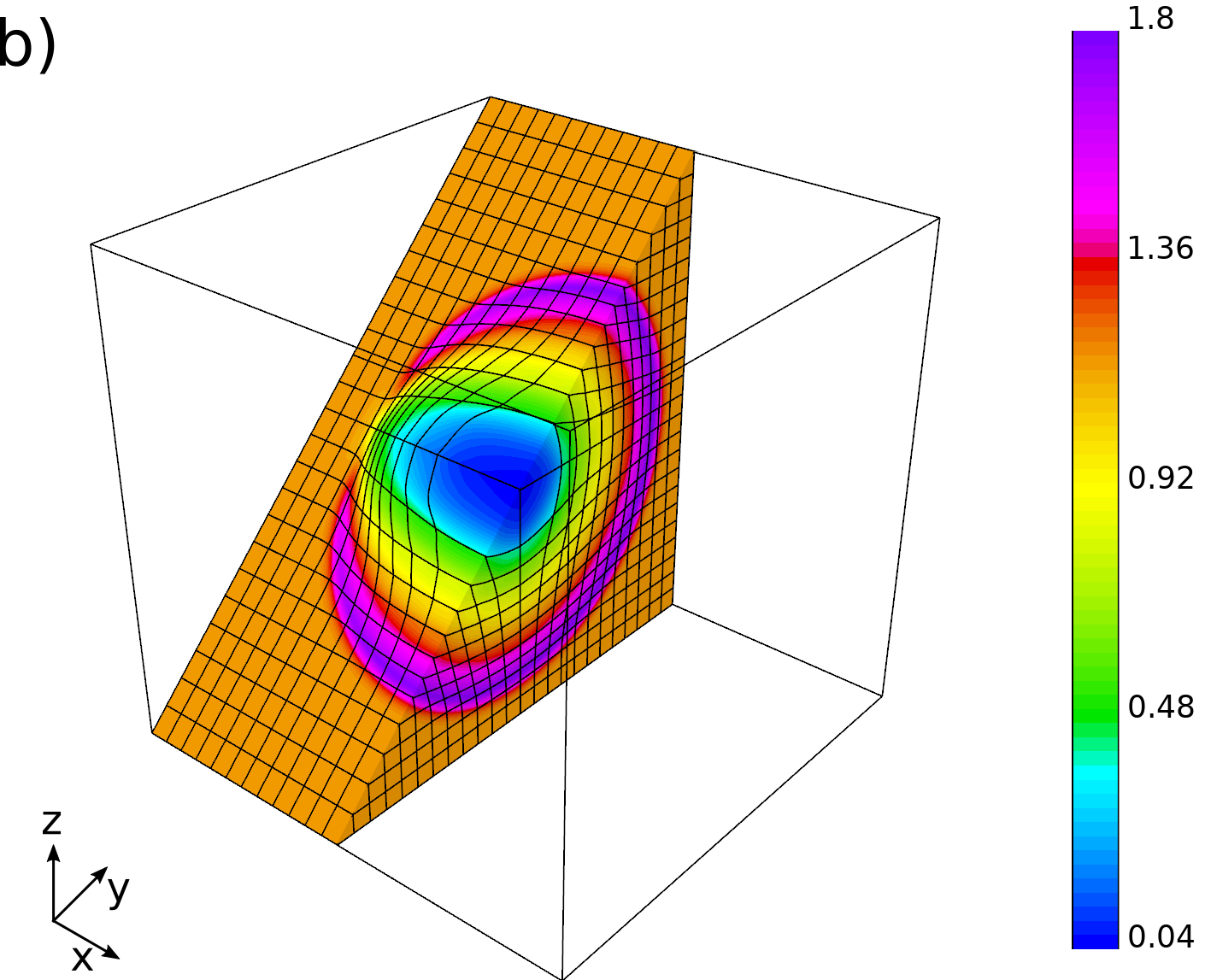}
\caption{The 3D simulations of MHD blast at the final time $t=0.25$: a) magnitude of velocity [relative units] b) magnitude of magnetic field [relative units]. The resolution of the computational mesh is $24\times24\times24$ finite elements of the $T2M2$ family.}
\label{fig_sim_blast_3D}
\end{figure}

The results of the simulations are shown in \figref{fig_sim_blast_3D}. The computation is performed on a mesh with $24\times24\times24$ finite elements of the $T2M2$ family. The scheme RK3hc(A.$\alpha$) is chosen for time integration to match the spatial order according to the findings of \secref{sub_sim_taylorgreen}. The CFL factor is set to $0.5$ and the identical model of artificial viscosity as in \secref{sub_sim_rotor} is applied with the linear coefficient equal to $0.5$ and the quadratic to $2$. From the numerical profiles, it can be observed that the Lagrangian formulation naturally leads to an increased resolution in the areas of the greatest compression near the fronts of the jets. Therefore, a good level of detail is obtained with a relatively low number of elements and even in 3D the geometric effects on the solution are minimal. The results show the ability of the scheme to treat challenging physical scenarios in multiple dimensions and of relevance to real problems in ICF \cite{Perkins2017} and astrophysics \cite{Pudritz2012}.


\section{Discussion}
\label{sec_disc}

The proposed numerical scheme for conservative Lagrangian magneto-hydrodynamics is presented in \secref{sec_num}, which starts from the general weak formulation (\secref{sub_num_weak}) and proceeds towards the fully discrete model (\secref{sub_num_disc}). To give a better insight into the construction and outline the future direction of research, multiple aspects of the scheme are worth of a further discussion, also in the light of the simulation results in \secref{sec_sim}.

The weak formulation of \secref{sub_num_weak} established an appealing form of the magneto-hydrodynamic equations \eqref{eqn_phys_int} for numerical solution. The symmetries between the momentum equation \eqref{eqn_num_weak_mom} and the energy equations \eqref{eqn_num_weak_eps}, \eqref{eqn_num_weak_epsB} and between the magnetic field equation \eqref{eqn_num_weak_B} and the energy equations already predetermine the system for a conservative discretization. The specific choices of the functional spaces in \secref{sub_num_semi} and the applied transformation rules guarantee conservation of momentum, magnetic flux and internal energy. In addition, conservation of mass is already given by the geometric conservation law holding for the curvilinear finite elements. The only prize for the total energy conservation is the energy correction term $\vcb{e}_B^c$, which accounts for the simultaneous magnetic flux and energy conservation. It also justifies the formulation with the auxiliary equation for magnetic energy \eqref{eqn_phys_int_epsB}. Following the reasoning of \secref{subsub_num_semi_md}, the definitions of magnetic energy as $\vcb{1}^T\mat{M}_\spc{T}\vcb{e}_B$ and $\vcb{B}^T\mat{M}_\spc{M}\vcb{B}/(2\mu_0)$ coincide in absence of motion, but they start to slightly differ for the medium in motion, when the mass matrix $\mat{M}_\spc{M}$ is not constant in contrast to $\mat{M}_\spc{T}$. The correction term $\vcb{e}_B^c$ defined by \eqref{eqn_num_semi_enBc} is then introduced, which equals to the difference between the increments of the energies given by the two definitions. However, this is not the only possible approach to the problem, as the discrepancy between the energies is tracked due to existence of $\varepsilon_B$ and does not need to be explicitly compensated immediately. This idea is elaborated in the symmetrical semi-implicit (SSI) technique suggested in \cite{Wu2018}, which originates from the context of heat conduction \cite{Livne1985}. Since the formulation is already integral within FEM, the advantage of an explicit update is not of an use here (but note the mass matrix $\mat{M}_\spc{T}$ is block-diagonal according to \secref{subsub_num_semi_cons}). Moreover, SSI leads to violation of immediate conservation and energy is conserved only in the limit sense. Therefore, the direct subtraction is preferred.

The time integration procedure described in \ref{subsub_num_disc_time} employs the second order RK2-Average scheme or the higher order IMEX methods for the ideal magneto-hydrodynamic part of the scheme. All of them are symplectic, conserving kinetic energy and thus the total energy (as internal and magnetic energy are already conserved on the semi-discrete level, see \secref{sub_num_semi}). Moreover, the RK2-Average scheme enables to incorporate the magnetodynamic step directly, improving the coupling between the two. Unfortunately, this is not possible for the higher order IMEX methods, which offer sympletic integration without the necessity of implicit solution, but with a drawback of negative time steps appearing in the Butcher tables \cite{Sandu2021}. This makes the time stepping inapplicable to the diffusion equation of the electric field \eqref{eqn_num_disc_md_E}, since the problem would be ill-posed. For weakly coupled diffusion and convection ($\textup{Re}_m \ll 1$ or $\textup{Re}_m \gg 1$), the splitting approach, when the magnetodynamic and ideal magneto-hydrodynamic parts are solved separately, is sufficient, but becomes a limiting factor for tightly coupled problems ($\textup{Re}_m \sim 1$). Moreover, the magnetodynamic time integration is limited to the second order of convergence. Extension of this part of the scheme to the higher orders  is foreseeable by means of singly diagonally implicit Runge-Kutta (SDIRK) methods \cite{Ferracina2008}, for example. However, conservation of magnetic energy would be violated this way, as the proof \eqref{eqn_num_disc_md_en} does not hold any longer. Similarly to the ideal MHD part, a symplectic formulation is needed in this case. Thus, conservative high-order methods for the magnetodynamic part and their coupling with the ideal MHD part are topics of the future research.

Despite that the numerical scheme is designed in a rather general and multi-dimensional manner from the beginning, \secref{subsub_num_disc_spc} narrows the choice of the finite elements to quadrilaterals in 2D and hexahedrals in 3D (note that all elements are curvilinear in general due to the isoparametric mapping). However, this is not required by the construction and is presented only for the sake of brevity and simplicity. An advantage is that the quadrilateral/hexahedral elements are tensorial, where the definitions in the lower and higher dimensions are completely analogous, but simplexes or more complex shapes could be used instead. As far as the conservation properties are considered, the schemes would posses the same features. However, the numerical dispersion relations might differ considerably \cite{Warren1996}. On the other hand, simplexes enable conformal \textit{h}-refinement of the computational mesh, which can increase resolution locally. However, conservation of all thermodynamic and kinematic quantities presents a non-linear problem, which becomes challenging for Lagrangian meshes \cite{Anderson2004}. Another approach is to perform \textit{p}-refinement, where the polynomial order of the elements is increased locally \cite{Karanam2008}. As \secref{sec_sim} shows, the order of convergence is proportional to the order of the elements, indicating that the procedure of \textit{p}-refinement would bear its fruits. However, the  problem of mesh entangling inherent to all Lagrangian models is greatly reduced by the use of the curvilinear finite elements already. The problem of a MHD rotor in \secref{sub_sim_rotor} showed that numerical simulations remain feasible even with considerably distorted meshes. For even more dramatic deformations of the mesh, like turbulences or instabilities, the arbitrary Lagrangian--Eulerian (ALE) methods \cite{Barlow2016} can be employed. Such methods exist for the high-order Lagrangian hydrodynamics \cite{Hajduk2020,Anderson2018}, but must be non-trivially extended to the proposed MHD model, which presents one of the topics of the future work.


\section{Conclusions}
\label{sec_concl}

Numerical magneto-hydrodynamics is a well-established field of computational physics, but conservative high-order multi-dimensional methods are scarce (with some exceptions for Eulerian methods \cite{Balsara2015,Susanto2013}). The presented Lagrangian numerical scheme based on curvilinear finite elements offers these features. The high-order elements equipped with the isoparametric mapping can track the motion of the simulated magnetized fluid for a long time without getting entangled, as the results of \secref{sec_sim} show. The simulations are also resilient with respect to mesh deformation, exhibiting only minimal mesh imprint. Moreover, the order of spatial convergence is proportional to the polynomial order of the finite elements for smooth problems. Together with the high-order time integration methods of \ref{subsub_num_disc_time}, theoretically an arbitrary order of overall convergence can be attained for ideal MHD. In all cases, exact conservation of mass, momentum, magnetic flux and the total energy is guaranteed by the construction of the elements and scheme according to \secref{sec_num}. Furthermore, the magnetic field remains divergence-free for the whole time span of the simulation.

Altogether, the novel method can push the frontiers of research in ICF \cite{Perkins2017}, astrophysics \cite{Pudritz2012}, prepulse effects of ultra-intense lasers \cite{Nikl2019spie,Holec2018b} or laser ion acceleration beamlines \cite{Arefiev2016,Psikal2021,Psikal2019spie,Batani2010} and other areas, where the Lagrangian approach is favored for a strong expansion/compression of the matter and a multi-dimensional treatment is needed. These applications can then benefit from the rapid convergence, flexibility of the mesh design and robustness of the method.

Despite all the attractive features of the scheme, the discussion of \secref{sec_disc} pointed out some limitations of the current design. Among other, it is the fact that the magnetodynamic part of the scheme is not fully coupled with the ideal hydrodynamic part for the higher orders of time integration and remains limited to the second order of convergence. Hence, development of conservative high-order time integration methods for the fully coupled resistive MHD system is a topic of future work. Similarly, advantage of the formulation generality can be taken and methods of \textit{hp}-adaptivity or an ALE model can be proposed. Furthermore, the physical model can be extended by the Nernst effect, thermoelectric effect and others to what is known as the extended MHD model \cite{Walsh2020}, where possible non-locality of the transport should be taken into account \cite{Holec2018,Ridgers2008}. Finally, the topic of self-generated magnetic fields through the process of the Biermann battery can be addressed \cite{Nikl2021eps,Tzeferacos2015}.


\section*{Acknowledgments}
\label{sec_ack}

Portions of this research were carried out at ELI Beamlines, a European user facility operated by the Institute of Physics of the Academy of Sciences of the Czech Republic.
Supported by CAAS project CZ.02.1.01/0.0/0.0/16\_019/0000778 from European Regional Development Fund; Czech Technical University grant SGS19/191/OHK4/3T/14 and Czech Science Foundation project 19-24619S. The computations were performed using computational resources funded from the CAAS project. This work has received funding from the Eurofusion Enabling Research Project No. CfP-FSD-AWP21-ENR-01-CEA-02.

Stefan Weber was supported by the project Advanced research using high intensity laser
 produced photons and particles (ADONIS) (CZ.02.1.01/0.0/0.0/16\_019/0000789)  and by the project High Field Initiative (HiFI) (CZ.02.1.01/0.0/0.0/15\_003/0000449), 
 both from European Regional Development Fund.


\begin{appendix}

\section{Equations of magneto-hydrodynamics in 1D and 2D}
\label{app_phys_dif}

The differential formulation of the magneto-hydrodynamic equations \eqref{eqn_phys_dif} reduces and splits for the transverse components subscripted by $\perp$ and coplanar/collinear components denoted by the subscript $\parallel$ in 1D and 2D Cartesian geometry. The equations of mass \eqref{eqn_phys_dif_mass} and momentum \eqref{eqn_phys_dif_mom}, which do not involve the fields directly, are not affected formally and only the definition of the magnetic stress tensor $\tens{\sigma}_B$ must be modified accordingly.

The 1D formulation requires to define the transverse curl $\nabla_\perp\times$
operating on the transverse components of the fields\footnote{
$\nabla_\perp\times=
\left(\begin{array}{cc}
0 & -\partial_x\\
\partial_x & 0
\end{array}\right)$,
where $x$ is the 1D coordinate.}. 
The equations of the magnetic field \eqref{eqn_phys_dif_B} and energies \eqref{eqn_phys_dif_eps}, \eqref{eqn_phys_dif_epsB} can then be written as:
\begin{subequations}
\begin{align}
\dod{\vec{B}_\parallel}{t} &= 0
,\label{eqn_app_phys_dif_1D_Bs}\\
\dod{\vec{B}_\perp}{t} &=
-\nabla_\perp\times\vec{E}_\perp
+ (\vec{B}_\parallel\cdot\nabla) \vec{u}_\perp
,\label{eqn_app_phys_dif_1D_Bp}\\
\rho\dod{\varepsilon}{t} &= \tens{\sigma}:\nabla\vec{u} + \vec{j}_\perp\cdot\vec{E}_\perp
,\label{eqn_app_phys_dif_1D_eps}\\
\rho\dod{\varepsilon_B}{t} &= \tens{\sigma}_B:\nabla\vec{u} - \frac{1}{\mu_0}\vec{B}_\perp \cdot(\nabla_\perp\times\vec{E}_\perp)
.\label{eqn_app_phys_dif_1D_epsB}
\end{align}
\label{eqn_app_phys_dif_1D}
\end{subequations}

In 2D, the curl operators $\nabla_\parallel\times$ and $\nabla_\perp\times$, crossing between the coplanar and transverse components, are used\footnote{
$\nabla_\parallel\times=
\left(\begin{array}{cc}
\partial_y & -\partial_x
\end{array}\right)^T$ and 
$\nabla_\perp\times=
\left(\begin{array}{cc}
-\partial_y & \partial_x
\end{array}\right)$,
where $x$ and $y$ are the 2D coordinates.}.
The splitted part of the system \eqref{eqn_phys_dif} is then taking the form:
\begin{subequations}
\begin{align}
\dod{\vec{B}_\parallel}{t} &= 
-\nabla_\parallel\times\vec{E}_\perp
,\label{eqn_app_phys_dif_2D_Bs}\\
\dod{\vec{B}_\perp}{t} &=
-\nabla_\perp\times\vec{E}_\parallel
+ (\vec{B}_\parallel\cdot\nabla) \vec{u}_\perp
,\label{eqn_app_phys_dif_2D_Bp}\\
\rho\dod{\varepsilon}{t} &= 
\tens{\sigma}:\nabla\vec{u}
+ \vec{j}_\parallel\cdot\vec{E}_\parallel
+ \vec{j}_\perp\cdot\vec{E}_\perp
,\label{eqn_app_phys_dif_2D_eps}\\
\rho\dod{\varepsilon_B}{t} &=
\tens{\sigma}_B:\nabla\vec{u}
- \frac{1}{\mu_0}\vec{B}_\parallel \cdot \nabla_\parallel \times \vec{E}_\perp
- \frac{1}{\mu_0}\vec{B}_\perp \cdot (\nabla_\perp \times \vec{E}_\parallel)
.\label{eqn_app_phys_dif_2D_epsB}
\end{align}
\label{eqn_app_phys_dif_2D}
\end{subequations}


\section{Weak formulation in 1D and 2D}
\label{app_num_weak}

The weak formulation following \secref{sub_num_weak} in 1D and 2D requires to define the functional spaces for the transverse and coplanar/collinear components introduced along with the governing equations in \ref{app_phys_dif}. The space of scalar thermodynamic potentials $\spc{T}\subset L_2(\Omega)$ remains unaffected, while the space of kinematic quantities $\spc{K}\subset (H^1(\Omega))^3$ is divided into the coplanar/collinear part $\spc{K}_\parallel\subset (H^1(\Omega))^d$ and the transverse part $\spc{K}_\perp\subset (H^1(\Omega))^{3-d}$, where $d\in\lbrace 1,2 \rbrace$ is the dimension. The space $\spc{K}_\parallel$ is used for the coordinates $\vec{x}$, whereas the velocities $\vec{u}$ are taken from the full space $\spc{K}$ in \eqref{eqn_num_weak_mom}, but only the collinear/coplanar components enter the update of coordinates $\dif{\vec{x}}/\dif{t}=\vec{u}_\parallel$.

\subsection{Weak formulation in 1D}
\label{app_num_weak_1D}

In one dimension, the collinear component of electric field does not exist (with the exception of a non-interacting homogeneous field) and the rest of the spaces is defined as follows:
\begin{itemize}
\item collinear magnetic ($\spc{M}_\parallel$) -- $\spc{M}_\parallel \subset L_2(\Omega)$,
\item transverse magnetic ($\spc{M}_\perp$) -- $\spc{M}_\perp \subset (L_2(\Omega))^2$,
\item transverse electric ($\spc{E}_\perp$) -- $\spc{E}_\perp \subset (H^1(\Omega))^2$.
\end{itemize}

The weak formulation of the system \eqref{eqn_app_phys_dif_1D} (without trivial \eqref{eqn_app_phys_dif_1D_Bs}) together with Ohm's law $\vec{E}_\perp=\eta\vec{j}_\perp=\eta/\mu_0\nabla_\perp\times\vec{B}_\perp$ then takes the form:
\begin{subequations}
\begin{align} 
\dint_\Omega \dod{\vec{B}_\perp}{t}\cdot \vec{\Xi}_\perp \dif{V} &=
- \dint_\Omega\left( ((\nabla_\perp\times\vec{E}_\perp) \cdot \vec{\Xi}_\perp 
+ ((\vec{B}_\parallel\cdot\nabla) \vec{u}_\perp) \cdot \vec{\Xi}_\perp \right)\dif{V}
, & \forall\vec{\Xi}_\perp \in\spc{M}_\perp
,\label{eqn_app_num_weak_1D_Bp}\\
\dint_\Omega \frac{1}{\eta} \vec{E}_\perp \cdot \vec{\xi}_\perp \dif{V} &=
\dint_\Omega \frac{1}{\mu_0} \vec{B}_\perp \cdot(\nabla_\perp\times\vec{\xi}_\perp) \dif{V}
-\int_{\Gamma_B} \frac{1}{\mu_0}\vec{B}^\tau_\perp \cdot \vec{\xi}_\perp \dif{S}
, & \forall\vec{\xi}_\perp\in\spc{E}_\perp
,\label{eqn_app_num_weak_1D_Ep}\\
\dint_\Omega \rho \dod{\varepsilon}{t} \varphi \dif{V} &=
\dint_\Omega\left( \tens{\sigma}:\nabla\vec{u} \varphi
+ \frac{1}{\mu_0}\vec{B}_\perp \cdot(\nabla_\perp \times\vec{E}_\perp) \varphi
+ \frac{1}{\mu_0} (\vec{E}_\perp \times\vec{B}_\perp) \cdot\nabla\varphi \right)\dif{V}
+\notag\\ & \phantom{=}
+\int_{\Gamma_E} \frac{1}{\mu_0} \vec{E}^\tau_\perp \cdot \tr{}_\perp{\vec{B}_\perp} ~\tr{\varphi} \dif{S}
-\int_{\Gamma_B} \frac{1}{\mu_0} \vec{E}_\perp \cdot \vec{B}^\tau_\perp ~\tr{\varphi} \dif{S}
, & \forall\varphi\in\spc{T}
,\label{eqn_app_num_weak_1D_eps}\\
\dint_\Omega \rho\dod{\varepsilon_B}{t} \varphi \dif{V} &= 
\dint_\Omega\left( \tens{\sigma}_B:\nabla\vec{u}\varphi
- \frac{1}{\mu_0}\vec{B}_\perp \cdot(\nabla_\perp \times\vec{E}_\perp) \varphi \right)\dif{V}
, & \forall\varphi\in\spc{T}
,\label{eqn_app_num_weak_1D_epsB}
\end{align}
\label{eqn_app_num_weak_1D}
\end{subequations}
where $\tr{}_\perp$ is the trace of functions from $\spc{M}_\perp$ on the boundary.

The boundary conditions are analogous to \eqref{eqn_num_weak_bc}, where the conditions for the fields reduce to the transverse components only.

\subsection{Weak formulation in 2D}
\label{app_num_weak_2D}

The two-dimensional formulation involves all components of the fields, which are taken from the following functional spaces:
\begin{itemize}
\item coplanar magnetic ($\spc{M}_\parallel$) -- $\spc{M}_\parallel\subset H_{div}(\Omega)$,
\item transverse magnetic ($\spc{M}_\perp$) -- $\spc{M}_\perp\subset L_2(\Omega)$,
\item coplanar electric ($\spc{E}_\parallel$) -- $\spc{E}_\parallel\subset H_{curl}(\Omega)$,
\item transverse electric ($\spc{E}_\perp$) -- $\spc{E}_\perp\subset H^1(\Omega)$,
\end{itemize}

The system \eqref{eqn_app_phys_dif_2D} along with Ohm's law $\vec{E}_\parallel=\eta\vec{j}_\parallel=\eta/\mu_0\nabla_\parallel\times\vec{B}_\perp$ and $\vec{E}_\perp=\eta\vec{j}_\perp=\eta/\mu_0\nabla_\perp\times\vec{B}_\parallel$ can be formulated in the weak sense as:
\begin{subequations}
\begin{align} 
\dint_\Omega \dod{\vec{B}_\parallel}{t}\cdot \vec{\Xi}_\parallel \dif{V} &=
- \dint_\Omega (\nabla_\parallel\times\vec{E}_\perp) \cdot \vec{\Xi}_\parallel \dif{V}
, & \forall\vec{\Xi}_\parallel \in\spc{M}_\parallel
,\label{eqn_app_num_weak_2D_Bs}\\
\dint_\Omega \dod{\vec{B}_\perp}{t}\cdot \vec{\Xi}_\perp \dif{V} &=
- \dint_\Omega\left( (\nabla_\perp\times\vec{E}_\parallel) \cdot \vec{\Xi}_\perp
+ ((\vec{B}_\parallel\cdot\nabla) \vec{u}_\perp) \cdot \vec{\Xi}_\perp \right)\dif{V}
, & \forall\vec{\Xi}_\perp \in\spc{M}_\perp
,\label{eqn_app_num_weak_2D_Bp}\\
\dint_\Omega \frac{1}{\eta} \vec{E}_\parallel \cdot \vec{\xi}_\parallel \dif{V} &=
\dint_\Omega \frac{1}{\mu_0} \vec{B}_\perp \cdot(\nabla_\perp\times\vec{\xi}_\parallel) \dif{V}
-\int_{\Gamma_B} \frac{1}{\mu_0}\vec{B}^\tau_\perp \cdot \vec{\xi}_\parallel \dif{S}
, & \forall\vec{\xi}_\parallel \in\spc{E}_\parallel
,\label{eqn_app_num_weak_2D_Ep}\\
\dint_\Omega \frac{1}{\eta} \vec{E}_\perp \cdot \vec{\xi}_\perp \dif{V} &=
\dint_\Omega \frac{1}{\mu_0} \vec{B}_\parallel \cdot(\nabla_\parallel \times\vec{\xi}_\perp) \dif{V}
-\int_{\Gamma_B} \frac{1}{\mu_0}\vec{B}^\tau_\parallel \cdot \vec{\xi}_\perp \dif{S}
, & \forall\vec{\xi}_\perp \in\spc{E}_\perp
,\label{eqn_app_num_weak_2D_Es}\\
\dint_\Omega \rho \dod{\varepsilon}{t} \varphi \dif{V} &=
\dint_\Omega\left( \tens{\sigma}:\nabla\vec{u} \varphi
+ \frac{1}{\mu_0}\vec{B}_\parallel \cdot(\nabla_\parallel \times\vec{E}_\perp) \varphi \dif{V}
+ \frac{1}{\mu_0}\vec{B}_\perp \cdot(\nabla_\perp \times\vec{E}_\parallel) \varphi
+\right.\notag\\ & \phantom{=}\left.
+\frac{1}{\mu_0} (\vec{E}_\perp \times\vec{B}_\parallel) \cdot \nabla\varphi
+\frac{1}{\mu_0} (\vec{E}_\parallel \times\vec{B}_\perp) \cdot \nabla\varphi
\right)\dif{V}
+\notag\\ & \phantom{=}
+\int_{\Gamma_E}\left(
\frac{1}{\mu_0} \vec{E}^\tau_\perp \cdot \tr{}^\tau_\parallel\vec{B}_\parallel ~\tr{\varphi}
+ \frac{1}{\mu_0} \vec{E}^\tau_\parallel \cdot \tr{}_\perp{\vec{B}_\perp} ~\tr{\varphi}
\right)\dif{S}
-\notag\\ & \phantom{=}
-\int_{\Gamma_B}\left(
\frac{1}{\mu_0} \vec{E}_\perp \cdot \vec{B}^\tau_\parallel ~\tr{\varphi}
+ \frac{1}{\mu_0} \vec{E}_\parallel \cdot \vec{B}^\tau_\perp ~\tr{\varphi}
\right)\dif{S}
, & \forall\varphi\in\spc{T}
,\label{eqn_app_num_weak_2D_eps}\\
\dint_\Omega \rho\dod{\varepsilon_B}{t} \varphi \dif{V} &= 
\dint_\Omega\left( \tens{\sigma}_B:\nabla\vec{u}\varphi
- \frac{1}{\mu_0}\vec{B}_\parallel \cdot(\nabla_\parallel \times\vec{E}_\perp) \varphi 
- \frac{1}{\mu_0}\vec{B}_\perp \cdot(\nabla_\perp \times\vec{E}_\parallel) \varphi \right)\dif{V}
, & \forall\varphi\in\spc{T}
,\label{eqn_app_num_weak_2D_epsB}
\end{align}
\label{eqn_app_num_weak_2D}
\end{subequations}
where $\tr{}_\perp$ is the trace of functions from $\spc{M}_\perp$ on the boundary and $\tr{}^\tau_\parallel$ is the trace of the tangential component of functions from $\spc{M}_\parallel$.

The boundary conditions are analogous to \eqref{eqn_num_weak_bc}, but the components of the fields must be distinguished. The coplanar components of the boundary field values define the transverse boundary data and vice versa, following \eqref{eqn_num_weak_bc_E}.

\section{Definitions of the discrete vectors, matrices and tensors}
\label{app_num_semi_mats}

The transition from the weak formulation \eqref{eqn_num_weak} to the semi-discrete model \eqref{eqn_num_semi} leads to the definition of the (bi-/tri-)linear forms on the given finite element spaces and their associated vectors/matrices/tensors. In all cases, the integrals are formulated in the real space, i.e., on the moving domain $\Omega(t)$ (and its boundary), and the base functions there are understood as compound functions of the basis function and the reverse map of $\vec{x}(t,\vec{X})$ denoted as $\vec{X}(t,\vec{x})$, i.g. $\varphi(t,\vec{X})=\varphi(t,\vec{X}(t,\vec{x}))$. However, the integrals are numerically evaluated on $\Omega_0$ for practical purposes, so the inversion of the mapping is not required, but the transformation is obvious and not presented for brevity. The vectors/matrices/tensors are defined as follows:
\begin{align}
(\mat{M}_\spc{T})_{ij} &= \int_{\Omega(t)} \rho \varphi_i \varphi_j \dif{V}
, & 
(\mat{M}_\spc{K})_{ij} &= \int_{\Omega(t)} \rho \vec{\psi}_i \cdot \vec{\psi}_j \dif{V} 
,\label{eqn_app_num_semi_mats_MtMk}\\
(\mat{M}_\spc{M})_{ij} &= \int_{\Omega(t)} \vec{\Xi}_i \cdot \vec{\Xi}_j \dif{V}
, &
(\mat{M}_\spc{E})_{ij} &= \int_{\Omega(t)} \eta^{-1} \vec{\xi}_i \cdot \vec{\xi}_j \dif{V} 
,\label{eqn_app_num_semi_mats_MmMe}\\
\mat{F}_{ij} &= \int_{\Omega(t)} \sigma:\nabla\vec{\psi}_i ~\varphi_j \dif{V}
, &
(\mat{F}_B)_{ij} &= \int_{\Omega(t)} \sigma_B:\nabla\vec{\psi}_i ~\varphi_j \dif{V}
,\label{eqn_app_num_semi_mats_FFb}\\
\mat{C}_{ijk} &= \int_{\Omega(t)} (\nabla\times \vec{\xi}_i) \cdot \vec{\Xi}_j ~\varphi_k \dif{V}
, &
\mat{D}_{ij} &= \int_{\Omega(t)} (\nabla\times\vec{\xi}_i) \cdot (\nabla\times\vec{\xi}_j) \dif{V}
,\label{eqn_app_num_semi_mats_CD}\\
(\mat{X}_E)_{ij} &= \int_{\Gamma_E(t)} \frac{1}{\mu_0} \vec{E}^\tau \cdot \vec{\Xi}_j ~\varphi_i \dif{S}
, &
(\mat{X}_B)_{ij} &= -\int_{\Gamma_B(t)} \frac{1}{\mu_0} \vec{\xi}_j \cdot \vec{B}^\tau ~\varphi_i \dif{S}
,\label{eqn_app_num_semi_mats_XEXB}\\
(\vcb{b}_\sigma)_i &= \int_{\Gamma_\sigma(t)} \vec{\sigma}_n \cdot \vec{\psi}_i \dif{S}
,\label{eqn_app_num_semi_mats_bbB}
\end{align}
where the parametric time dependency is omitted for brevity and indexes iterate over all base functions of the given kind. Finally, the tensor of Poynting vector divergence is defined as:
\begin{equation}
\mat{S}_{ijk} = \sum_{\Omega_e\in \Sigma_h}\int_{\Omega_e(t)} \frac{1}{\mu_0}(\vec{\xi}_i \times \vec{\Xi}_j) \cdot\nabla\varphi_k \dif{V} 
+ \sum_{\Gamma\in \Upsilon_h}\int_{\Gamma(t)} \frac{1}{\mu_0}((\vec{\xi}_i \times \vec{n})\times \lbrace\vec{\Xi}_j\times\vec{n}\rbrace)\cdot\vec{n} [\varphi_k]\dif{S}
.\label{eqn_app_num_semi_mats_S}
\end{equation}
The notation utilizes the operators $\lbrace \cdot \rbrace$ for the mean value at the edge and $[\cdot]$ for the normal jump of the quantity across the edge.


\section{Time integration in 1D and 2D}
\label{app_num_disc_time}

The procedure of time integration in 1D and 2D is completely analogous to \secref{subsub_num_disc_time}. In the case of the IMEX schemes, the magnetic field is added to the $Y$ state, as mentioned in \secref{subsub_num_disc_visc}. The only notable difference appears in the case of the RK2-Average scheme, which couples together the magnetodynamic and ideal magnetohydrodynamic models. The equations of the transverse magnetic field \eqref{eqn_app_num_weak_1D_Bp} and \eqref{eqn_app_num_weak_2D_Bp} involve the magnetic dynamo term $(\vec{B}_\parallel\cdot\nabla)\vec{u}_\perp$, which contributes to the magnetic field within the ideal MHD step. Therefore, the first step of the RK2-Average scheme must consider both contribution, resistive and convective. It takes the following form in 1D:
\begin{subequations}
\begin{align}
\vcb{u}^{n+1/2} &= \vcb{u}^{n} - \frac{\Delta t}{2} \mat{M}_\spc{K}^{-1}((\mat{F}^n + \mat{F}_B^n)\vcb{1} + \vcb{b}_\sigma^n)
,\label{eqn_app_num_disc_time_rkavg_1_u}\\
\vcb{x}^{n+1/2} &= \vcb{x}^{n} + \frac{\Delta t}{2} \vcb{u}^{n+1/2}_\parallel
,\label{eqn_app_num_disc_time_rkavg_1_x}\\
\hat{\vcb{B}}^{n+1/2}_\perp &= \vcb{B}^n_\perp + \frac{\Delta t}{2}\mat{G}_{B_\parallel} \vcb{u}^{n+1/2}_\perp
,\label{eqn_app_num_disc_time_rkavg_1_Bh}\\
\vcb{E}^{n+1/2}_\perp
&= \left( \mat{M}_{\spc{E}_\perp} + \frac{\Delta t/2}{\mu_0} \mat{D}_\perp \right)^{-1} \left(\frac{1}{\mu_0} (\mat{C}_\perp)_{\cdot jk} (\hat{\vcb{B}}_\perp)_j^{n+1/2} \vcb{1}_k
+ \mat{X}_{B_\perp}^T\vcb{1}\right)
,\label{eqn_app_num_disc_time_rkavg_1_E}\\
\vcb{B}^{n+1/2}_\perp &= 
\hat{\vcb{B}}^{n+1/2}_\perp
-\Delta t\mat{C}_{D} \vcb{E}^{n+1/2}_\perp
,\label{eqn_app_num_disc_time_rkavg_1_B}\\
\vcb{e}^{n+1/2} &= \vcb{e}^{n} + \frac{\Delta t}{2}\mat{M}_\spc{T}^{-1}
(\mat{F}^n)^T\vcb{u}^{n+1/2}
+ \frac{\Delta t}{2} \eval[2]{\dod{\vcb{e}}{t}}_{\text{Joule}}^{n,\Delta t/2,\alpha=1}
,\label{eqn_app_num_disc_time_rkavg_1_e}\\
\vcb{e}_B^{n+1/2} &= \vcb{e}_B^{n} + \frac{\Delta t}{2}\mat{M}_\spc{T}^{-1}
(\mat{F}_B^n)^T\vcb{u}^{n+1/2}
+ \frac{\Delta t}{2} \eval[2]{\dod{\vcb{e}_B}{t}}_{\text{Joule}}^{n,\Delta t/2,\alpha=1}
.\label{eqn_app_num_disc_time_rkavg_1_eB}
\end{align}
\label{eqn_app_num_disc_time_rkavg_1}
\end{subequations}
Afterwards, the second part of the scheme is calculated as follows:
\begin{subequations}
\begin{align}
\vcb{u}^{n+1} &= \vcb{u}^{n} - \Delta t \mat{M}_\spc{K}^{-1}((\mat{F}^{n+1/2} + \mat{F}_B^{n+1/2})\vcb{1} + \vcb{b}_\sigma^{n+1/2})
,\label{eqn_app_num_disc_time_rkavg_2_u}\\
\vcb{x}^{n+1} &= \vcb{x}^{n} + \Delta t \bar{\vcb{u}}_\parallel^{n+1/2}
,\label{eqn_app_num_disc_time_rkavg_2_x}\\
\hat{\vcb{B}}^{n+1}_\perp &= \vcb{B}^n_\perp + \Delta t\mat{G}_{B_\parallel} \bar{\vcb{u}}^{n+1/2}_\perp
,\label{eqn_app_num_disc_time_rkavg_2_Bh}\\
\bar{\vcb{E}}^{n+1/2}_\perp
&= \left( \mat{M}_{\spc{E}_\perp} + \frac{\Delta t/2}{\mu_0} \mat{D}_\perp \right)^{-1} \left(\frac{1}{\mu_0} (\mat{C}_\perp)_{\cdot jk} (\hat{\bar{\vcb{B}}}_\perp)_j^{n+1/2} \vcb{1}_k
+ \mat{X}_{B_\perp}^T\vcb{1}\right)
,\label{eqn_app_num_disc_time_rkavg_2_E}\\
\vcb{B}^{n+1}_\perp &= 
\hat{\vcb{B}}^{n+1}_\perp
-\Delta t\mat{C}_{D} \bar{\vcb{E}}^{n+1/2}_\perp
,\label{eqn_app_num_disc_time_rkavg_2_B}\\
\vcb{e}^{n+1} &= \vcb{e}^{n} + \Delta t\mat{M}_\spc{T}^{-1}
(\mat{F}^{n+1/2})^T\bar{\vcb{u}}^{n+1/2}
+ \Delta t \eval[2]{\dod{\vcb{e}}{t}}_{\text{Joule}}^{n+1/2,\Delta t,\alpha=1/2}
+ \vcb{e}^c_B
,\label{eqn_app_num_disc_time_rkavg_2_e}\\
\vcb{e}_B^{n+1} &= \vcb{e}_B^{n} + \Delta t\mat{M}_\spc{T}^{-1}
(\mat{F}_B^{n+1/2})^T\bar{\vcb{u}}^{n+1/2}
+ \Delta t \eval[2]{\dod{\vcb{e}_B}{t}}_{\text{Joule}}^{n+1/2,\Delta t,\alpha=1/2}
.\label{eqn_app_num_disc_time_rkavg_2_eB}
\end{align}
\label{eqn_app_num_disc_time_rkavg_2}
\end{subequations}
The vectors, matrices and tensors used are defined analogously to \ref{app_num_semi_mats}, where the transverse and longitudinal components are distinguished in this case. 
Similarly to the velocity $\bar{\vcb{u}}^{n+1/2}=1/2(\vcb{u}^{n+1} + \vcb{u}^{n+1})$, the average magnetic fields $\hat{\bar{\vcb{B}}}^{n+1/2}_\perp = 1/2(\hat{\vcb{B}}^{n+1}_\perp+\vcb{B}^{n}_\perp)$ and $\bar{\vcb{B}}^{n+1/2}_\perp=1/2(\vcb{B}^{n+1}_\perp+\vcb{B}^{n}_\perp)$ are considered. The latter together with the electric field $\bar{\vcb{E}}^{n+1/2}_\perp$ are used in the Joule heating terms in \eqref{eqn_app_num_disc_time_rkavg_2_e} and \eqref{eqn_app_num_disc_time_rkavg_2_eB}.
In addition, the matrix $\mat{G}_{B_\parallel}$ corresponds to the discrete operator of the aforementioned magnetic dynamo effect, which is constructed along the lines of \secref{subsub_num_semi_cons}.

The construction in 2D is completely analogous to the 1D case and is not presented for the sake of brevity. The explicit convective update is performed before the implicit resitive one takes place.


\section{Specific spatial discretization in 1D and 2D}
\label{app_num_disc_spc}

Following \secref{subsub_num_disc_spc}, specific choices of the finite element spaces are made in 1D and 2D. In all cases, they are conforming with the corresponding functional spaces used in the weak formulation of \ref{app_num_weak} and an isoparametric mapping is applied on them. 

The finite element spaces in 1D are defined as follows:
\begin{itemize}
\item thermodynamic -- $\spc{T}_h = \lbrace \varphi \in \spc{T} ~|~ 
\eval[0]{\varphi}_{\Omega_e} \in P^{p}(\Omega_e)
\quad \forall \Omega_e \in \Sigma_h \rbrace$,
\item collinear kinematic -- $\spc{K}_{\parallel h} = \lbrace \vec{\psi} \in \spc{K}_\parallel ~|~ 
\eval[0]{\vec{\psi}}_{\Omega_e} \in P^{p+1}(\Omega_e)
\quad \forall \Omega_e \in \Sigma_h \rbrace$,
\item transverse kinematic -- $\spc{K}_{\perp h} = \lbrace \vec{\psi} \in \spc{K}_\perp ~|~ 
\eval[0]{\vec{\psi}}_{\Omega_e} \in (P^{p+1}(\Omega_e))^2
\quad \forall \Omega_e \in \Sigma_h \rbrace$,
\item collinear magnetic -- $\spc{M}_{\parallel h} = \lbrace \vec{\Xi} \in \spc{M}_\parallel ~|~ 
\eval[0]{\vec{\Xi}}_{\Omega_e} \in {P}^{q}(\Omega_e) \quad \forall \Omega_e \in \Sigma_h \rbrace$,
\item transverse magnetic -- $\spc{M}_{\perp h} = \lbrace \vec{\Xi} \in \spc{M}_\perp ~|~ 
\eval[0]{\vec{\Xi}}_{\Omega_e} \in ({P}^{q}(\Omega_e))^2 \quad \forall \Omega_e \in \Sigma_h \rbrace$,
\item transverse electric -- $\spc{E}_{\perp h} = \lbrace \vec{\xi} \in \spc{E}_\perp ~|~ 
\eval[0]{\vec{\xi}}_{\Omega_e} \in ({P}^{q+1}(\Omega_e))^{2}
\quad \forall \Omega_e \in \Sigma_h \rbrace$.
\end{itemize}

In two dimensions, the definitions are following:
\begin{itemize}
\item thermodynamic -- $\spc{T}_h = \lbrace \varphi \in \spc{T} ~|~ 
\eval[0]{\varphi}_{\Omega_e} \in Q^{p}(\Omega_e) \quad \forall \Omega_e \in \Sigma_h \rbrace$,
\item coplanar kinematic -- $\spc{K}_{\parallel h} = \lbrace \vec{\psi} \in \spc{K}_\parallel ~|~ 
\eval[0]{\vec{\psi}}_{\Omega_e} \in (Q^{p+1}(\Omega_e))^2, 
\eval[0]{\vec{\psi}}_{\Gamma_e} \in (P^{p+1}(\Gamma_e))^2
\quad \forall \Omega_e \in \Sigma_h, \Gamma_e \subset \partial \Omega_e \rbrace$,
\item transverse kinematic -- $\spc{K}_{\perp h} = \lbrace \vec{\psi} \in \spc{K}_\perp ~|~ 
\eval[0]{\vec{\psi}}_{\Omega_e} \in Q^{p+1}(\Omega_e), 
\eval[0]{\vec{\psi}}_{\Gamma_e} \in P^{p+1}(\Gamma_e)
\quad \forall \Omega_e \in \Sigma_h, \Gamma_e \subset \partial \Omega_e \rbrace$,
\item coplanar magnetic -- $\spc{M}_{\parallel h} = \lbrace \vec{\Xi} \in \spc{M}_\parallel ~|~ 
\eval[0]{\vec{\Xi}}_{\Omega_e} \in {RT}^{q+1}_{2D}(\Omega_e) \quad \forall \Omega_e \in \Sigma_h \rbrace$,
\item transverse magnetic -- $\spc{M}_{\perp h} = \lbrace \vec{\Xi} \in \spc{M}_\perp ~|~ 
\eval[0]{\vec{\Xi}}_{\Omega_e} \in {Q}^{q}(\Omega_e) \quad \forall \Omega_e \in \Sigma_h \rbrace$,
\item coplanar electric -- $\spc{E}_{\parallel h} = \lbrace \vec{\xi} \in \spc{E}_\parallel ~|~ 
\eval[0]{\vec{\xi}}_{\Omega_e} \in {N\!D}^{q+1}_{2D}(\Omega_e) \quad \forall \Omega_e \in \Sigma_h \rbrace$,
\item transverse electric -- $\spc{E}_{\perp h} = \lbrace \vec{\xi} \in \spc{E}_\perp ~|~ 
\eval[0]{\vec{\xi}}_{\Omega_e} \in {Q}^{q+1}(\Omega_e),
\eval[0]{\vec{\xi}}_{\Gamma_e} \in P^{q+1}(\Gamma_e)
\quad \forall \Omega_e \in \Sigma_h, \Gamma_e \subset \partial \Omega_e \rbrace$,
\end{itemize}
where $P^p$ are polynomials up to the order $p$. The polynomials $RT_{2D}^{p+1}$ and $N\!D_{2D}^{p+1}$ are Raviart--Thomas $H_{div}$ conforming and Nédélec $H_{curl}$ conforming finite elements in 2D \cite{Lesaint1974,Nedelec1980}. The sets $\Gamma_e$ represent the edges of the element $\Omega_e$ in this context.

\end{appendix}



\bibliography{manuscript}

\begin{thebibliography}{10}
\expandafter\ifx\csname url\endcsname\relax
  \def\url#1{\texttt{#1}}\fi
\expandafter\ifx\csname urlprefix\endcsname\relax\def\urlprefix{URL }\fi
\expandafter\ifx\csname href\endcsname\relax
  \def\href#1#2{#2} \def\path#1{#1}\fi

\bibitem{Perkins2017}
L.~J. Perkins, D.~D. Ho, B.~G. Logan, G.~B. Zimmerman, M.~A. Rhodes, D.~J.
  Strozzi, D.~T. Blackfield, S.~A. Hawkins, The potential of imposed magnetic
  fields for enhancing ignition probability and fusion energy yield in
  indirect-drive inertial confinement fusion, Physics of Plasmas 24~(6) (2017)
  062708.
\newblock \href {https://doi.org/10.1063/1.4985150}
  {\path{doi:10.1063/1.4985150}}.

\bibitem{Clark2015}
D.~S. Clark, M.~M. Marinak, C.~R. Weber, D.~C. Eder, S.~W. Haan, B.~A. Hammel,
  D.~E. Hinkel, O.~S. Jones, J.~L. Milovich, P.~K. Patel, H.~F. Robey, J.~D.
  Salmonson, S.~M. Sepke, C.~A. Thomas, Radiation hydrodynamics modeling of the
  highest compression inertial confinement fusion ignition experiment from the
  {N}ational {I}gnition {C}ampaign, Physics of Plasmas 22~(2) (2015) 1--19.
\newblock \href {https://doi.org/10.1063/1.4906897}
  {\path{doi:10.1063/1.4906897}}.

\bibitem{Wu2018}
F.~Wu, R.~Ramis, Z.~Li, {A conservative MHD scheme on unstructured Lagrangian
  grids for Z-pinch hydrodynamic simulations}, Journal of Computational Physics
  357 (2018) 206--229.
\newblock \href {https://doi.org/10.1016/j.jcp.2017.12.014}
  {\path{doi:10.1016/j.jcp.2017.12.014}}.

\bibitem{Livne2004}
E.~Livne, A.~Burrows, R.~Walder, I.~Lichtenstadt, T.~A. Thompson,
  Two‐dimensional, time‐dependent, multigroup, multiangle radiation
  hydrodynamics test simulation in the core‐collapse supernova context, The
  Astrophysical Journal 609~(1) (2004) 277--287.
\newblock \href {https://doi.org/10.1086/421012} {\path{doi:10.1086/421012}}.

\bibitem{Pudritz2012}
R.~E. Pudritz, M.~J. Hardcastle, D.~C. Gabuzda, Magnetic fields in
  astrophysical jets: From launch to termination, Space Science Reviews
  169~(1-4) (2012) 27--72.
\newblock \href {https://doi.org/10.1007/s11214-012-9895-z}
  {\path{doi:10.1007/s11214-012-9895-z}}.

\bibitem{Nikl2019spie}
J.~Nikl, M.~Jirka, M.~Kucha\v{r}\'{i}k, M.~Holec, M.~Vranic, S.~Weber, The
  effect of pre-plasma formed under the non-local transport conditions on the
  interaction of the ultra-high intensity laser with a solid target, in:
  Research using Extreme Light: Entering New Frontiers with Petawatt-Class
  Lasers IV, Vol. 11039 of Proceedings of SPIE, 2019, p. 110391E.
\newblock \href {https://doi.org/10.1117/12.2522450}
  {\path{doi:10.1117/12.2522450}}.

\bibitem{Holec2018b}
M.~Holec, J.~Nikl, M.~Vranic, S.~Weber, The effect of pre-plasma formation
  under nonlocal transport conditions for ultra-relativistic laser-plasma
  interaction, Plasma Physics and Controlled Fusion 60~(4) (2018) 044019.
\newblock \href {https://doi.org/10.1088/1361-6587/aab05a}
  {\path{doi:10.1088/1361-6587/aab05a}}.

\bibitem{Psikal2021}
J.~Psikal, {Laser-driven ion acceleration from near-critical Gaussian plasma
  density profile}, Plasma Physics and Controlled Fusion 63~(6) (2021) 064002.
\newblock \href {https://doi.org/10.1088/1361-6587/abf448}
  {\path{doi:10.1088/1361-6587/abf448}}.

\bibitem{Psikal2019spie}
J.~Psikal, V.~Horny, M.~Zakova, M.~Matys, Comparison of ion acceleration from
  nonexpanded and expanded thin foils irradiated by ultrashort petawatt laser
  pulse, in: E.~Esarey, C.~B. Schroeder, J.~Schreiber (Eds.), Laser
  Acceleration of Electrons, Protons, and Ions V, Vol. 11037 of Proceedings of
  SPIE, 2019, pp. 7 -- 13.
\newblock \href {https://doi.org/10.1117/12.2520278}
  {\path{doi:10.1117/12.2520278}}.

\bibitem{Batani2010}
D.~Batani, R.~Jafer, M.~Veltcheva, R.~Dezulian, O.~Lundh, F.~Lindau,
  A.~Persson, K.~Osvay, C.~G. Wahlstr{\"{o}}m, D.~C. Carroll, P.~McKenna,
  A.~Flacco, V.~Malka, Effects of laser prepulses on laser-induced proton
  generation, New Journal of Physics 12 (2010).
\newblock \href {https://doi.org/10.1088/1367-2630/12/4/045018}
  {\path{doi:10.1088/1367-2630/12/4/045018}}.

\bibitem{Weber2017}
S.~Weber, S.~Bechet, S.~Borneis, L.~Brabec, M.~Bu\u{c}ka, E.~Chacon-Golcher,
  M.~Ciappina, M.~DeMarco, A.~Fajstavr, K.~Falk, E.-R. Garcia, J.~Grosz, Y.-J.
  Gu, J.-C. Hernandez, M.~Holec, P.~Jane\u{c}ka, M.~Janta\u{c}, M.~Jirka,
  H.~Kadlecova, D.~Khikhlukha, O.~Klimo, G.~Korn, D.~Kramer, D.~Kumar,
  T.~Lastovi\u{c}ka, P.~Lutoslawski, L.~Morejon, V.~Ol\u{s}ovcov\'a, M.~Rajdl,
  O.~Renner, B.~Rus, S.~Singh, M.~\u{S}mid, M.~Sokol, R.~Versaci, R.~Vr\'ana,
  M.~Vranic, J.~Vysko\u{c}il, A.~Wolf, Q.~Yu, {P3: an installation for
  high-energy density plasma physics and ultra-high intensity laser-matter
  interaction at ELI-Beamlines}, {Matter and Radiation at Extremes} 2 (2017)
  149.

\bibitem{ELIBeamlines}
{The~Extreme Light Infrastructure project: ELI Beamlines},
  \url{http://www.eli-beams.eu}.

\bibitem{Danson2019}
C.~N. Danson, C.~Haefner, J.~Bromage, T.~Butcher, J.-C.~F. Chanteloup, E.~A.
  Chowdhury, A.~Galvanauskas, L.~A. Gizzi, J.~Hein, D.~I. Hillier, N.~W. Hopps,
  Y.~Kato, E.~A. Khazanov, R.~Kodama, G.~Korn, R.~Li, Y.~Li, J.~Limpert, J.~Ma,
  C.~H. Nam, D.~Neely, D.~Papadopoulos, R.~R. Penman, L.~Qian, J.~J. Rocca,
  A.~A. Shaykin, C.~W. Siders, C.~Spindloe, S.~Szatm{\'{a}}ri, R.~M. G.~M.
  Trines, J.~Zhu, P.~Zhu, J.~D. Zuegel, {Petawatt and exawatt class lasers
  worldwide}, High Power Laser Science and Engineering 7 (2019) e54.
\newblock \href {https://doi.org/10.1017/hpl.2019.36}
  {\path{doi:10.1017/hpl.2019.36}}.

\bibitem{Balsara2015}
D.~S. Balsara, M.~Dumbser, Divergence-free {MHD} on unstructured meshes using
  high order finite volume schemes based on multidimensional {R}iemann solvers,
  Journal of Computational Physics 299 (2015) 687--715.
\newblock \href {https://doi.org/10.1016/j.jcp.2015.07.012}
  {\path{doi:10.1016/j.jcp.2015.07.012}}.

\bibitem{Susanto2013}
A.~Susanto, L.~Ivan, H.~{De Sterck}, C.~Groth, High-order central {ENO}
  finite-volume scheme for ideal {MHD}, Journal of Computational Physics 250
  (2013) 141--164.
\newblock \href {https://doi.org/10.1016/j.jcp.2013.04.040}
  {\path{doi:10.1016/j.jcp.2013.04.040}}.

\bibitem{Rieben2007}
R.~N. Rieben, D.~A. White, B.~K. Wallin, J.~M. Solberg, {An arbitrary
  Lagrangian–Eulerian discretization of MHD on 3D unstructured grids},
  Journal of Computational Physics 226~(1) (2007) 534--570.
\newblock \href {https://doi.org/10.1016/j.jcp.2007.04.031}
  {\path{doi:10.1016/j.jcp.2007.04.031}}.

\bibitem{Toth2000}
G.~T{\'{o}}th, The {$\nabla${\textperiodcentered}B=0} constraint in
  shock-capturing magnetohydrodynamics codes, Journal of Computational Physics
  161~(2) (2000) 605--652.
\newblock \href {https://doi.org/10.1006/jcph.2000.6519}
  {\path{doi:10.1006/jcph.2000.6519}}.

\bibitem{Dobrev2010}
V.~A. Dobrev, T.~V. Kolev, R.~N. Rieben, High-order curvilinear finite element
  methods for {L}agrangian hydrodynamics, SIAM Journal on Scientific Computing
  34~(5) (2012) B606--B641.
\newblock \href {https://doi.org/10.1137/120864672}
  {\path{doi:10.1137/120864672}}.

\bibitem{Nikl2018eps}
J.~Nikl, M.~Kucha\v{r}\'{i}k, M.~Holec, S.~Weber, Curvilinear high-order
  {L}agrangian hydrodynamic code for the laser-target interaction, in: S.~Coda,
  J.~Berndt, G.~Lapenta, M.~Mantsinen, C.~Michaut, S.~Weber (Eds.), Europhysics
  Conference Abstracts -- 45th EPS Conference on Plasma Physics, Vol. 42A,
  European Physical Society, 2018, p. P1.2019.

\bibitem{Nikl2018}
J.~Nikl, M.~Holec, M.~Zeman, M.~Kucha\v{r}\'{i}k, J.~Limpouch, S.~Weber,
  Macroscopic laser-plasma interaction under strong non-local transport
  conditions for coupled matter and radiation, Matter and Radiation at Extremes
  3 (2018) 110--126.
\newblock \href {https://doi.org/10.1016/j.mre.2018.03.001}
  {\path{doi:10.1016/j.mre.2018.03.001}}.

\bibitem{Holec2018}
M.~Holec, J.~Nikl, S.~Weber, Nonlocal transport hydrodynamic model for laser
  heated plasmas, Physics of Plasmas 25~(3) (2018) 032704.
\newblock \href {https://doi.org/10.1063/1.5011818}
  {\path{doi:10.1063/1.5011818}}.

\bibitem{Lax1976}
M.~Lax, D.~F. Nelson, Maxwell equations in material form, Physical Review B 13
  (1976) 1777--1784.
\newblock \href {https://doi.org/10.1103/PhysRevB.13.1777}
  {\path{doi:10.1103/PhysRevB.13.1777}}.

\bibitem{Arnold2006}
D.~N. Arnold, R.~S. Falk, R.~Winther, Differential complexes and stability of
  finite element methods {I}. {T}he de {R}ham complex, in: D.~N. Arnold, P.~B.
  Bochev, R.~B. Lehoucq, R.~A. Nicolaides, S.~M. (Eds.), Compatible Spatial
  Discretizations, Springer New York, New York, NY, 2006, pp. 23--46.
\newblock \href {https://doi.org/10.1007/0-387-38034-5_2}
  {\path{doi:10.1007/0-387-38034-5_2}}.

\bibitem{Rieben2006}
R.~N. Rieben, D.~A. White, Verification of high-order mixed finite-element
  solution of transient magnetic diffusion problems, IEEE Transactions on
  Magnetics 42~(1) (2006) 25--39.
\newblock \href {https://doi.org/10.1109/TMAG.2005.860127}
  {\path{doi:10.1109/TMAG.2005.860127}}.

\bibitem{Brezzi1991}
F.~Brezzi, M.~Fortin, Mixed and Hybrid Finite Element Methods, Vol.~15 of
  Springer Series in Computational Mathematics, Springer, New York, NY, 1991.
\newblock \href {https://doi.org/10.1007/978-1-4612-3172-1}
  {\path{doi:10.1007/978-1-4612-3172-1}}.

\bibitem{Boffi2013}
D.~Boffi, F.~Brezzi, M.~Fortin, Mixed finite element methods and applications,
  Vol.~44, Springer, 2013.

\bibitem{Thomas1979}
P.~D. Thomas, C.~K. Lombard, Geometric conservation law and its application to
  flow computations on moving grids, AIAA Journal 17~(10) (1979) 1030--1037.
\newblock \href {https://doi.org/10.2514/3.61273} {\path{doi:10.2514/3.61273}}.

\bibitem{Abgrall2017}
R.~Abgrall, S.~Tokareva, Staggered grid residual distribution scheme for
  {L}agrangian hydrodynamics, SIAM Journal on Scientific Computing 39~(5)
  (2017) A2317--A2344.
\newblock \href {https://doi.org/10.1137/16M1078781}
  {\path{doi:10.1137/16M1078781}}.

\bibitem{Abgrall2020}
R.~Abgrall, K.~Lipnikov, N.~Morgan, S.~Tokareva, Multidimensional staggered
  grid residual distribution scheme for {L}agrangian hydrodynamics, SIAM
  Journal on Scientific Computing 42~(1) (2020) A343--A370.
\newblock \href {https://doi.org/10.1137/18M1223939}
  {\path{doi:10.1137/18M1223939}}.

\bibitem{Kolev2009b}
T.~V. Kolev, P.~S. Vassilevski, Parallel auxiliary space {AMG} for {H(Curl)}
  problems, Journal of Computational Mathematics 27~(5) (2009) 604--623.
\newblock \href {https://doi.org/10.4208/jcm.2009.27.5.013}
  {\path{doi:10.4208/jcm.2009.27.5.013}}.

\bibitem{Henson2002}
V.~E. Henson, U.~M. Yang, {BoomerAMG}: A parallel algebraic multigrid solver
  and preconditioner, Applied Numerical Mathematics 41~(1) (2002) 155--177.
\newblock \href {https://doi.org/10.1016/S0168-9274(01)00115-5}
  {\path{doi:10.1016/S0168-9274(01)00115-5}}.

\bibitem{Ruge1987}
J.~W. Ruge, K.~St\"uben, Algebraic multigrid (AMG), SIAM, 1987, Ch.~4, pp.
  73--130.
\newblock \href {https://doi.org/10.1137/1.9781611971057.ch4}
  {\path{doi:10.1137/1.9781611971057.ch4}}.

\bibitem{Sandu2021}
A.~Sandu, V.~Tomov, L.~Cervena, T.~Kolev, Conservative high-order time
  integration for {L}agrangian hydrodynamics, SIAM Journal on Scientific
  Computing 43~(1) (2021) A221--A241.
\newblock \href {https://doi.org/10.1137/20M1314495}
  {\path{doi:10.1137/20M1314495}}.

\bibitem{Torrilhon2003}
M.~Torrilhon, Uniqueness conditions for {R}iemann problems of ideal
  magnetohydrodynamics, Journal of Plasma Physics 69~(3) (2003) 253–276.
\newblock \href {https://doi.org/10.1017/S0022377803002186}
  {\path{doi:10.1017/S0022377803002186}}.

\bibitem{Caramana1998a}
E.~J. Caramana, M.~J. Shashkov, P.~P. Whalen, Formulations of artificial
  viscosity for multi-dimensional shock wave computations, Journal of
  Computational Physics 144~(1) (1998) 70--97.
\newblock \href {https://doi.org/10.1006/jcph.1998.5989}
  {\path{doi:10.1006/jcph.1998.5989}}.

\bibitem{Kolev2009}
T.~V. Kolev, R.~N. Rieben, A tensor artificial viscosity using a finite element
  approach, Journal of Computational Physics 228~(22) (2009) 8336 -- 8366.
\newblock \href {https://doi.org/https://doi.org/10.1016/j.jcp.2009.08.010}
  {\path{doi:https://doi.org/10.1016/j.jcp.2009.08.010}}.

\bibitem{Nedelec1980}
J.~C. Nedelec, Mixed finite elements in {R3}, Numerische Mathematik 35~(3)
  (1980) 315--341.
\newblock \href {https://doi.org/10.1007/BF01396415}
  {\path{doi:10.1007/BF01396415}}.

\bibitem{Glaubitz2019}
J.~Glaubitz, Shock capturing by {B}ernstein polynomials for scalar conservation
  laws, Applied Mathematics and Computation 363 (2019) 124593.
\newblock \href {https://doi.org/https://doi.org/10.1016/j.amc.2019.124593}
  {\path{doi:https://doi.org/10.1016/j.amc.2019.124593}}.

\bibitem{Abgrall2010}
R.~Abgrall, J.~Trefil{\'{i}}k, An example of high order {R}esidual distribution
  scheme using non-{L}agrange elements, Journal of Scientific Computing
  45~(1-3) (2010) 3--25.
\newblock \href {https://doi.org/10.1007/s10915-010-9405-y}
  {\path{doi:10.1007/s10915-010-9405-y}}.

\bibitem{mfem-library}
{MFEM}: Modular finite element methods library, \url{https://mfem.org}.
\newblock \href {https://doi.org/10.11578/dc.20171025.1248}
  {\path{doi:10.11578/dc.20171025.1248}}.

\bibitem{Anderson2019}
R.~Anderson, J.~Andrej, A.~Barker, J.~Bramwell, J.-S. Camier, J.~Cerveny,
  V.~Dobrev, Y.~Dudouit, A.~Fisher, T.~Kolev, W.~Pazner, M.~Stowell, V.~Tomov,
  I.~Akkerman, J.~Dahm, D.~Medina, S.~Zampini, {MFEM}: A modular finite element
  methods library, Computers \& Mathematics with Applications (2020).
\newblock \href {https://doi.org/10.1016/j.camwa.2020.06.009}
  {\path{doi:10.1016/j.camwa.2020.06.009}}.

\bibitem{glvis-tool}
{GLVis}: {OpenGL} finite element visualization tool, \url{https://glvis.org}.
\newblock \href {https://doi.org/10.11578/dc.20171025.1249}
  {\path{doi:10.11578/dc.20171025.1249}}.

\bibitem{Torrilhon2002}
M.~Torrilhon, Exact solver and uniqueness conditions for riemann problems of
  ideal magnetohydrodynamics, Research report 2002-06, ETH Zurich (2002).
\newblock \href {https://doi.org/10.3929/ETHZ-A-004339390}
  {\path{doi:10.3929/ETHZ-A-004339390}}.

\bibitem{Bochev2003}
P.~B. Bochev, J.~J. Hu, A.~C. Robinson, R.~S. Tuminaro, Towards robust {3D}
  {Z}-pinch simulations: discretization and fast solvers for magnetic diffusion
  in heterogeneous conductors, Electronic Transactions on Numerical Analysis 15
  (2003) 186--210.

\bibitem{White2009}
D.~White, Nonphysical reverse currents in transient finite-element magnetics
  simulation, IEEE transactions on magnetics 45~(4) (2009) 1973--1989.

\bibitem{Mouschovias1980}
T.~C. Mouschovias, E.~V. Paleologou, Magnetic braking of an aligned rotator
  during star formation: An exact, time-dependent solution, The Astrophysical
  Journal 237 (1980) 877--899.

\bibitem{Balsara1999}
D.~S. Balsara, D.~S. Spicer, A staggered mesh algorithm using high order
  {G}odunov fluxes to ensure solenoidal magnetic fields in magnetohydrodynamic
  simulations, Journal of Computational Physics 149~(2) (1999) 270--292.
\newblock \href {https://doi.org/10.1006/jcph.1998.6153}
  {\path{doi:10.1006/jcph.1998.6153}}.

\bibitem{Zachary1994}
A.~L. Zachary, A.~Malagoli, P.~Colella, A higher-order {G}odunov method for
  multidimensional ideal magnetohydrodynamics, SIAM Journal on Scientific
  Computing 15~(2) (1994) 263--284.
\newblock \href {https://doi.org/10.1137/0915019} {\path{doi:10.1137/0915019}}.

\bibitem{Kossl1990}
D.~K{\"{o}}ssl, E.~M{\"{u}}ller, W.~Hilldebrandt, Numerical simulations of
  axially symmetric magnetized jets. {I} : The influence of equipartition
  magnetic fields, Astronomy and astrophysics 229~(2) (1990) 378--396.

\bibitem{Sedov1993}
L.~I. Sedov, {Similarity and dimensional methods in mechanics}, CRC press,
  1993.

\bibitem{Livne1985}
E.~Livne, A.~Glasner, {A finite difference scheme for the heat conduction
  equation}, Journal of Computational Physics 58~(1) (1985) 59--66.
\newblock \href {https://doi.org/10.1016/0021-9991(85)90156-1}
  {\path{doi:10.1016/0021-9991(85)90156-1}}.

\bibitem{Ferracina2008}
L.~Ferracina, M.~N. Spijker, Strong stability of singly-diagonally-implicit
  {R}unge–{K}utta methods, Applied Numerical Mathematics 58~(11) (2008)
  1675--1686.
\newblock \href {https://doi.org/10.1016/j.apnum.2007.10.004}
  {\path{doi:10.1016/j.apnum.2007.10.004}}.

\bibitem{Warren1996}
G.~S. Warren, W.~R. Scott, Numerical dispersion of higher order nodal elements
  in the finite-element method, IEEE Transactions on Antennas and Propagation
  44~(3) (1996) 317--320.
\newblock \href {https://doi.org/10.1109/8.486299}
  {\path{doi:10.1109/8.486299}}.

\bibitem{Anderson2004}
R.~W. Anderson, N.~S. Elliott, R.~B. Pember, An arbitrary
  {L}agrangian–{E}ulerian method with adaptive mesh refinement for the
  solution of the {E}uler equations, Journal of Computational Physics 199~(2)
  (2004) 598--617.
\newblock \href {https://doi.org/10.1016/j.jcp.2004.02.021}
  {\path{doi:10.1016/j.jcp.2004.02.021}}.

\bibitem{Karanam2008}
A.~K. Karanam, A p-adaptive stabilized finite element method for fluid
  dynamics, Ph.D. thesis, Rensselaer Polytechnic Institute (2008).

\bibitem{Barlow2016}
A.~J. Barlow, P.-H. Maire, W.~J. Rider, R.~N. Rieben, M.~J. Shashkov, Arbitrary
  {L}agrangian–{E}ulerian methods for modeling high-speed compressible
  multimaterial flows, Journal of Computational Physics 322 (2016) 603--665.
\newblock \href {https://doi.org/10.1016/j.jcp.2016.07.001}
  {\path{doi:10.1016/j.jcp.2016.07.001}}.

\bibitem{Hajduk2020}
H.~Hajduk, D.~Kuzmin, T.~Kolev, V.~Tomov, I.~Tomas, J.~N. Shadid, {Matrix-free
  subcell residual distribution for Bernstein finite elements: Monolithic
  limiting}, Computers \& Fluids 200 (2020) 104451.
\newblock \href {https://doi.org/10.1016/j.compfluid.2020.104451}
  {\path{doi:10.1016/j.compfluid.2020.104451}}.

\bibitem{Anderson2018}
R.~W. Anderson, V.~A. Dobrev, T.~V. Kolev, R.~N. Rieben, V.~Z. Tomov,
  High-order multi-material {ALE} hydrodynamics, SIAM Journal on Scientific
  Computing 40~(1) (2018) B32--B58.
\newblock \href {https://doi.org/10.1137/17M1116453}
  {\path{doi:10.1137/17M1116453}}.

\bibitem{Arefiev2016}
A.~Arefiev, T.~Toncian, G.~Fiksel, Enhanced proton acceleration in an applied
  longitudinal magnetic field, New Journal of Physics 18~(10) (2016) 105011.
\newblock \href {https://doi.org/10.1088/1367-2630/18/10/105011}
  {\path{doi:10.1088/1367-2630/18/10/105011}}.

\bibitem{Walsh2020}
C.~A. Walsh, J.~P. Chittenden, D.~W. Hill, C.~Ridgers,
  Extended-magnetohydrodynamics in under-dense plasmas, Physics of Plasmas
  27~(2) (2020) 022103.
\newblock \href {https://doi.org/10.1063/1.5124144}
  {\path{doi:10.1063/1.5124144}}.

\bibitem{Ridgers2008}
C.~P. Ridgers, R.~J. Kingham, A.~G.~R. Thomas, Magnetic cavitation and the
  reemergence of nonlocal transport in laser plasmas, Physical Review Letters
  100~(7) (2008) 075003.
\newblock \href {https://doi.org/10.1103/PhysRevLett.100.075003}
  {\path{doi:10.1103/PhysRevLett.100.075003}}.

\bibitem{Nikl2021eps}
J.~Nikl, M.~Kucha\v{r}\'{i}k, S.~Weber, Self-generated magnetic fields
  modelling within high-order {L}agrangian magneto-hydrodynamics, in:
  Europhysics Conference Abstracts -- 47th EPS Conference on Plasma Physics,
  European Physical Society, 2021, p. P1.2022.

\bibitem{Tzeferacos2015}
P.~Tzeferacos, M.~Fatenejad, N.~Flocke, C.~Graziani, G.~Gregori, D.~Q. Lamb,
  D.~Lee, J.~Meinecke, A.~Scopatz, K.~Weide, {FLASH} {MHD} simulations of
  experiments that study shock-generated magnetic fields, High Energy Density
  Physics 17 (2015) 24--31.
\newblock \href {https://doi.org/10.1016/j.hedp.2014.11.003}
  {\path{doi:10.1016/j.hedp.2014.11.003}}.

\bibitem{Lesaint1974}
P.~Lasaint, P.~A. Raviart, On a finite element method for solving the neutron
  transport equation, in: Mathematical Aspects of Finite Elements in Partial
  Differential Equations, Vol.~33, Elsevier, 1974, pp. 89--123.
\newblock \href {https://doi.org/10.1016/B978-0-12-208350-1.50008-X}
  {\path{doi:10.1016/B978-0-12-208350-1.50008-X}}.

\end{thebibliography}

\end{document}